\documentclass[a4paper,fleqn]{cas-sc}

\usepackage[authoryear]{natbib}
\usepackage[version=4]{mhchem}
\usepackage{multirow}
\usepackage{chemfig}

\usepackage{chemmacros}
\usechemmodule{reactions}
\usepackage{lineno}

\chemsetup[reactions]{
  before-tag = R ,
  tag-open = ( ,
  tag-close = )
}

\newcommand\aj{\ref@jnl{AJ}}
\newcommand\psj{\ref@jnl{PSJ}}
\newcommand\araa{\ref@jnl{ARA\&A}}
\newcommand\apj{\ref@jnl{ApJ}}
\newcommand\apjl{\ref@jnl{ApJL}}     
\newcommand\apjs{\ref@jnl{ApJS}}
\newcommand\ao{\ref@jnl{ApOpt}}
\newcommand\apss{\ref@jnl{Ap\&SS}}
\newcommand\aap{\ref@jnl{A\&A}}
\newcommand\aapr{\ref@jnl{A\&A~Rv}}
\newcommand\aaps{\ref@jnl{A\&AS}}
\newcommand\azh{\ref@jnl{AZh}}
\newcommand\baas{\ref@jnl{BAAS}}
\newcommand\icarus{\ref@jnl{Icarus}}
\newcommand\jaavso{\ref@jnl{JAAVSO}}  
\newcommand\jrasc{\ref@jnl{JRASC}}
\newcommand\memras{\ref@jnl{MmRAS}}
\newcommand\mnras{\ref@jnl{MNRAS}}
\newcommand\pra{\ref@jnl{PhRvA}}
\newcommand\prb{\ref@jnl{PhRvB}}
\newcommand\prc{\ref@jnl{PhRvC}}
\newcommand\prd{\ref@jnl{PhRvD}}
\newcommand\pre{\ref@jnl{PhRvE}}
\newcommand\prl{\ref@jnl{PhRvL}}
\newcommand\pasp{\ref@jnl{PASP}}
\newcommand\pasj{\ref@jnl{PASJ}}
\newcommand\qjras{\ref@jnl{QJRAS}}
\newcommand\skytel{\ref@jnl{S\&T}}
\newcommand\solphys{\ref@jnl{SoPh}}
\newcommand\sovast{\ref@jnl{Soviet~Ast.}}
\newcommand\ssr{\ref@jnl{SSRv}}
\newcommand\zap{\ref@jnl{ZA}}
\newcommand\nat{\ref@jnl{Nature}}
\newcommand\iaucirc{\ref@jnl{IAUC}}
\newcommand\aplett{\ref@jnl{Astrophys.~Lett.}}
\newcommand\apspr{\ref@jnl{Astrophys.~Space~Phys.~Res.}}
\newcommand\bain{\ref@jnl{BAN}}
\newcommand\fcp{\ref@jnl{FCPh}}
\newcommand\gca{\ref@jnl{GeoCoA}}
\newcommand\grl{\ref@jnl{Geophys.~Res.~Lett.}}
\newcommand\jcp{\ref@jnl{JChPh}}
\newcommand\jgr{\ref@jnl{J.~Geophys.~Res.}}
\newcommand\jqsrt{\ref@jnl{JQSRT}}
\newcommand\memsai{\ref@jnl{MmSAI}}
\newcommand\nphysa{\ref@jnl{NuPhA}}
\newcommand\physrep{\ref@jnl{PhR}}
\newcommand\physscr{\ref@jnl{PhyS}}
\newcommand\planss{\ref@jnl{Planet.~Space~Sci.}}
\newcommand\procspie{\ref@jnl{Proc.~SPIE}}

\newcommand\actaa{\ref@jnl{AcA}}
\newcommand\caa{\ref@jnl{ChA\&A}}
\newcommand\cjaa{\ref@jnl{ChJA\&A}}
\newcommand\jcap{\ref@jnl{JCAP}}
\newcommand\na{\ref@jnl{NewA}}
\newcommand\nar{\ref@jnl{NewAR}}
\newcommand\pasa{\ref@jnl{PASA}}
\newcommand\rmxaa{\ref@jnl{RMxAA}}

\newcommand\maps{\ref@jnl{M\&PS}}
\newcommand\aas{\ref@jnl{AAS Meeting Abstracts}}
\newcommand\dps{\ref@jnl{AAS/DPS Meeting Abstracts}}

\def\tsc#1{\csdef{#1}{\textsc{\lowercase{#1}}\xspace}}
\tsc{WGM}
\tsc{QE}

\begin{document}
\let\WriteBookmarks\relax
\def\floatpagepagefraction{1}
\def\textpagefraction{.001}

\shorttitle{Hydrocarbon shielding and prebiotic chemistry in exoplanet atmospheres}    

\shortauthors{}  

\title [mode = title]{Hydrocarbon complexity and photochemical shielding of prebiotic feedstock molecules in exoplanet atmospheres}  



%

\author[1]{Marrick Braam}[orcid=
https://orcid.org/0000-0002-9076-2361]

\cormark[1]


\ead{marrick.braam@unibe.ch}


\credit{Conceptualisation, Methodology, Formal analysis, Investigation, Writing - Original Draft, Visualisation}

\affiliation[1]{organization={Center for Space and Habitability, University of Bern},
            addressline={Gesellschaftsstrasse 6}, 
            city={Bern},
            postcode={3012}, 
            country={Switzerland}}

\author[2]{Ellery Gopaoco}


\credit{Conceptualisation, Methodology, Formal analysis, Investigation, Writing - Original Draft}

\affiliation[2]{organization={Centre for Exoplanet Science, University of Edinburgh},
            city={Edinburgh},
            postcode={EH9 3FD}, 
            country={UK}}

\author[3,4]{Shang-Min Tsai}[orcid=https://orcid.org/0000-0002-8163-4608]




\credit{Methodology, Investigation, Formal analysis, Writing - Review \& Editing}

\affiliation[3]{organization={Department of Earth and Planetary Sciences, University of California Riverside},
            city={Riverside},
            state={CA},
            country={USA}}

\affiliation[4]{organization={Institute of Astronomy \& Astrophysics, Academia Sinica, Taipei 10617, Taiwan},
            city={Taipei},
            postcode={10617}, 
            country={Taiwan}}
            
\author[5,2]{Gergely Friss}[orcid=https://orcid.org/0009-0008-5099-5698]




\credit{Methodology, Validation, Writing - Review \& Editing}

\affiliation[5]{organization={School of Physics and Astronomy, University of Edinburgh},
            city={Edinburgh},
            postcode={EH9 3FD}, 
            country={UK}}
            
\author[6,2]{Paul I. Palmer}[orcid=
https://orcid.org/0000-0002-1487-0969]




\credit{Conceptualisation, Methodology, Writing- Reviewing and Editing, Funding acquisition}
\affiliation[6]{organization={School of GeoSciences, University of Edinburgh},
            city={Edinburgh},
            postcode={EH9 3FF}, 
            country={UK}}

\author[7]{Paul B. Rimmer}[orcid=
https://orcid.org/0000-0002-7180-081X]




\credit{Methodology, Investigation, Writing- Reviewing and Editing}
\affiliation[7]{organization={Cavendish Laboratory, University of Cambridge},
            addressline={JJ Thomson Avenue}, 
            city={Cambridge},
            postcode={CB3 0HE}, 
            country={UK}}
            
\author[7]{Skyla B. White}[orcid=
https://orcid.org/0009-0005-0027-7426]



\credit{Conceptualisation, Writing - Review \& Editing}

\cortext[1]{Corresponding author}



\begin{abstract}
The potential of prebiotic chemistry to initiate and propagate on an exoplanet fundamentally depends on whether the atmospheric physical and chemical conditions can facilitate the production of prebiotic feedstock molecules. Photochemical simulations of exoplanet atmospheres can be used to explore this potential atmospheric synthesis, but require a comprehensive chemical network for the conditions considered. We present the implementation of the Consistent Reduced Atmospheric Hybrid Chemical Network with Oxygen extension (CRAHCN-O), constructed to simulate the formation of feedstock molecules such as hydrogen cyanide (HCN), formaldehyde (H$_2$CO), and simple hydrocarbons, into the VULCAN photochemical kinetics code. We investigate the photochemical production of feedstock molecules driven by M-star radiation and compare these to predictions by the N-C-H-O network in VULCAN, for N$_2$-dominated atmospheres with C/O ratios between 0.5--1.5 from CO$_2$ and CH$_4$ abundances. Predicted abundances are similar for C/O${=}$0.5 in the absence of CH$_4$. As soon as CH$_4$ is included (i.e., for C/O${>}$0.5), the abundance profiles diverge in the photochemical regions. By analysing the attenuation of UV radiation, we find that hydrocarbon photochemical shielding causes the diverging profiles. CRAHCN-O accumulates ethane (C$_2$H$_6$), while sinks to higher-order hydrocarbons in N-C-H-O accumulate C$_4$H$_3$ (the 1-butene-3-yne-1-yl radical) and  C$_3$H$_4$ (allene). Importantly, C$_2$H$_6$ is photochemically active whereas C$_4$H$_3$ and C$_3$H$_4$ are assumed inactive. With mixing ratios up to a few percent in CRAHCN-O, C$_2$H$_6$ shields CH$_4$ and CO$_2$ from photodissociation. Therefore, these species survive to lower pressures, in turn weakening the destruction of HCN and H$_2$CO. Maximum HCN mixing ratios reach 1000~ppm for the highest CH$_4$ abundances with CRAHCN-O compared to only 3~ppm with N-C-H-O. Other feedstock molecules like cyanoacetylene (HC$_3$N) and acetylene (C$_2$H$_2$) form more efficiently in N-C-H-O. The shielding mechanism and its impact on feedstock molecules persist for radiation from distinct M-star types. We present the key pathways to feedstock molecules, along with photochemical cycles and net reactions to hydrocarbons for each network. These findings can be used to prioritise experimental rate coefficient determination, to reconcile network differences, and to develop more complex climate-chemistry modeling. The results demonstrate the crucial role of chemical kinetics in understanding prebiotic chemistry in exoplanet atmospheres, including important considerations for the construction and applicability of chemical networks.
\end{abstract}


\begin{highlights}
\item The production of prebiotic feedstock molecules varies with exoplanet atmospheric environments and chemical networks.
\item Hydrocarbon shielding significantly affects prebiotic chemical processes.
\item Key pathways identified to guide experimental rate measurements, chemical network reconciliation, and reduction strategies for complex climate–chemistry models.
\item Crucial role of photochemical network structure and kinetics in understanding prebiotic feedstock molecules in exoplanet atmospheres.
\end{highlights}


\maketitle

\section{Introduction} \label{sec:intro}
Evidence for the first appearance of life on Earth dates back to 3.7 Ga, while the planet may have been habitable as early as 4.3~Ga \citep{schopf_first_2006, zahnle_emergence_2007, pearce_constraining_2018, catling_archean_2020}. Early studies on the origin of life proposed that the complex organic chemistry required for its emergence would have necessitated a more reducing environment than Earth's present oxidising atmosphere \citep{haldane1929origin, oparin1938}. This idea formed the basis for the Miller-Urey experiments \citep[][]{miller_production_1953}, which demonstrated that a reducing gas mixture (H$_2$, H$_2$O, CH$_4$, and NH$_3$) could yield simple prebiotic molecules such as hydrogen cyanide (HCN) and formaldehyde (H$_2$CO), when subjected to electrical discharges. In aqueous solution, the simple feedstock molecules can synthesise the building blocks of life \citep[see also][]{bada_new_2013}. Candidate environments to facilitate this synthesis on Early Earth include shallow ponds undergoing wet-dry cycles that concentrate feedstock molecules \citep[e.g.,][]{pearce_origin_2017, damer_hot_2020}, given sufficient atmospheric production and delivery, such as rainout and gravitational settling. Other theories for the origin of life include exogenous delivery of prebiotic feedstock molecules \citep[e.g.,][]{chyba_endogenous_1992} and hydrothermal vent synthesis of prebiotic molecules \citep[][]{shock_geochemical_1990}. The increasing diversity of known rocky exoplanets (e.g., in terms of irradiation by and orbital distance from the host star) encourages investigations of the atmospheric synthesis of feedstock molecules in these exoplanet environments \citep[][]{ranjan_influence_2016, rimmer_origin_2018}. Therefore, understanding the atmospheric chemical and physical conditions that allow for the formation of feedstock molecules and facilitate subsequent increasing chemical complexity is central to exploring the origin of life on Earth and beyond.

Several feedstock molecules are crucial for the synthesis of complex biomolecules. HCN can produce amino acids in the presence of NH$_3$ and aldehydes via Strecker synthesis \citep[][]{strecker1854ueber}. Furthermore, HCN can directly polymerise into nucleobases such as adenine and guanine \citep[][]{ferris1966unusual, sanchez_studies_1967, bada_how_2004} and is a key driver of the cyanosulfidic protometabolism, which has shown that RNA, protein, and lipid precursors can emerge simultaneously \citep[][]{ritson_prebiotic_2012, patel_common_2015, xu_photochemical_2018, rimmer_origin_2018}. Cyanoacetylene (HC$_3$N) is another simple nitrile that plays an essential role in the synthesis of RNA precursors \citep[][]{powner2007prebiotic, powner_synthesis_2009}. H$_2$CO is a simple aldehyde that can aid Strecker synthesis and forms sugars like ribose, which forms the backbone of RNA \citep[][]{butlerow1861formation, cleaves_prebiotic_2008, kim_synthesis_2011, benner_when_2020}. Lastly, acetylene (C$_2$H$_2$) and ethylene (C$_2$H$_6$) can form long hydrocarbon chains, potentially important constituents of primitive cell membranes \citep[][]{balucani_elementary_2009, ruiz-mirazo_prebiotic_2014}. On Titan, increasingly complex organic molecules aggregate to non-volatile solids or hazes that can produce amino acids and nucleobases through hydrolysis \citep[e.g.,][]{khare_amino_1986, poch_production_2012, horst_formation_2012}. Therefore, the formation of HCN, HC$_3$N, H$_2$CO, and hydrocarbons such as C$_2$H$_2$ and C$_2$H$_6$ in representative atmospheric conditions gives insights into the prebiotic potential of the planet in question.

However, the formation of any of these molecules will strongly depend on the environmental conditions \citep[e.g.,][]{bada_like_2002}, as also shown by the Miller-Urey experiments \citep[][]{miller_organic_1959}. Many experiments of atmospheric prebiotic synthesis followed, also considering a background atmosphere dominated by N$_2$ and CO$_2$ \citep[e.g.,][]{miller_atmosphere_1983, miyakawa_prebiotic_2002, cleaves_reassessment_2008}. The experimental studies were complemented by photochemical models of prebiotic synthesis, simulating the photochemical balance of feedstock molecules in a single atmospheric column for the Early Earth \citep[][]{zahnle_photochemistry_1986, tian_revisiting_2011, airapetian_prebiotic_2016, pearce_towards_2022}, Titan \citep[][]{yung_photochemistry_1984, lavvas_coupling_2008, krasnopolsky_photochemical_2009, hebrard_neutral_2012, dobrijevic_1d-coupled_2016, loison_neutral_2015, willacy_new_2016, vuitton_simulating_2019, pearce_hcn_2020}, and rocky exoplanets \citep[][]{rimmer_origin_2018, rimmer_hydrogen_2019}. An important outcome of these is the clear link between HCN production and CH$_4$ abundances, which was further quantified by \citet{pearce_experimental_2022}. HCN abundances for N$_2$-dominated rocky planets vary considerably with C/O ratio, with \citet{rimmer_hydrogen_2019} finding low HCN production for C/O${\leq}$0.5, mixing ratios between 1--1000~ppm for 0.5${<}$C/O${\leq}$1.5, and mixing ratios exceeding 0.1\% for C/O${>}$1.5. The abundances also depend strongly on the UV irradiance, particularly short wavelengths: 50--100~nm and 100--200~nm photons drive HCN production (N$_2$ photolysis) and destruction (H$_2$O photolysis), respectively \citep[][]{rimmer_hydrogen_2019}. Lightning can be an additional source of feedstock molecules like HCN \citep[e.g.,][]{chameides_rates_1981, barth_effect_2024}. Lightning is advantageously a tropospheric phenomenon, providing a more straightforward source of feedstock molecules to aqueous surface environments where once again the UV environment dictates further complexity \citep[e.g.,][]{ranjan_influence_2016}.

Considering modern Earth, we see considerable spatial and temporal variations in chemical abundances (e.g., ozone and methane distributions), caused by a combination of photochemistry, atmospheric circulation, and spatially varying emissions. For rocky exoplanets, diverse orbital configurations and host stars affect the climate \citep[e.g.,][]{joshi_simulations_1997, boutle_exploring_2017, del_genio_habitable_2019, turbet_trappist-1_2022}, atmospheric circulation \citep[e.g.,][]{haqq-misra_demarcating_2018, carone_stratosphere_2018, sergeev_bistability_2022}, and photochemically active regions \citep[e.g.,][]{proedrou_characterising_2016, chen_biosignature_2018, yates_ozone_2020, braam_earth-like_2025}. Additionally, the effects of energetic processes such as stellar flares from M-dwarfs \citep[][]{segura_effect_2010, airapetian_prebiotic_2016, chen_persistence_2021, ridgway_3d_2023} and lightning \citep[][]{ardaseva_lightning_2017, braam_lightning-induced_2022, barth_effect_2024} will affect the atmospheric chemistry with spatial and temporal variations. The complex interplay between all these processes is studied using 3D coupled climate-chemistry models and, in turn, determines the appearance of spectroscopic observations of exoplanet atmospheres \citep[e.g.,][]{chen_persistence_2021, cooke_variability_2023, braam_earth-like_2025}. We can use the predicted environmental conditions of exoplanets to understand the atmospheric formation of feedstock molecules and determine the circumstellar abiogenesis zone, the zone in which life can originate for a given scenario \citep[][]{rimmer_origin_2018, rimmer_origins_2023}. \citet{claringbold_prebiosignature_2023} showed that many of these feedstock molecules have absorption signatures in mid-infrared wavelengths, likely below current detection limits but relevant for future missions specifically targeting terrestrial exoplanets, such as the Large Interferometer For Exoplanets \citep[][]{quanz_atmospheric_2022}. Solar System detections of nitriles include HCN and HC$_3$N by Voyager in the IR on Titan \citep{hanel_infrared_1981, kunde_c4h2_1981}, as well as HCN on Neptune \citep{rosenqvist_millimeter-wave_1992, marten_first_1993}, Jupiter \citep{marten_collision_1995}, and Saturn \citep{benmahi_first_2022}.

To anticipate future exoplanet atmospheric observations and connect atmospheric synthesis of feedstock molecules to known exoplanet parameters (irradiation, orbit), photochemical models are applied. These models (1D or 3D) fundamentally depend on the chemical kinetics and the implemented chemical network (the set of photo- and thermochemical reactions). Therefore, our understanding of the underlying chemical processes and knowledge of reaction rate coefficients is crucial to these photochemical studies \citep{winiberg_how_2025}. Uncertainties in a single or multiple measurements of rate coefficients, as well as extrapolation outside the measured temperature and pressure ranges, affect predictions from a chemical network. Limitations on network sizes, including the lumping together of multi-step reactions into a single one, further influence network predictions. The scenario of interest is also relevant: chemical networks of Solar System atmospheres are topologically distinct, with unique features for Earth's network presenting a potential biosignature \citep[][]{sole_large-scale_2004, wong_toward_2023, fisher_complex_2025}. For oxidised environments like modern Earth, chemical networks were constructed decades ago and regularly improved \citep[e.g.,][]{chapman_xxxv_1930, bates_photochemistry_1950, crutzen_influence_1970, logan_tropospheric_1981, jacob_heterogeneous_2000, archibald_description_2020}. Chemical networks were also developed to explain observed disequilibrium chemical abundances in Jupiter's atmosphere \citep[e.g.,][]{fegley_chemical_1994, moses_photochemistry_2005, visscher_deep_2010} and for the high temperatures of hot Jupiters \citep[e.g.,][]{zahnle_atmospheric_2009, moses_disequilibrium_2011, venot_chemical_2012}. The reducing environment of Early Earth has similarities to Titan, and thus chemical networks for the Early Earth \citep[][]{zahnle_photochemistry_1986, pavlov_uv_2001, pearce_crahcn-o_2020} incorporate relevant reactions from chemical networks for Titan \citep[][]{yung_photochemistry_1984, hebrard_photochemical_2007, lavvas_coupling_2008, krasnopolsky_photochemical_2009, loison_neutral_2015, dobrijevic_1d-coupled_2016, willacy_new_2016, vuitton_simulating_2019}. Other networks were constructed to treat reducing to oxidising conditions \citep[][]{rimmer_chemical_2016, tsai_comparative_2021}. \citet{rimmer_chemical_2016} collect all the available rate coefficient data into a large network, which was benchmarked for several scenarios. The Consistent Reduced Atmospheric Hybrid Chemical Network with Oxygen extension (CRAHCN-O) was constructed to simulate the formation of simple feedstock molecules in planetary atmospheres and is a relatively small network of a few hundred reactions \citep[][]{pearce_consistent_2019, pearce_crahcn-o_2020, pearce_hcn_2020}. Quantum chemical calculations were used to explore previously unknown reactions and determine rate coefficients, and the network was applied to the atmospheres of Titan and Early Earth \citep[e.g.,][]{pearce_hcn_2020, pearce_towards_2022}. Regardless of choices in network construction, the uncertainties in rate coefficients present an issue for chemical networks and their applications \citep[][]{lira-barria_darwen_2024}.

Specifically for the implementation in 3D atmosphere models, the computational cost rises rapidly with a larger number of species and/or reactions, making small networks like CRAHCN-O attractive. Several approaches for network reduction, which aim to simplify large chemical networks without compromising accuracy, have been applied in the context of exoplanets. These include a pathway analysis to inform the network reduction \citep[][]{tsai_toward_2018, tsai_mini-chemical_2022, lee_mini-chemical_2023}, detailed kinetic analyses with commercial software \citep[][]{venot_reduced_2019}, and a genetic algorithm to quantify network sensitivity and find optimal solutions for reduction \citep[][]{lira-barria_darwen_2024}. Pathway analyses find the shortest path to form species $b$ from species $a$, which depend on the local pressure, temperature, and actinic flux \citep[][]{tsai_toward_2018}. Beyond network reductions, pathway analyses enable the comparison of different chemical networks when focusing on a particular species of interest.

In this paper, we present the implementation of the CRAHCN-O chemical network into the VULCAN photochemical kinetics code \citep[][]{tsai_vulcan_2017, tsai_comparative_2021}. We test the atmospheric production of hydrocarbons and feedstock molecules for planets orbiting M stars and compare this to predictions from the standard N-C-H-O network in VULCAN. We present our simulation setup and describe both chemical networks in Section~\ref{sec:methods}. In Section~\ref{sec:results}, we compare the abundances predicted by both networks, describe the role of photochemical shielding, and identify the key pathways and chemical mechanisms affecting feedstock molecules. In Section~\ref{sec:disc_conc}, we demonstrate the effect of the shielding mechanism, put our results into context, and conclude our study.

\section{Methods} \label{sec:methods}
\subsection{Modelling framework}\label{sec:modelframework}
VULCAN is an open-source 1D photochemical kinetics code that solves the mass continuity equations for the temporal evolution of the number density of chemical species \citep[][]{tsai_vulcan_2017, tsai_comparative_2021}. The reactions in the chosen chemical network define the interactions between different species and are driven by the background temperature-pressure and vertical mixing profiles, a set of boundary conditions, and the incoming stellar radiation. In what follows, we describe the details of our simulation setup and changes to the chemical network and refer the reader to \citet{tsai_vulcan_2017, tsai_comparative_2021} for an extensive description of VULCAN. 

\begin{figure}
\centering
\includegraphics[width=0.6\columnwidth]{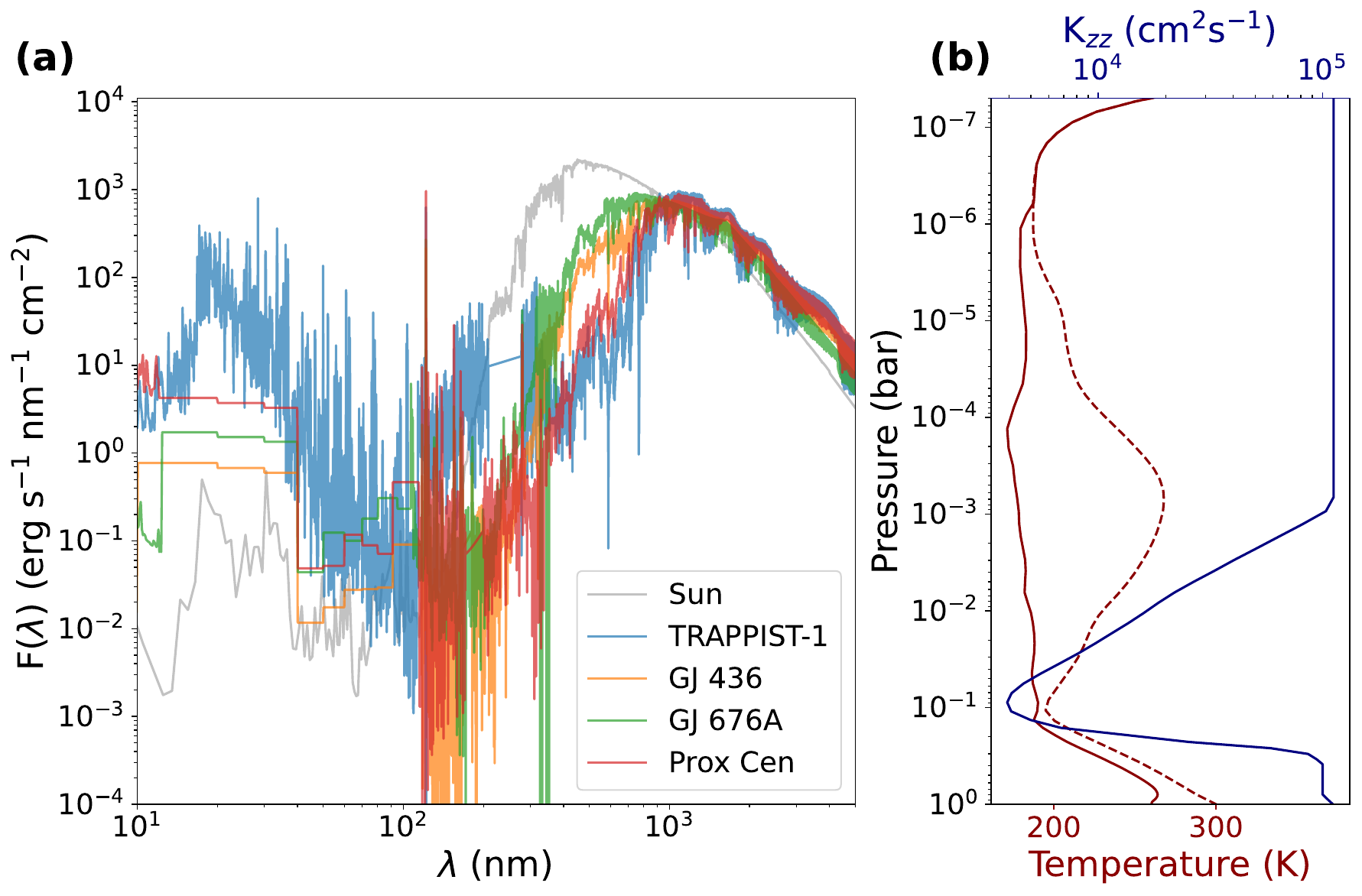}
\caption{Background stellar and planetary environments: (a) the spectral radiant flux density as received by Proxima Centauri b and planets around different M-type stars receiving the same irradiance, compared to the Solar flux at 1~AU. Panel (b) shows vertical profiles of the temperature from \citet{braam_lightning-induced_2022} (solid line) compared to Earth's temperature profile (dashed line) and eddy diffusion ($K_{zz}$), following \citet{tsai_comparative_2021} and \citet{massie1981stratospheric}. }
\label{methods-fig:pcb_ptkzz_flux}
\end{figure}

In this study, we simulate the atmosphere of Proxima Centauri b \citep{anglada2016terrestrial}, the nearest exoplanet to Earth at 1.3~pc, to benchmark production of hydrocarbons and feedstock molecules. Proxima Centauri b is expected to be tidally locked to its host star at an orbital distance of 0.0485~AU and with a period of 11.186~days \citep[e.g.,][]{goldreich1966spin, barnes2017tidal}. Following \citet{turbet2016habitability}, we use a radius of 1.1~R$_\oplus$ and gravity of 10.90~m~s$^{-2}$. For the stellar spectrum of Proxima Centauri (or GJ 551), we use version 2.2 from the MUSCLES spectral library\footnote{\href{https://archive.stsci.edu/prepds/muscles/}{https://archive.stsci.edu/prepds/muscles/}} \citep[][]{france2016muscles, youngblood2016muscles, loyd2016muscles}{}{} as shown in Figure~\ref{methods-fig:pcb_ptkzz_flux}a. Comparing the spectral radiant flux density $F(\lambda)$ received by Proxima Centauri b with that received by Earth, we see that the M5.5V spectrum peaks at higher wavelengths and that Proxima Centauri b receives substantially larger fluxes in the near-UV and far-UV wavelength regions \citep[e.g.,][]{ribas2017full}{}{}. We consider the quiescent stellar spectrum as the driver of photochemistry, but note that stellar activity enhances the stellar UV and X-ray fluxes \citep[][]{vida2019flaring}{}{}, potentially affecting the space weather and atmospheric chemistry \citep[][]{garraffo2016space, airapetian2020impact, ridgway20233d}{}{} of the planet. We include additional simulations using the stellar spectra of GJ 436 (M2.5V) from the MUSCLES survey, as well as GJ 676 A (M0V) and TRAPPIST-1 (M8V) from the Mega-MUSCLES survey \citep[][]{wilson_mega-muscles_2021, wilson_mega-muscles_2024}, to explore the sensitivity of network differences to different spectral energy distributions. For each star, we simulate the photochemistry at the distance for which the irradiance equals that received by Proxima Centauri b, whilst keeping all the other parameters the same. The irradiance criterion results in semi-major axes of 0.35 AU (GJ 676 A), 0.18 AU (GJ 436), and 0.028 AU (TRAPPIST-1) and gives the spectral radiant flux densities in Figure~\ref{methods-fig:pcb_ptkzz_flux}a.

The temperature profile follows from simulations with a 3D coupled climate-chemistry model as presented in \citet{braam_lightning-induced_2022}, assuming a surface pressure of 1 bar. We calculate the mean dayside temperature -- since this is the region with active photochemistry -- and follow the temperature profile of Earth for pressures lower than 6$\times$10$^{-7}$~bar (see the solid red line in Figure~\ref{methods-fig:pcb_ptkzz_flux}b). For comparison, Figure~\ref{methods-fig:pcb_ptkzz_flux}b also shows the mean temperature profile of Earth's atmosphere. The troposphere is visible for P${>}$0.1~bar (the tropopause). Between 0.1--10$^{-5}$~bar, we see the relatively stable stratosphere and mesosphere. For P${<}10^{-6}$~bar, we see the rising temperatures characteristic of the thermosphere. Except for the absence of a stratospheric inversion -- due to varying strengths of radiative heating with stellar type \citep[][]{godolt20153d, kozakis2022ozone, deluca2024impact}{}{} -- the temperature profile is qualitatively similar to Earth. The eddy diffusion coefficient ($K_{zz}$) parametrises vertical mixing as a 1D diffusion process and is shown as the navy line in Figure~\ref{methods-fig:pcb_ptkzz_flux}b. Since no universal derivation of the $K_{zz}$ profile currently exists \citep[see e.g.][]{baeyens2021grid}{}{} and because of the similarities with the Earth's temperature profile, we opt for the Earth-like $K_{zz}$ profile following \citet{tsai_comparative_2021} and \citet{massie1981stratospheric}.

We consider N$_2$-dominated atmospheres of 1~bar containing: 1) 400~ppm CO$_2$, 2) 1\%~CO$_2$ based on \citet{arney_pale_2017}, and 3) 10\%~CO$_2$ following \citet{rimmer_hydrogen_2019}, as shown in Table~\ref{tab:merged_comp}. For each CO$_2$ scenario, we vary the amount of CH$_4$ to probe the carbon-to-oxygen (C/O) ratio from 0.5 to 1.5, based on the reported sensitivity of HCN chemistry to C/O ratios within this range \citep[][]{rimmer_hydrogen_2019}. The initial mixing ratios are constant with pressure, except for H$_2$O, which is fixed at the surface to a mixing ratio of 0.00894 (corresponding to 25\% relative humidity). Except for the fixed surface H$_2$O, all initial mixing ratios evolve chemically in the simulations. Given the H$_2$O and CO$_2$ profiles, the 400~ppm CO$_2$ case is the most radiatively consistent for the temperature profile in Figure~\ref{methods-fig:pcb_ptkzz_flux}. Note that we keep the temperature and $K_{zz}$ profiles fixed to focus on changes in the chemistry. Because of the adjusted N$_2$ mixing ratios for the 1\% and 10\% CO$_2$ cases, the carbon-to-nitrogen (C/N) ratio also changes, as shown in Table~\ref{tab:merged_comp}. The variations in C/N ratio are most prominent for the 10\%~CO$_2$ case, since CO$_2$ and CH$_4$ account for up to 20\% of the atmosphere. C/N variations are small for the 1\%~CO$_2$ and 400~ppm~CO$_2$ case, as N$_2$ mixing ratios do not go below 96\% and 99.9\%, respectively. Hence, beyond comparing the effect of C/O ratios, the contrast between the three setups can show the effect of varying C/N ratios on the final chemical abundances. Since the initial abundance of CH$_4$ is varied with C/O ratio, we also change the H/O ratio. For the photochemistry around M stars other than Proxima Centauri, we use the same setups but only consider C/O ratios of 0.5, 1.0, and 1.5. The runs exclude surface emissions or deposition of chemical species, to focus on differences caused by employing distinct chemical networks.

\begin{table}
\centering
\caption{Initial compositions for all CO$_2$ cases (400ppm, 1\%, and 10\% CO$_2$). The 400 ppm case is most radiatively consistent with Figure~\ref{methods-fig:pcb_ptkzz_flux}, 1\%~CO$_2$ case is based on \citet{arney_pale_2017}, and the 10\%~CO$_2$ case on \citet{rimmer_hydrogen_2019}}
\label{tab:merged_comp}
\begin{tabular}{l|l|l|l|l|l|l|l|l}
\hline
\textbf{CO$_2$ Level} & \textbf{C/O} & \textbf{C/N} & \textbf{H/O} & \textbf{N$_2$} & \textbf{H$_2$O} & \textbf{O$_2$} & \textbf{CO$_2$} & \textbf{CH$_4$} \\ \hline

\multirow{7}{*}{400ppm} 
& 0.5 & 2e-4 & 2e-4& 0.999 & 1e-6 & 0.0 & 4e-4 & 0.0 \\
& 0.75 & 3e-4 & 1.001 & 0.999 & 1e-6 & 0.0 & 4e-4 & 2e-4 \\
& 1.0 & 4e-4 & 2.000 & 0.999 & 1e-6 & 0.0 & 4e-4 & 4e-4 \\
& 1.25 & 5e-4 & 2.999 & 0.999 & 1e-6 & 0.0 & 4e-4 & 6e-4 \\
& 1.5 & 6e-4 & 3.998 & 0.999 & 1e-6 & 0.0 & 4e-4 & 8e-4 \\ \hline

\multirow{7}{*}{1\%} 
& 0.5 & 5.1e-3 & 1e-4 & 0.98 & 1e-6 & 1e-8 & 0.01 & 0.0 \\
& 0.75 & 7.7e-3 & 1.000 & 0.98 & 1e-6 & 1e-8 & 0.01 & 0.005 \\
& 1.0 & 1.03e-2 & 2.000 & 0.97 & 1e-6 & 1e-8 & 0.01 & 0.01 \\
& 1.25 & 1.29e-2 & 3.000 & 0.97 & 1e-6 & 1e-8 & 0.01 & 0.015 \\
& 1.5 & 1.56e-2 & 4.000 & 0.96 & 1e-6 & 1e-8 & 0.01 & 0.02 \\ \hline

\multirow{7}{*}{10\%} 
& 0.5 & 5.62e-2 & 1e-5 & 0.89 & 1e-6 & 0.0 & 0.1 & 0.0 \\
& 0.75 & 8.93e-2 & 1.000 & 0.84 & 1e-6 & 0.0 & 0.1 & 0.05 \\
& 1.0 & 0.127 & 2.000 & 0.79 & 1e-6 & 0.0 & 0.1 & 0.1 \\
& 1.25 & 0.169 & 3.000 & 0.74 & 1e-6 & 0.0 & 0.1 & 0.15 \\
& 1.5 & 0.217 & 4.000 & 0.69 & 1e-6 & 0.0 & 0.1 & 0.2 \\ \hline
\end{tabular}
\end{table}

\subsection{Chemical Networks}\label{subsec:networks}
\subsubsection{VULCAN N-C-H-O}
The first version of VULCAN included the C-H-O network \citep{tsai_vulcan_2017}, which \citet{tsai_comparative_2021} extended to also include nitrogen and sulfur chemistry. Here, we use the full version of the N-C-H-O network, covering reducing to oxidising conditions and including network updates from \citet{tsai_biogenic_2024}. This version of N-C-H-O contains a total of 417 (gas-phase) chemical reactions that connect 73 species, including 349 bimolecular and 68 termolecular reactions. Additionally, 57 photodissociation reactions are considered. All the thermochemical reactions are reversed using an equilibrium constant \citep[][]{heng_atmospheric_2016, tsai_vulcan_2017}. Generally, reaction rate coefficients are taken from the NIST database\footnote{\href{https://kinetics.nist.gov/kinetics/}{https://kinetics.nist.gov/kinetics/}}, the KIDA database\footnote{\href{https://kida.astrochem-tools.org/}{https://kida.astrochem-tools.org/}}, and literature sources including \citet{moses_photochemistry_2005, lavvas_coupling_2008, moses_disequilibrium_2011}, and \citet{zahnle_photolytic_2016}. Many thermochemical reactions are valid for a wide temperature range (300--2500~K), although the experimental data for some are limited to narrow temperature ranges. The condensation of H$_2$O is also included. The N-C-H-O network has been evaluated for hot Jupiters HD 189733 b and HD 209458 b (using other photochemical models) and for measured abundances in the atmospheres of Jupiter and present-day Earth. 

Hydrocarbons are generally truncated at 2 carbon atoms due to limitations of the kinetics data \citep[][]{venot_new_2015}, except for select higher-order species up to C6 that act as sinks for C2 hydrocarbons and potential haze precursors, particularly benzene \citep[][]{tsai_comparative_2021}. The top panel of Figure~\ref{fig:network_comparison} shows the species in N-C-H-O as nodes and the reactions between them as edges. Atomic hydrogen is clearly central to this network and involved in the most reactions (86). HCN is involved in 39 reactions. H$_2$CO is less central to the chemical processes in the network, with a total of 31 reactions. We see the inclusion of some higher-order hydrocarbons -- those with three or more carbon atoms -- such as allene (C$_3$H$_4$ or CH$_2$CCH$_2$), diacetylene (C$_4$H$_2$), 1-butene-3-yne-1-yl radical (C$_4$H$_3$), benzene (C$_6$H$_6$). However, the peripheral location and limited number of reaction connections (up to 8) of most of these higher-order hydrocarbons indicate their role as sinks for the C2 species. Several hydrocarbons with fewer than three carbon atoms are more central to the chemical network and participate in a significantly larger number of reactions: examples are CH$_4$ with 43 reactions, methyl radical (CH$_3$) with 79, methylene (CH$_2$) with 46, methylidyne radical (CH) with 36, acetylene (C$_2$H$_2$) with 49, and ethane (C$_2$H$_6$) with 15.

\subsubsection{CRAHCN-O}
The CRAHCN-O network\footnote{\label{fn:crahcno} \href{https://github.com/bennski/CRAHCN}{https://github.com/bennski/CRAHCN}}  is a relatively small chemical network developed to model the production of feedstock molecules in planetary atmospheres \citep[][]{pearce_consistent_2019, pearce_crahcn-o_2020, pearce_hcn_2020}. The coupling between computational quantum chemistry with canonical variational transition state theory and Rice-Ramsperger-Kassel-Marcus/master equation theory allows the exploration of previously unknown reactions and the consistent calculation of reaction rate coefficients for thermochemical reactions \citep[][]{pearce_consistent_2019, pearce_crahcn-o_2020, pearce_hcn_2020}. The resulting reactions are collected in the CRAHCN-O network that focuses on the accurate simulation of HCN and H$_2$CO production in CO$_2$-, N$_2$-, H$_2$O-, CH$_4$-, and H$_2$-dominated atmospheres and are valid for temperatures of 50--400~K. CRAHCN-O contains experimental data where available and is supplemented by the consistently calculated quantum chemical rate coefficients where experimental data is missing \citep[][]{pearce_crahcn-o_2020}. Updates to the CRAHCN-O were presented in later work by \citet{pearce_towards_2022} and \citet{pearce_experimental_2022}. The CRAHCN-O network has been used to model HCN on Titan \citep[][]{pearce_hcn_2020} and HCN and H$_2$CO on the early Earth \citep[][]{pearce_experimental_2022, pearce_towards_2022}.

\begin{figure*}
    \centering
    \includegraphics[width=0.65\linewidth]{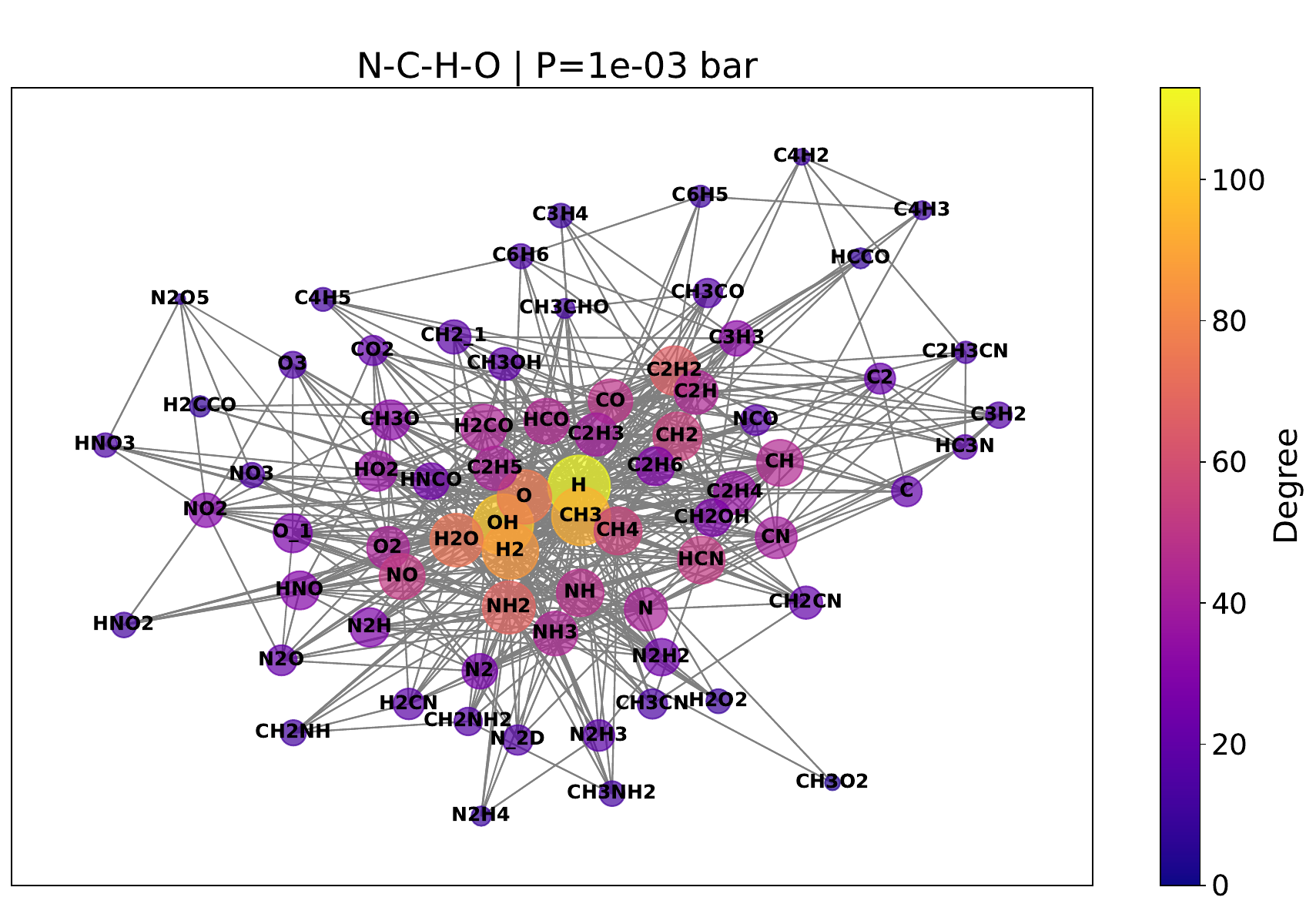}
    \includegraphics[width=0.65\linewidth]{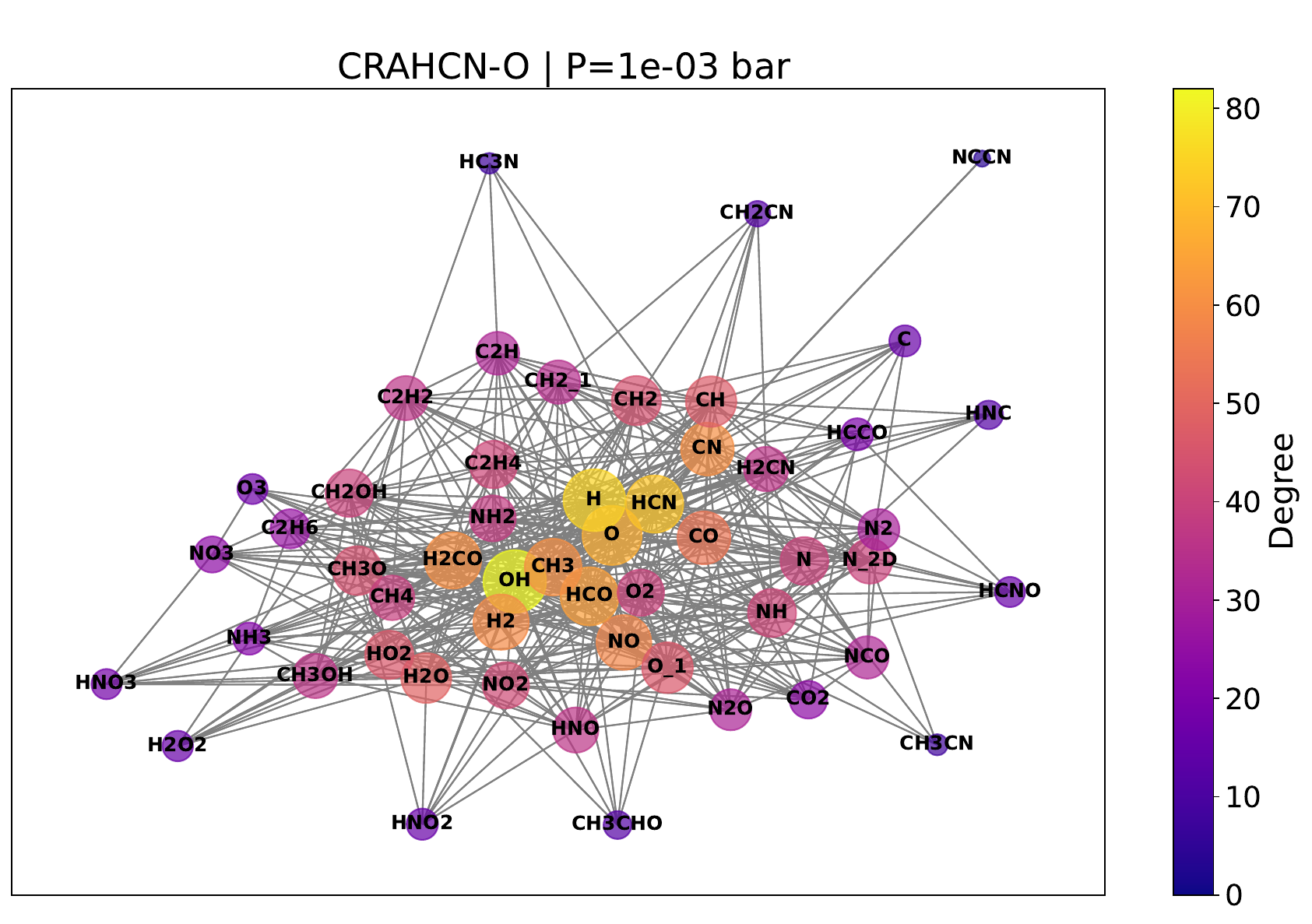}
    \caption{Graph networks comparing the chemical networks used in this study: N-C-H-O (top) and CRAHCN-O (bottom), for a reference pressure of $1\times10^{-3}$ bar. Chemical species are represented as nodes, whereas reactions between species are shown as edges with their length corresponding to the inverse reaction rate. The clustering and position of species give a relative measure of their connectivity. The colour of the species relates to their degree, or the number of reaction connections to other species. The size of nodes represents the eigenvector centrality, which measures the importance of a species by taking into account the number of reactions it is involved with, the rates of these reactions, and the connections to reactive species. CRAHCN-O has an additional slow pathway from C$_2$H$_4$ to C$_2$H$_3$, which was omitted from the plot for readability.}
    \label{fig:network_comparison}
\end{figure*}

Here, we use the publicly available version 3 of CRAHCN-O$^{\ref{fn:crahcno}}$ that includes further updates. In total, the network contains 399 forward gas-phase reactions, consisting of 336 bimolecular and 63 termolecular reactions. All of these are reversed in VULCAN. Figure~\ref{fig:network_comparison} shows that OH (involved in 72 reactions) and H (62 reactions) are the most important in the CRAHCN-O network. CRAHCN-O has more reactions involving HCN (59) and H$_2$CO (41), both of which also have a central position in the network. Because of its focus on simple feedstock molecules, CRAHCN-O simulates hydrocarbons only up to two carbon atoms, resulting in fewer reaction connections for species like CH$_4$ (31), CH$_3$ (48), CH$_2$ (37), and C$_2$H$_2$ (26), as compared to N-C-H-O. CH (40) and C$_2$H$_6$ (13) are exceptions with a higher and similar number of reactions, respectively. 

\subsubsection{Implementing CRAHCN-O into VULCAN}
The rate coefficients in CRAHCN-O follow the Arrhenius law at a reference temperature of 300~K:
\begin{equation}\label{eq:rrates_crahcno}
    k = A^{'}~ \left( \frac{T}{300}\right) ^b~exp\left(-\frac{E_a^{'}}{T}\right).
\end{equation}
Since VULCAN uses the Arrhenius law at a reference temperature of 1~K, we need to incorporate the reference temperature into $A=A^{'}/300^{b}$, which was done for both the bimolecular and termolecular reactions. A total of 49 bimolecular reactions that were initially added in \citet{pearce_experimental_2022} are presented inconsistently for a reference temperature of 1~K and with $E^{'}_a$ in cal~mol$^{-1}$, whereas $E_a$ in VULCAN is already normalised by the gas constant and thus has units $K$. For these 49 reactions, we normalise $E^{'}_a$ from CRAHCN-O by the specific gas constant in the appropriate units ($1.987204258641$~cal~mol$^{-1}$~K$^{-1}$) to have $E_a$ in units K and do not need to correct $A$ for the reference temperature. CRAHCN-O specifies the low-pressure parameters of the Arrhenius equation for termolecular reactions using N$_2$, CO$_2$, or H$_2$ as a third body, which makes the network valid for a range of atmospheric compositions. Since our experiments mainly concern N$_2$-dominated atmospheres, we implemented the rate coefficients for N$_2$ as a third body. 

HNC (hydrogen isocyanide, an isomer of HCN) was included as a new species for CRAHCN-O. The reactions in CRAHCN-O involving HCOO, HCOOH, and CH$_2$OO were omitted (13 in total), since they lack the NASA polynomial data required for reversing the rates \citep[see Appendix~E of][]{tsai_vulcan_2017}. For the sake of network comparison, we consider the same photodissociation reactions in both networks where possible. Since several species that are included in the N-C-H-O network are not present in CRAHCN-O, this results in the removal of the photolysis of HNCO, N$_2$H$_4$ (2 branches), N$_2$O$_5$ (2 branches), C$_6$H$_6$ (2 branches), and C$_4$H$_2$ for the CRAHCN-O implementation. 

VULCAN calculates the reverse reaction rates from the forward rate coefficients and equilibrium constant \citep[][]{tsai_vulcan_2017}. In CRAHCN-O, several reverse reactions have specified rate coefficients, as listed in Table~\ref{tab:duplicate_reactions}. For these duplicate reactions, we compared the rate coefficients with those in the NIST Chemical Kinetics Database \footnote{Standard Reference Database 17, Version 7.1 (Web Version), Release 1.6.8: \href{https://kinetics.nist.gov/kinetics/}{https://kinetics.nist.gov/kinetics/}}, and chose the coefficients from the most recent review that were measured in the relevant temperature range (180--400~K). The chosen coefficients are shown in bold and their respective references are also listed in Table~\ref{tab:duplicate_reactions}. Initial testing of the CRAHCN-O network with standard VULCAN settings for convergence parameters revealed numerical instabilities in the chemical solver when using the CRAHCN-O network. This was especially observed for the 100~ppm CO$_2$ cases, suggesting an important role for trace gases in the chemical evolution in CRAHCN-O. Therefore, we reduced the absolute tolerance to $10^{-8}$, slowing down the runs but increasing their accuracy. Previous simulations with the ChemKM photochemical kinetics model and CRAHCN-O use an absolute tolerance of $10^{-99}$ \citep[][]{pearce_towards_2022}. The relative tolerance is set to 0.05, in line with typical values \citep[][]{tsai_vulcan_2017}.

\begin{table}
    \centering
    \caption{Duplicate reactions in CRAHCN-O and, where available, their coefficients as given in N-C-H-O. Mostly, these reactions are represented by a set of rate coefficients for the forward (first column) and reverse reaction (second column). Occasionally, the reaction is simply duplicated, with the same coefficients in both columns. The H$_2$CN coefficients are specified for two distinct temperature ranges. The reaction used for the VULCAN adaptation is highlighted in bold. References are indicated by numbers in the last column and listed as footnotes below the table.}
    \tiny
    \label{tab:duplicate_reactions}
    \begin{tabular}{l|l|l|l|l|l|l|c}
        \hline
        Reaction & \multicolumn{3}{c|}{[$A$ (cm$^{3}$s$^{-1}$ or cm$^{6}$s$^{-1}$), $b$, $E_a$ (K)]} & \multicolumn{3}{c|}{(k$_{298}$) (cm$^{3}$s$^{-1}$ or cm$^{6}$s$^{-1}$)} & Ref \\
         & CRAHCN-O & CRAHCN-O 2 & N-C-H-O & CRAHCN-O & CRA. 2 & N-C-H-O & \\ \hline
        \ce{O + H2CN -> HCNO + H} & \textbf{[3.13e-17, 1.72, 788]} & [9.8e-11, 0, 0] & [N/A] & 4.01e-14 & 9.8e-11 & N/A & 1 \\
        \ce{HCN + O -> NCO + H} & \textbf{[5.44e-14, 0.96, 4039.8]} & [4.15e-11, 0.2, 5743.2] & [4.5e-15, 1.58, 13400] & 1.67e-17 & 5.53e-19 & 1.08e-30 & 2 \\
        \ce{H2O + CH -> OH + CH2} & \textbf{[6.64e-14, 0.38, 2178]} & [7.6e-13, 0, 0] & [1.43e-18, 2.02, 3410]$^{\dagger}$ & 3.87e-16 & 7.6e-13 & 1.53e-18 & 3 \\
        \ce{O + CH4 -> CH3 + OH } & \textbf{[1.5e-15, 1.56, 4269.8]} & [4.97e-12, -0.29, 4139] & [1.16E-19, 2.2, 2240]$^{\dagger}$ & 6.5e-18 & 8.85e-19 & 1.75e-17 & 4 \\
        \ce{O + H2 -> OH + H} & [\textbf{8.51e-20, 2.67, 3163.2}] & [8.11e-21, 2.8, 1950.0] & [8.52E-20, 2.67, 3160] & 8.44e-18 & 9.88e-17 & 8.54e-18 & 5 \\
        \ce{O(^1D) + H2CN -> HCNO + H}$^*$ & [2.13e-10, -0.39,1070] & \textbf{[1.41e-26, 5.3, -356]} & [N/A] & 6.37e-13 & 6.04e-13 & N/A  & 6 \\
        \ce{O(^1D) + H2CN -> OH + HCN}$^*$ & [1.52e-9, -0.81, 1386] & \textbf{[5.57e-25, 4.56, -46]} & [N/A] & 1.44e-13 & 1.25e-13 & N/A & 7 \\
        \ce{HCN + H -> HNC + H} & \textbf{[9.59e-14, 1.2, 6249]} & [9.28e-13, 0.8, 649] & [N/A] & 6.98e-20 & 1.00e-11 & N/A & 8 \\        
        \ce{CH4 + CH2 -> CH3 + CH3} & \textbf{[1.9e-31, 7.45, 3401]} & [5.3e-11, 0, 0] & [4.09e-18, 2, 4162] & 5.69e-18 & 5.3e-11 & 3.12e-19 & 9 \\
        \ce{H + CH4 -> CH3 + H2} & \textbf{[2.19e-20, 3, 4045]} & [1.14e-20, 2.74, 4740] & [2.2e-20, 3, 4040] & 7.38e-19 & 8.48e-21 & 7.54e-19 & 10 \\
        \ce{CH2 + H2 -> H + CH3} & [3.28e-36, 9.04, 1450] & \textbf{[3.68e-14, 1.15, 6529]} & [7.32E-19, 2, 3699] & 5.88e-16 & 7.87e-21 & 2.64e-19 & 11 \\
        \ce{H + CH2 -> CH + H2} & \textbf{[2e-10, 0, 0]} & [3.1e-10, 0, 1650] & [1.1e-11, 0, -900] & 2e-10 & 1.22e-12 & 2.25e-10 & 12 \\
        \ce{NH2 + OH -> NH3 + O} & [1.83e-21, 2.6, 869.6] & \textbf{[1.6e-11, 0, 3669]} & [1.6e-11, 0, 3669]$^{\dagger}$ & 2.68e-16 & 7.2e-17 & 7.20e-17 & 13 \\     
        \ce{CH3OH + N -> CH3 + HNO} & \textbf{[3.99e-10, 0, 4330]} & [4e-10, 0, 4330] & [N/A] & 1.95e-16 & 1.95e-16 & N/A & 14 \\
        \ce{CH3OH + OH -> H2CO + H2O + H} & \textbf{[3.01e-16, 1.44, 56]} & [2.98e-16, 1.44, 56] & [N/A] & 9.12e-13 & 9.03e-13 & N/A & 15 \\
        \ce{HO2 + NO -> OH + NO2}& [1.2e-13, 0, 0] & \textbf{[3.01e-11, 0, 3361]} & [3.6e-12, 0, -270] & 1.2e-13 & 3.81e-16 & 8.91e-12 & 16 \\
        \ce{NO2 + N -> N2O + O} & \textbf{[5.8e-12, 0, -220]} & [5.8e-12, 0, -220] & [5.8e-12, 0, -220] & 1.21e-11 & 1.21e-11 & 1.21e-11 & 17 \\
        \ce{NO3 + CH3O -> H2CO + HNO3 } & \textbf{[1.5e-12, 0, 0]} & [1.5e-12, 0, 0] & [N/A] & 1.5e-12 & 1.5e-12 & N/A & 18 \\
        \ce{NO2 + CH3O -> H2CO + HNO2 } & \textbf{[3e-13, 0, 0]} & [3e-13, 0, 0] & [N/A] & 3e-13 & 3e-13 & N/A & 19 \\
        \ce{CH3 + OH -> CH2OH + H} & [1.2e-12, 0, 2760] & \textbf{[1.6e-10, 0, 0]} & [1.6e-10, 0, 0]$^{\dagger}$ & 1.14e-16 & 1.6e-10 & 1.6e-10 & 20 \\
        \ce{CH3O + CH3OH} & \textbf{[5e-13, 0, 2050]} & [5e-13, 0, 2050] & [N/A] & 5.14e-16 & 5.14e-16 & N/A & 21 \\
        \ce{-> CH2OH + CH3OH} & & & & & & & \\
        \hline
    \end{tabular}

    {    [1]{\citealt{pearce_crahcn-o_2020}}
    [2]{\citealt{pearce_crahcn-o_2020}}
    [3]{\citealt{pearce_crahcn-o_2020}}
    [4]{\citealt{baulch_evaluated_1992}}
    [5]{\citealt{baulch_evaluated_1992}}
    [6]{\citealt{pearce_crahcn-o_2020}}
    [7]{\citealt{pearce_crahcn-o_2020}}
    [8]{\citealt{pearce_consistent_2019}}
    [9]{\citealt{pearce_consistent_2019}}
    [10]{\citealt{baulch_evaluated_1992}}
    [11]{\citealt{pearce_consistent_2019}}
    [12]{\citealt{pearce_crahcn-o_2020}}
    [13]{\citealt{baulch_evaluated_1992}}
    [14]{\citealt{roscoe_reactions_1973}}
    [15]{\citealt{srinivasan_high-temperature_2007}}
    [16]{\citealt{tsang_chemical_1991}}
    [17]{\citealt{demore_1997}}
    [18]{\citealt{kukui1995aldehyde}}
    [19]{\citealt{atkinson_evaluated_1989}}
    [20]{\citealt{tsang_chemical_1987}}
    [21]{\citealt{tsang_chemical_1987}} \\
    $^*$ The two sets of rate coefficients represent two temperature ranges, 50--200~K and 200--400~K, respectively.\\
    $^\dagger$ The coefficients in the N-C-H-O network represent the reverse reaction.}
\end{table}

\subsection{Pathway Analysis}\label{methods:pathways}
The conversion of one chemical species to another usually involves a multi-step reaction with intermediate species, together forming a pathway or cycle between them. Examples of such pathways on Earth are the Chapman mechanism forming ozone from molecular oxygen \citep[][]{chapman_xxxv_1930} or the catalytic cycles destroying ozone back into oxygen \citep[][]{bates_photochemistry_1950, crutzen_influence_1970}. Other Solar System examples include the production of hydrocarbons on Titan \citep[][]{yung_photochemistry_1984, lavvas_coupling_2008, krasnopolsky_photochemical_2009, hebrard_photochemistry_2013, dobrijevic_1d-coupled_2016, vuitton_simulating_2019}, seasonal CH$_4$ variations on Mars potentially explained by heterogeneous HCl chemistry \citep[][]{taysum_observed_2024}, and the CO--CH$_4$ conversion on Jupiter \citep[][]{prinn_carbon_1977, visscher_deep_2010}. Chemical cycles are also important for the CO-CH$_4$ conversion on brown dwarfs \citep[][]{zahnle_methane_2014} or hot Jupiters \citep[][]{moses_disequilibrium_2011} and can explain the observed SO$_2$ abundances for WASP-39b \citep[][]{tsai_photochemically_2023}. 

Here, we follow \citet{tsai_toward_2018} and perform a pathway analysis based on Dijkstra's algorithm \citep[][]{dijkstra_note_1959} to find the fastest paths in graph theory. In considering a chemical network as a graph, the chemical species represent nodes and reactions represent edges connecting the nodes, as depicted in Figure~\ref{fig:network_comparison}. A faster reaction implies a shorter edge, proportional to $1/r_j$, where $r_j$ is the rate of reaction $j$ in molecules~cm$^{-3}$s$^{-1}$. Based on the summed inverse rates between node A to node B (considering all intermediate nodes), Dijkstra's algorithm then finds the shortest paths between nodes A and B in the network. The chemical timescales for different intermediate steps in the pathways can differ by orders of magnitude, with the slowest reaction rate representing the bottleneck and thus the rate-limiting step \citep[][]{moses_disequilibrium_2011} of a pathway. The simulated reaction rates depend on rate coefficients, temperature, pressure, actinic flux, and chemical abundances, which are reflected in the pathways and their rate-limiting steps \citep[][]{tsai_mini-chemical_2022}. A similar pathway analysis tool has been used to reduce the VULCAN N-C-H-O network for hot Jupiters to the rate-limiting steps of pathways as net reactions for an application in 3D climate-chemistry models \citep[][]{tsai_mini-chemical_2022, lee_mini-chemical_2023} and allowed to quantify photochemical pathways to SO$_2$ in the atmosphere of WASP-39 b \citep[][]{tsai_photochemically_2023}. Our study mainly focuses on pathways to produce feedstock molecules: from N$_2$, CH$_4$, and CO$_2$ to HCN and HC$_3$N, from CH$_4$ and CO$_2$ to H$_2$CO, and from CH$_4$ to hydrocarbons including C$_2$H$_2$, C$_2$H$_6$, C$_3$H$_4$, and C$_4$H$_3$. 

\section{Results} \label{sec:results}
We start this Section by qualitatively analysing the vertical distributions of chemical species in each of the simulations and demonstrating the effects of photochemical shielding. We then present the abundances of species as a function of C/O ratio and show that the main network differences hold for other M-star types. Lastly, we compare reaction pathways to important feedstock molecules and present the chemical mechanisms that explain the hydrocarbon distributions for both networks.

\subsection{Vertical Distributions}\label{sec:vertdistributions}
\begin{figure}
        \centering
        \includegraphics[width=\textwidth]{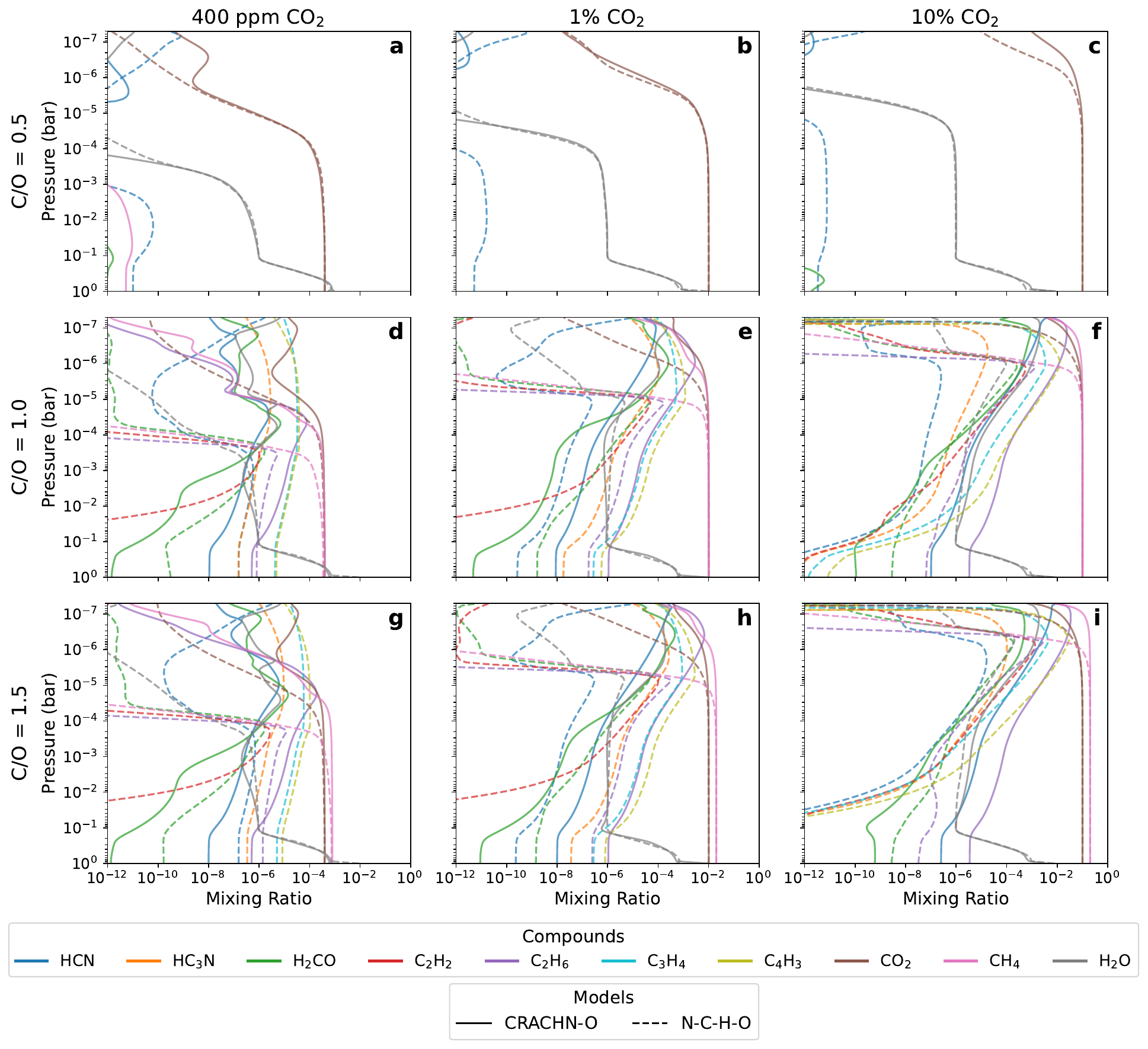}
    \caption{Grid of selected mixing ratios versus atmospheric pressure for VULCAN/CRAHCN-O and VULCAN/N-C-H-O simulations. The horizontal rows correspond to C/O=0.5 (a-c), 1.0 (d-f), and 1.5 (g-i), respectively. Each column shows one of the background CO$_2$ cases: 400~ppm (left), 1\% (middle), and 10\% (right). Solid lines correspond to the VULCAN/CRAHCN-O and dashed lines to the VULCAN/N-C-H-O simulations. The selected chemical compounds are part of the initialisation (CO$_2$, CH$_4$, H$_2$O) and prebiotically relevant (HCN, H$_2$CO, C$_2$H$_2$, C$_2$H$_6$, C$_4$H$_3$).}
    \label{fig:all_vert_distribs}
\end{figure}

Following the initialisation with N$_2$, H$_2$O, O$_2$, CO$_2$, and CH$_4$, we show the steady state abundance profiles from the simulations in Figure~\ref{fig:all_vert_distribs}, for different C/O ratios (rows) and the background scenarios of 400~ppm, 1~\%, and 10~\% CO$_2$ (columns). To compare the predictions from the CRAHCN-O network (solid) and N-C-H-O network (dashed) we selected initialisation species, feedstock molecules, and hydrocarbons based on the network comparison in Section~\ref{subsec:networks}.

For C/O=0.5, CO$_2$ is the main C-bearing species since CH$_4$ is initialised at zero abundance (see Table~\ref{tab:merged_comp}). The profiles of the most abundant species (CO$_2$ in brown and H$_2$O in grey) show good agreement between both chemical networks across the three CO$_2$ backgrounds. From the HCN profiles (blue), we see that the N-C-H-O network forms HCN for a range of pressures in the absence of CH$_4$. For CRAHCN-O, the upper atmosphere (${>}10^{-5}$~bar) shows HCN production. We further explore this inter-network difference in Section~\ref{subsec:pathways_nitriles}. Especially for the 400~ppm CO$_2$ cases, CO$_2$ abundances in the upper atmosphere are higher for CRAHCN-O. This is related to strong HCO formation, which allows regeneration of CO$_2$.

When CH$_4$ becomes significant at C/O${>}$0.5, the vertical profiles of most species begin to diverge between the two chemical networks. The maximum pressure of the CO$_2$ profile diverging scales with the initial CO$_2$ and CH$_4$ abundances. In the 400~ppm CO$_2$ scenario, CO$_2$ abundances are similar for P${>}10^{-4}$~bar. For lower pressures with N-C-H-O, CO$_2$ photodissociation causes decreasing abundances. Higher abundances of CO$_2$ persist for P${<}10^{-4}$~bar when using the CRAHCN-O network. In the 1\% and 10\% CO$_2$ scenarios, CO$_2$ vertical profiles are similar for P${>}10^{-5}$ and P${>}10^{-6}$~bar, respectively. For lower pressures, we again simulate fast photochemical destruction with the N-C-H-O network and significantly higher abundances with CRAHCN-O. Evidently, photochemical processes must be responsible for the divergence in CO$_2$ profiles. With the higher initial CO$_2$ (and CH$_4$) when moving from left to right columns in the plot, the enhanced attenuation of stellar radiation keeps the photochemical impact limited to smaller portions of the low-pressure upper atmosphere regions. The difference seems to indicate that a strong UV-absorbing species (at similar wavelengths to CO$_2$) forms in substantial abundances for CRAHCN-O when CH$_4$ is present at C/O${>}$0.5, shielding CO$_2$ from photodissociation. 

The inter-network differences in CH$_4$ (pink) profiles also depend on atmospheric pressure, with profiles that are similar for relatively high pressures and diverge strongly in the photochemically dominated low-pressure regions. For N-C-H-O, the photochemical destruction of CH$_4$ extends to higher pressures than that of CO$_2$. Again, the increased CH$_4$ abundance from left to right columns moves the photochemical regions upwards to lower pressure. However, the inter-network differences show that CH$_4$ abundances persist to much lower pressures for CRAHCN-O. Evidently, the UV-absorbing species that forms considerably from CH$_4$ with CRAHCN-O in turn shields CH$_4$ from the photochemical destruction seen with N-C-H-O. At first glance, H$_2$O profiles seem to show similar behaviour as CH$_4$. In both networks, the photochemical destruction of H$_2$O moves to lower pressures with increasing CH$_4$ abundances. Nevertheless, inter-network differences in H$_2$O are not just due to photochemical shielding: further investigation shows that CRAHCN-O has net production of H$_2$O in the upper atmosphere through the reaction $\ce{OH + CH2 -> H2O + CH}$, which increases with increasing CH$_4$ abundance. Additional tests (not shown) with the other rate coefficients for this reaction in Table~\ref{tab:duplicate_reactions} resulted in decreased H$_2$O abundances but negligible effects for other compounds in Figure~\ref{fig:all_vert_distribs} since CO$_2$ is the main O-bearing species in all cases.

The differences in CH$_4$ and CO$_2$ profiles, as well as the shielding species, affect the profiles of feedstock molecules such as HCN, HC$_3$N (orange), and H$_2$CO (green). First, the production of these feedstock molecules will depend on where the photochemistry of CH$_4$, N$_2$, and CO$_2$ take place. Second, as illustrated by the accumulation of HCN and H$_2$CO at low pressures (P${<}10^{-4}$ for the 400~ppm case, ${<}10^{-5}$ for the 1\% case, ${<}10^{-6}$~bar for the 10\% case) with CRAHCN-O, photochemically driven destruction processes of the feedstock molecules also weaken due to the attenuation of stellar radiation by photochemical shielding. Importantly for HCN, O and NO are much more abundant at low pressures for N-C-H-O, indicating that oxidising species from H$_2$O and CO$_2$ photolysis drive the destruction of HCN \citep[see e.g.,][]{rimmer_hydrogen_2019}. For the 10\% CO$_2$ case specifically, HCN abundances disappear for P${>}10^{-2}$~bar. For N-C-H-O at higher pressures (P${<}10^{-4}$ for the 400~ppm case, ${<}10^{-5}$ for the 1\% case, ${<}10^{-6}$~bar for the 10\% case), we see higher HCN mixing ratios than those in CRAHCN-O for 400~ppm CO$_2$ but lower mixing ratios for 1\% and 10\% CO$_2$. Notably different are the HC$_3$N profiles: it forms considerably in N-C-H-O, including at low pressures, but only reaches mixing ratios up to 10$^{-13}$ in CRAHCN-O. In Section~\ref{subsec:pathways_nitriles}, we investigate the causes of the distinct vertical profiles of nitriles.

The profiles of C$_2$H$_2$ (red) reveal stark inter-network differences. Generally, the N-C-H-O network produces significantly more C$_2$H$_2$ than CRAHCN-O, with the peak in the abundances at P${\sim}10^{-4}$ for the 400~ppm CO$_2$ case, ${\sim}10^{-5}$ for 1~\% CO$_2$, and ${\sim}10^{-6}$~bar for 10~\% CO$_2$. The pressure levels agree with the onset of photochemical CH$_4$ destruction. For the CRAHCN-O network, either C$_2$H$_2$ production pathways are very slow or C$_2$H$_2$ destruction pathways are much faster compared to N-C-H-O. A detailed analysis of these pathways will follow in Section~\ref{subsec:pathways_hydrocarbons}. C$_2$H$_6$ (purple) consistently forms more abundantly in the CRAHCN-O network, with abundances that are 1--2 orders of magnitude higher for P${>}10^{-4}$~bar (400~ppm CO$_2$ scenario), P${>}10^{-5}$~bar (1\% CO$_2$ scenario), and P${>}10^{-6}$~bar (10\% CO$_2$ scenario). The C$_2$H$_6$ abundances increase even further for lower pressures in the CRAHCN-O network, exceeding mixing ratios of 1\% in the 10\% CO$_2$ scenario, contrary to the strong decrease in the N-C-H-O network. Seemingly, differences in either C$_2$H$_6$ production or destruction allow for the upper atmospheric buildup with CRAHCN-O and strong depletion with N-C-H-O. A possible explanation is that CRAHCN-O simulates hydrocarbons only up to C$_2$H$_6$ (see Figure~\ref{fig:network_comparison}), allowing for its accumulation. N-C-H-O includes higher-order hydrocarbons such as C$_3$H$_3$, C$_3$H$_4$, C$_4$H$_2$,  C$_4$H$_3$, C$_4$H$_5$, or C$_6$H$_6$, albeit with limited coverage of their potential interactions (see Figure~\ref{fig:network_comparison}). C$_4$H$_3$ (yellow line in Figure~\ref{fig:all_vert_distribs}) and C$_3$H$_4$ (cyan) are the most abundant hydrocarbons with N-C-H-O. C$_4$H$_3$ and C$_3$H$_4$ abundance profiles follow a similar shape, with mixing ratios up to one order of magnitude higher for the former.

\begin{figure}
        \centering
        \includegraphics[width=0.6\textwidth]{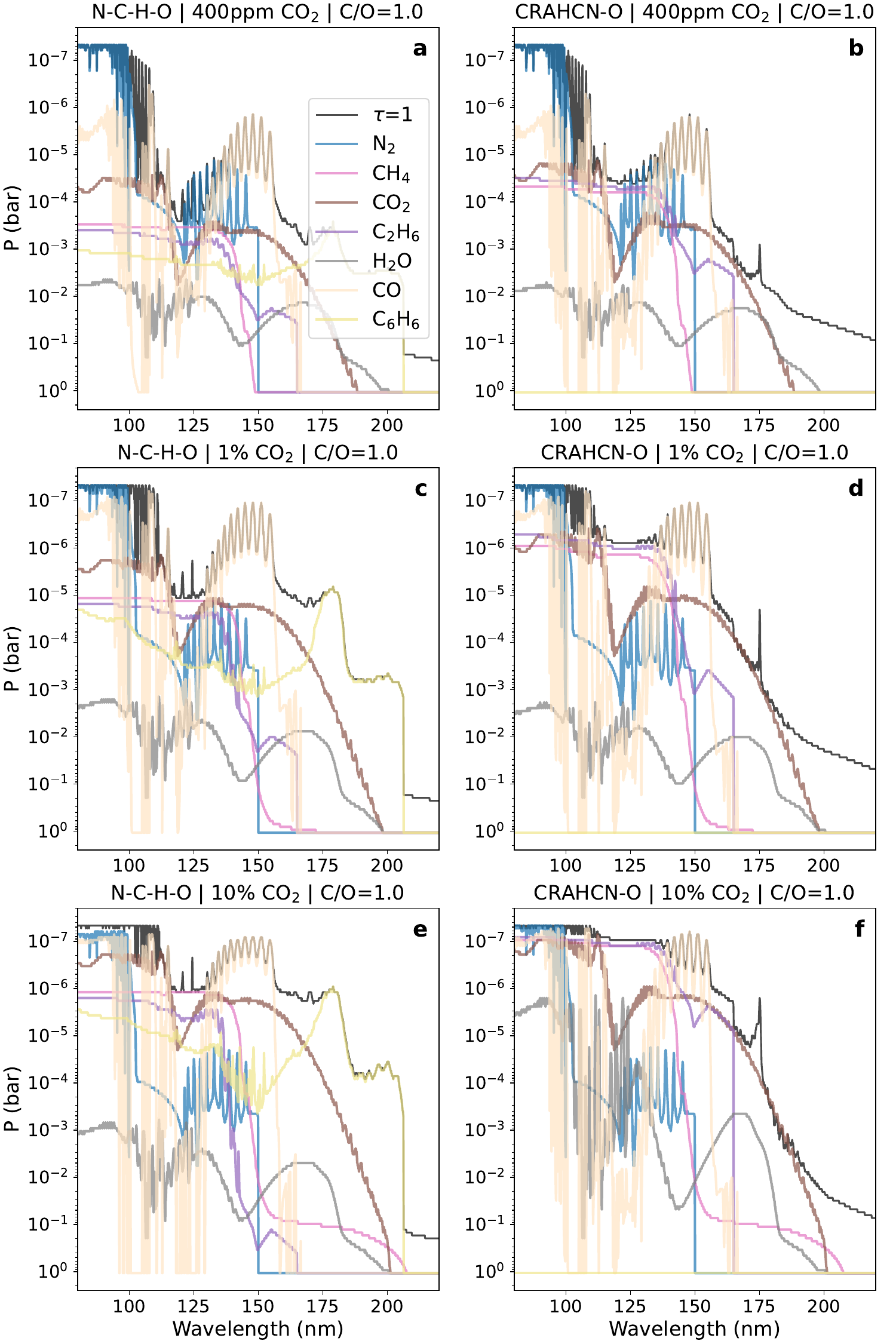}
    \caption{The UV photosphere, indicating the pressure level at which the atmosphere gets optically thick (optical depth $\tau=1$) to incoming radiation, along with the individual contributions by important absorbing species for these wavelength ranges and background composition. Shown are the simulations with the N-C-H-O (a, c, e) and CRAHCN-O (b, d, f) network for the three CO$_2$ scenarios at C/O=1.0.}
    \label{fig:all_photosphere}
\end{figure}

The C$_2$H$_6$ abundance in the upper atmosphere is crucial, due to its known UV shielding effects \citep[e.g.,][]{yoshida_self-shielding_2024}, potentially affecting the distributions of all the other species in Figure~\ref{fig:all_vert_distribs}. We further investigate the photochemical shielding by plotting the UV photosphere in Figure~\ref{fig:all_photosphere}. The black lines indicate the pressure levels at which the atmosphere becomes optically thick to stellar radiation at these wavelengths, and coloured lines denote the individual contributions of important absorbing species. First focusing on the top row, we see that photons with $\lambda{<}$115~nm are predominantly absorbed by N$_2$, CO, and CO$_2$. For the N-C-H-O network at 400~ppm (Figure~\ref{fig:all_photosphere}a), the same species dominate the absorption of photons with $\lambda{>}$115~nm until scattering takes over for $\lambda{>}$180~nm (not shown). However, some narrow wavelength regions are visible around 122 and 126~nm where CH$_4$ absorption dominates. Photons of these wavelengths can penetrate the atmosphere to ${\sim}5{\times}10^{-4}$~bar before being absorbed. We see that this agrees with the onset of the strong decrease in CH$_4$ abundances with decreasing pressure, from the CH$_4$ profiles (dashed pink) in Figures~\ref{fig:all_vert_distribs}d and g. For CRAHCN-O in Figure~\ref{fig:all_photosphere}b, N$_2$, CO, and CO$_2$ similarly dominate the photon absorption at most wavelengths. However, an important difference with N-C-H-O is the shift of C$_2$H$_6$ absorption (violet) to lower pressure levels. C$_2$H$_6$ and CH$_4$ absorb photons at similar wavelengths, and the enhanced abundance with CRAHCN-O drives C$_2$H$_6$ absorption to lower pressures than CH$_4$, indeed shielding CH$_4$ from photodissociation. The same shielding applies to photolysis of CO$_2$, decreasing the abundances of O($^3$P) and NO and thus reducing the oxidising capacity of the atmosphere at low pressures. Moreover, C$_2$H$_6$ absorption dominates the UV photosphere in a few narrow wavelength regions between 115--145~nm and shifts the UV photosphere to lower pressures. Again in Figure~\ref{fig:all_vert_distribs}d and g, the shielding by C$_2$H$_6$ explains why the onset of photodissociation starts at a lower pressure of ${\sim}5{\times}10^{-5}$~bar.

For the 1\% (Figure~\ref{fig:all_photosphere}c--d) and 10\% CO$_2$ (Figure~\ref{fig:all_photosphere}e--f) cases, N$_2$, CO, and CO$_2$ again dominate for a broad range of wavelengths, but the inter-network differences between 115--145~nm are evident. With N-C-H-O, CH$_4$ dominates photon absorption between 115--132~nm to ${\sim}10^{-5}$~bar (1\% CO$_2$, Figure~\ref{fig:all_photosphere}c) and ${\sim}10^{-6}$~bar (10\% CO$_2$, Figure~\ref{fig:all_photosphere}e). Once again, these pressures agree with the pressure level at which the CH$_4$ starts to decrease dramatically with decreasing pressure (Figure~\ref{fig:all_vert_distribs}). For CRAHCN-O, the photosphere is at a relatively low P${\sim}5{\times}10^{-6}$~bar (1\% CO$_2$, Figure~\ref{fig:all_photosphere}d) and ${\sim}2{\times}10^{-7}$~bar (10\% CO$_2$, Figure~\ref{fig:all_photosphere}f). At the same time, C$_2$H$_6$ absorption dominates photon absorption at lower pressures than CH$_4$, again shielding the latter from photodissociation and thus supporting higher CH$_4$ (as well as CO$_2$) abundances down to the pressure levels of the UV photosphere. In turn, the shielding of CH$_4$ and CO$_2$ affects the photochemical balance of feedstock molecules, whilst C$_2$H$_6$ (and the enhanced CH$_4$ and CO$_2$) also directly shields feedstock molecules from photochemical destruction.

In Appendix~\ref{app:ARGO_comp}, we present an additional comparison to simulations with ARGO \citep[][]{rimmer_chemical_2016, 2021PSJ.....2..133R}. ARGO solves the photochemistry-transport equation using a Lagrangian framework and includes the extensive STAND2020 chemical network of over 6000 reactions for 480 species, including ion-neutral chemistry. By comparing results for the 1\% and 10\% CO$_2$ scenarios, we find that the inclusion of ion chemistry in ARGO affects the photochemically active regions. Consequently, ARGO has strong CO$_2$ photochemical destruction down to higher pressures, and the effectiveness of photochemical shielding by hydrocarbons for C/O${>}$0.5 lies in between that of N-C-H-O and CRAHCN-O. Using ARGO, peak HCN abundances agree best with CRAHCN-O, HC$_3$N with N-C-H-O, and H$_2$CO abundances are lowest out of all models. Hydrocarbons accumulate near the top of the atmosphere as C$_3$H$_4$, which is not treated as photochemically active. These results further emphasise the need for enhanced understanding of hydrocarbon chemistry and the associated kinetic data in these scenarios.

\subsection{Trends with C/O Ratio} \label{sec:coratios}
\begin{figure}
\centering
\includegraphics[width=0.6\columnwidth]{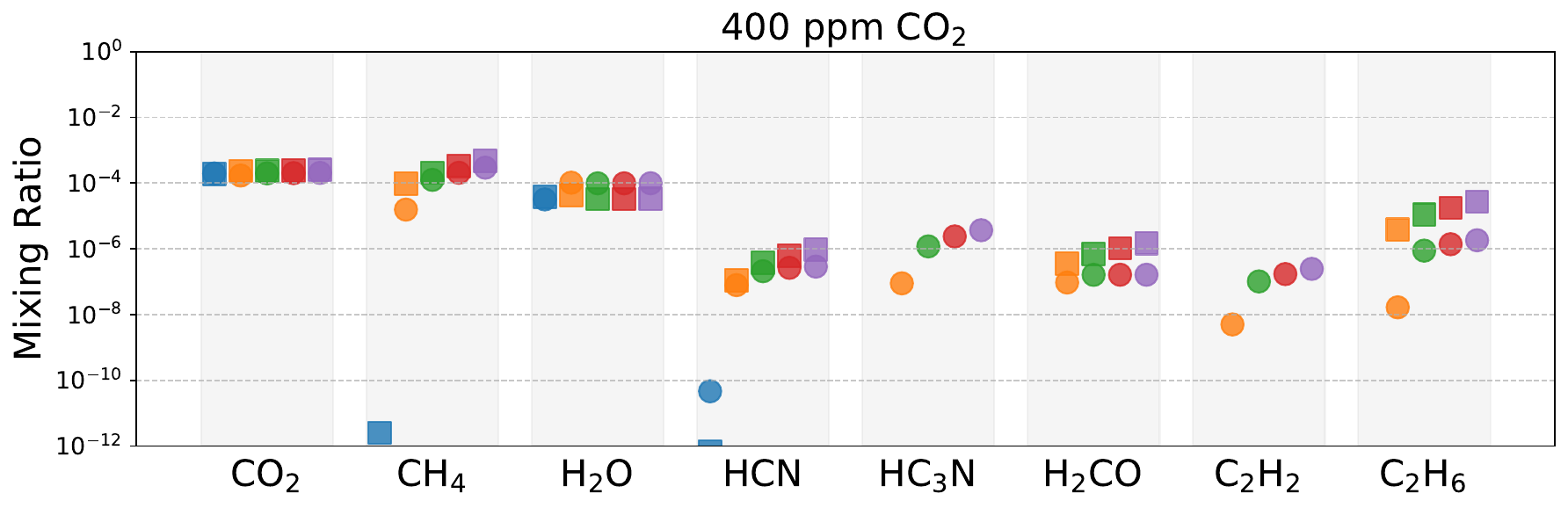}
\includegraphics[width=0.6\columnwidth]{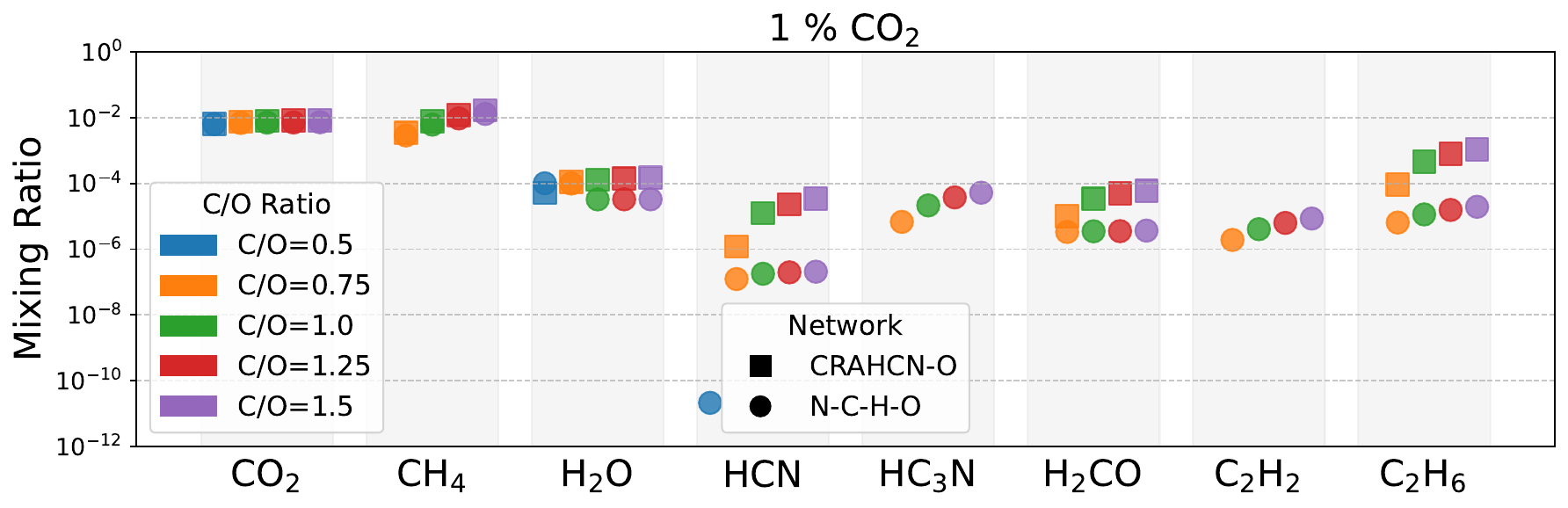}
\includegraphics[width=0.6\columnwidth]{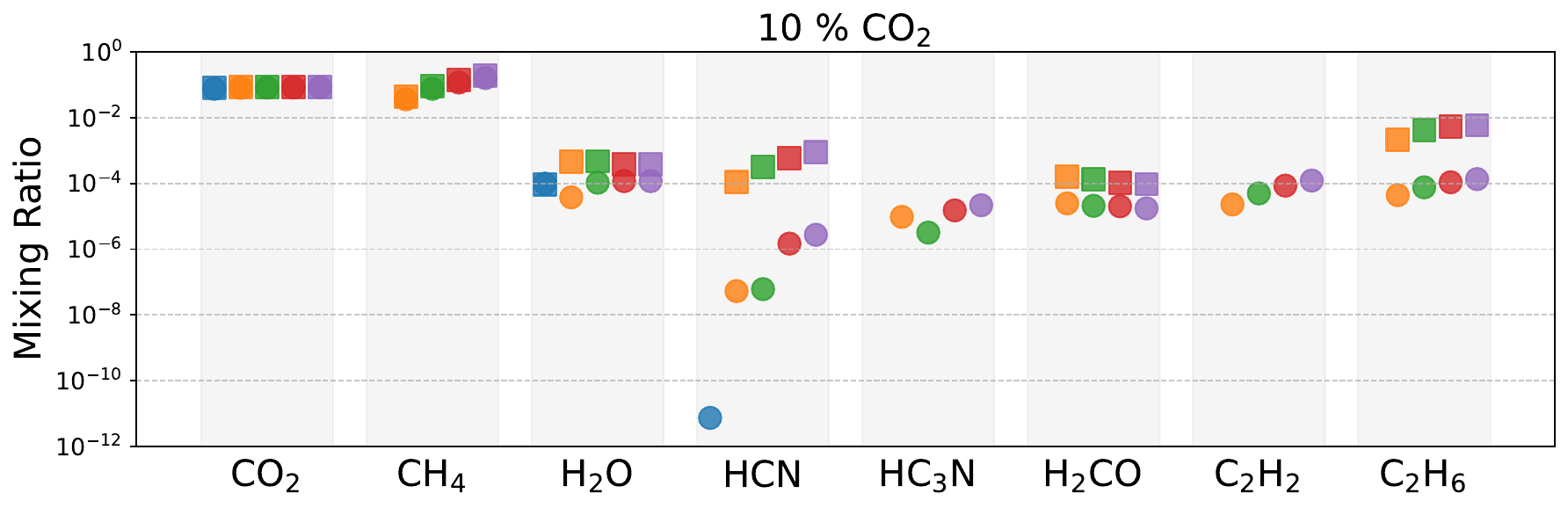}
\caption{Vertically averaged mixing ratios from simulations with CRAHCN-O (squares) and N-C-H-O (circles), for 400~ppm CO$_2$ (top), 1\% CO$_2$ (middle), and 10\% CO$_2$ (bottom). For each species, the mixing ratios are shown as a function of C/O ratio as specified in Table~\ref{tab:merged_comp}.}
\label{fig:species_avg_mixrat_all}
\end{figure}

We can infer the evolution of mixing ratios with C/O ratio for the individual networks from the vertically averaged mixing ratios in Figure~\ref{fig:species_avg_mixrat_all}, where each panel shows one of the background CO$_2$ scenarios. The vertical averages are not pressure-weighted, to focus on the shielding effects in the upper atmospheric regions. We briefly discuss the main outcomes. Except for the effects of photochemical shielding, the vertically averaged CO$_2$ mixing ratios remain comparatively constant, whereas CH$_4$ increases as a function of C/O. Figure~\ref{fig:species_avg_mixrat_all} shows that these trends persist regardless of network and CO$_2$ scenario. Hydrocarbons (C$_2$H$_2$ and C$_2$H$_6$) generally increase with C/O, although at rather low mixing ratios (${<}10^{-16}$) for C$_2$H$_2$ in CRAHCN-O. The increased availability of CH$_4$ provides more material to form HCN and HC$_3$N, explaining the increasing abundance with C/O ratio in Figure~\ref{fig:species_avg_mixrat_all} (note that HC$_3$N mixing ratios stay below ${<}10^{-12}$ for CRAHCN-O). In the 10\% CO$_2$ case with the CRAHCN-O network, H$_2$O already peaks at the top of the atmosphere for C/O=1.0 (see Figure~\ref{fig:all_vert_distribs}f), thus preventing a further increase in H$_2$O abundances with C/O ratio (bottom panel of Figure~\ref{fig:species_avg_mixrat_all}). H$_2$CO increases with C/O for the 400~ppm and 1\% CO$_2$ scenarios, but decreases for 10\% CO$_2$. This seems to indicate that its formation is CH$_4$-limited at lower abundances, but becomes O-limited with the decreased photodissociation of CO$_2$ and H$_2$O in the 10~\% CO$_2$ scenario. In Section~\ref{subsec:pathways_aldehydes}, we explore the existence of pathways from both CO$_2$ and CH$_4$.

\begin{figure}
\centering
\includegraphics[width=0.7\columnwidth]{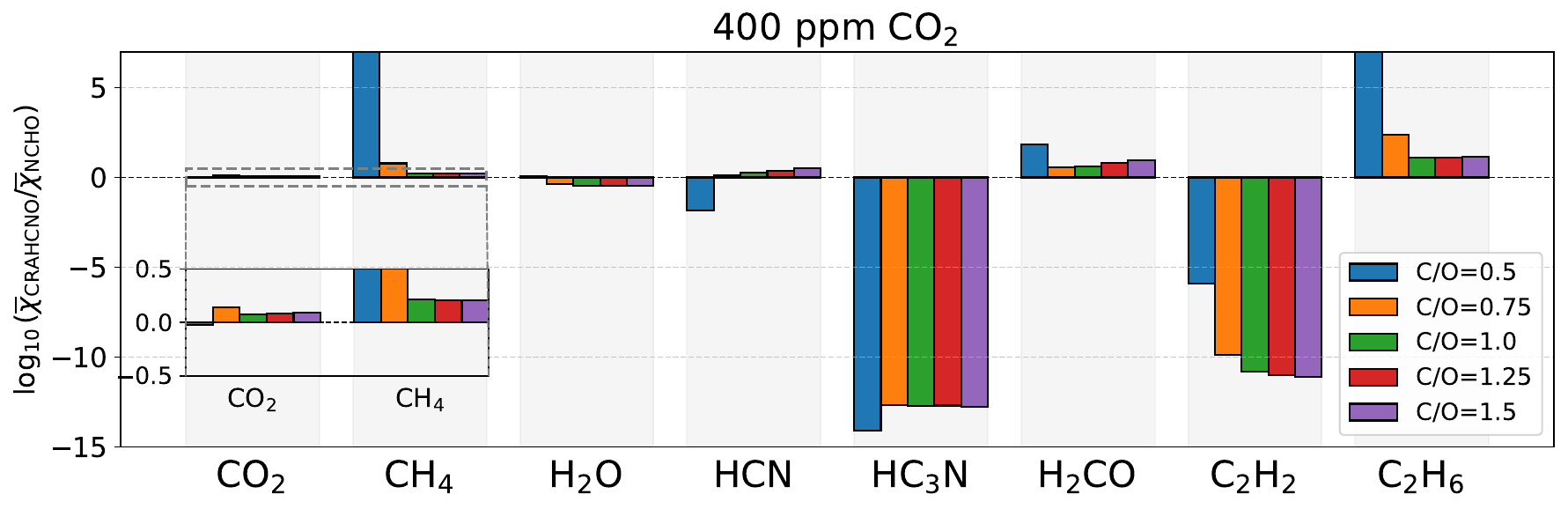}
\includegraphics[width=0.7\columnwidth]{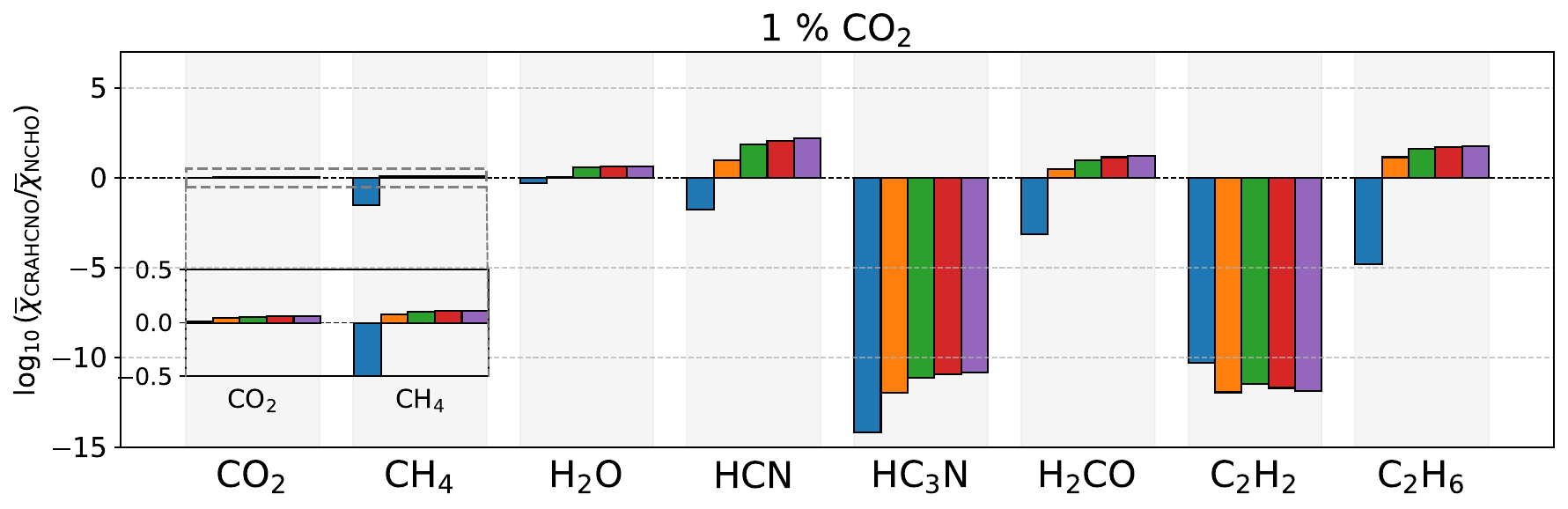}
\includegraphics[width=0.7\columnwidth]{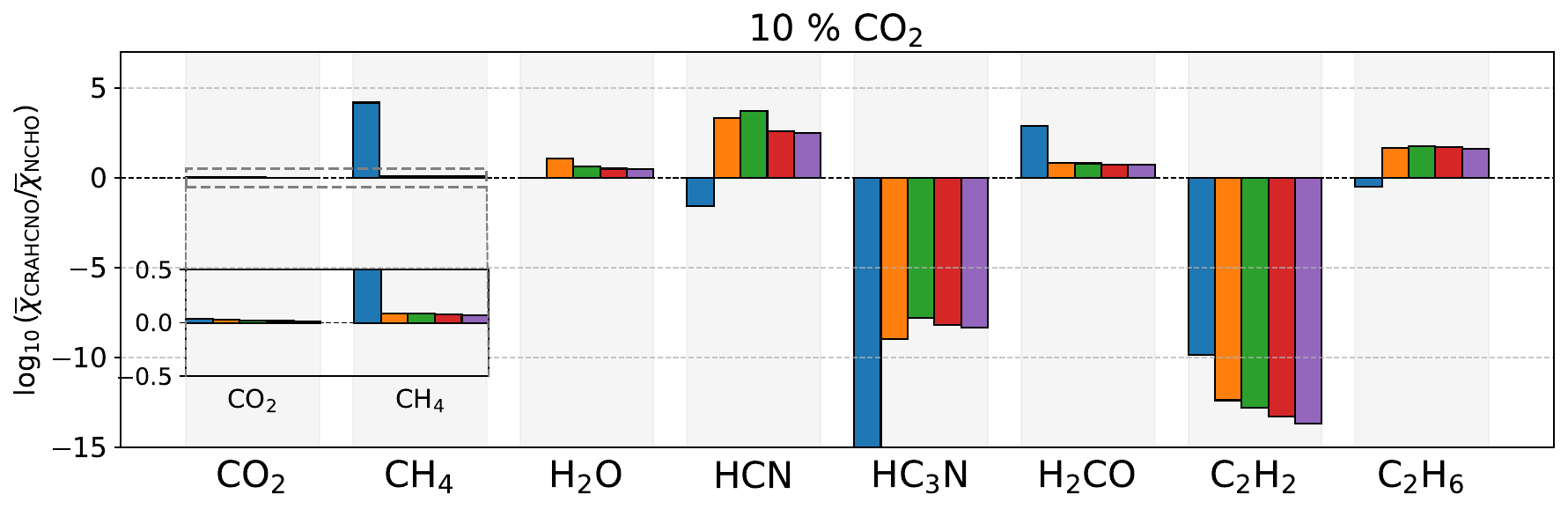}
\caption{Ratios of vertically averaged mixing ratios ($\chi$) from simulations with CRAHCN-O versus N-C-H-O, for 400~ppm CO$_2$ (top), 1\% (middle), and 10\% CO$_2$ (bottom). For each species, we compute the difference as a function of C/O ratio as specified in Table~\ref{tab:merged_comp}.}
\label{fig:species_ratios_all}
\end{figure}

Given these general trends, we can investigate how inter-network differences evolve with the C/O ratio. Figure~\ref{fig:species_ratios_all} compares the vertically averaged mixing ratios of the same set of species by calculating the ratio between CRAHCN-O and N-C-H-O predictions. Each panel indicates one of the background CO$_2$ scenarios. The first observation is that, for C/O${>}$0.5, CRAHCN-O produces enhanced mixing ratios of CO$_2$, CH$_4$, HCN, H$_2$CO, and C$_2$H$_6$, regardless of the scenario. This can be traced back to the accumulation of C$_2$H$_6$ and its resulting photochemical shielding. Second, the trends of inter-network differences of CO$_2$, CH$_4$, and C$_2$H$_6$ as a function of C/O agree: a maximum enhancement with CRAHCN-O at C/O=0.75 for 400~ppm CO$_2$, increasing enhancements as a function of C/O for 1\% CO$_2$, and decreasing enhancements as a function of C/O for 10\% CO$_2$. Third, and also for C/O${>}$0.5, HC$_3$N and C$_2$H$_2$ are much more abundant in N-C-H-O and stay below mixing ratios of 10$^{-12}$ for CRAHCN-O. Fourth, for C/O=0.5, the relative enhancement in N-C-H-O of HCN, HC$_3$N, H$_2$CO, and C$_2$H$_2$ (although, at small abundances ${<}10^{-14}$ for the latter three) shows that the pathways to feedstock molecules in the absence of CH$_4$ are faster in N-C-H-O. Lastly, H$_2$O forms in the upper atmosphere from photochemical products of CH$_4$ for CRAHCN-O, provided that sufficient CH$_4$ is present. This explains why the enhancement in CRAHCN-O is seen for the 1\% and 10\% cases and not for 400~ppm case.

For 1\% CO$_2$, increasing C$_2$H$_6$ abundances as a function of C/O with the CRAHCN-O network strengthens the photochemical shielding. C$_2$H$_6$ abundances increase more moderately with the N-C-H-O network, which causes the enhanced inter-network differences with the C/O ratio in Figure~\ref{fig:species_ratios_all}. On the other hand, for 10\% CO$_2$, C$_2$H$_6$ formation saturates in CRAHCN-O for a C/O ratio between 0.75 and 1.0, which can be seen in Figure~\ref{fig:all_vert_distribs}f with peak abundances of C$_2$H$_6$ already reaching the top of the atmosphere. With N-C-H-O, C$_2$H$_6$ increases are stronger with C/O ratio (Figure~\ref{fig:species_avg_mixrat_all}), leading to decreased inter-network differences. These contrasting trends are reflected in the distribution of inter-network differences of HCN and H$_2$O due to the shielding effects. Hence, even though the reactions included in each network do not change, the resulting inter-network differences still fundamentally depend on the background scenario.

\begin{figure}
\centering
\includegraphics[width=0.7\columnwidth]{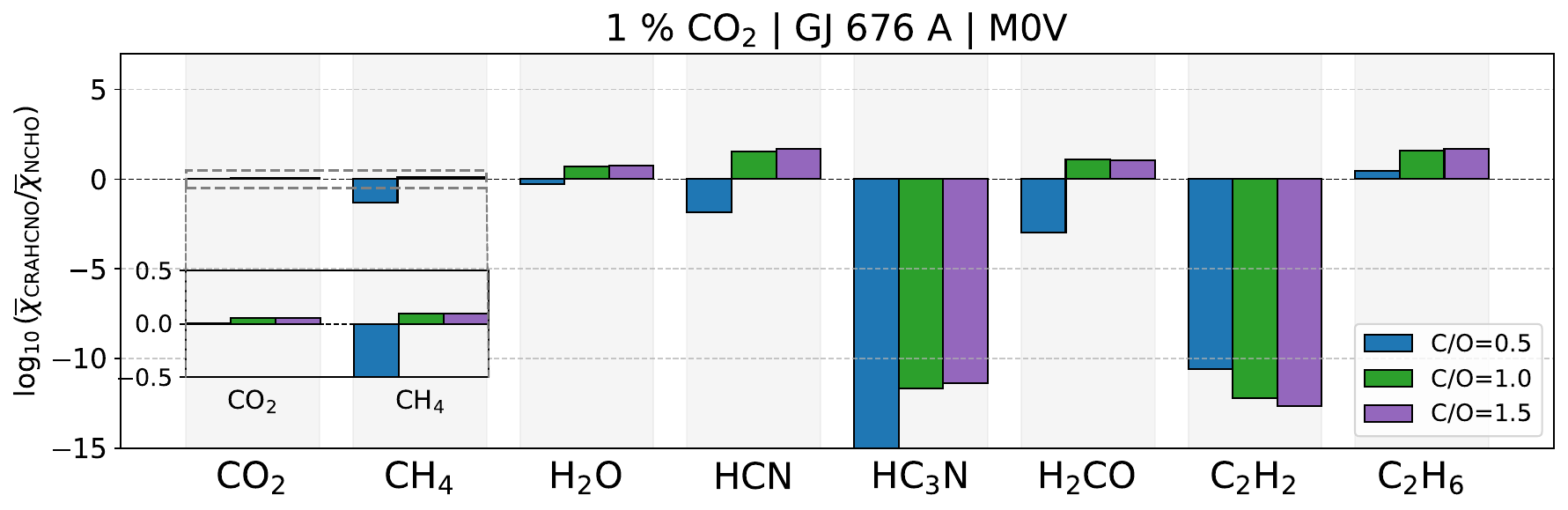}
\includegraphics[width=0.7\columnwidth]{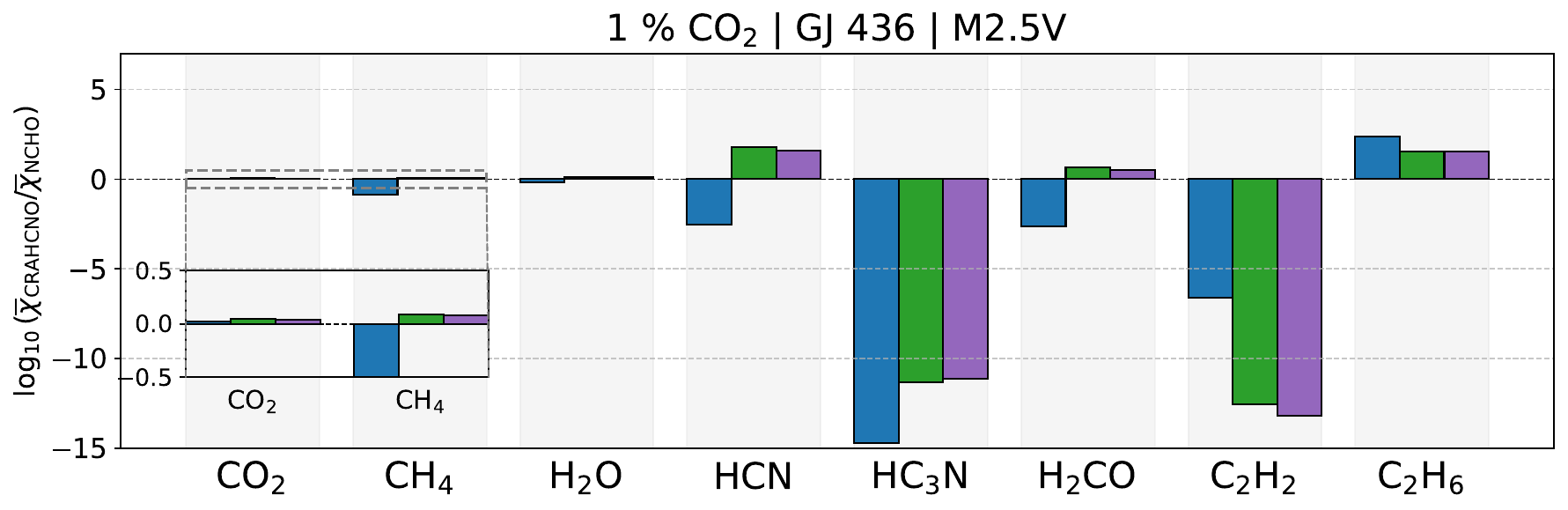}
\includegraphics[width=0.7\columnwidth]{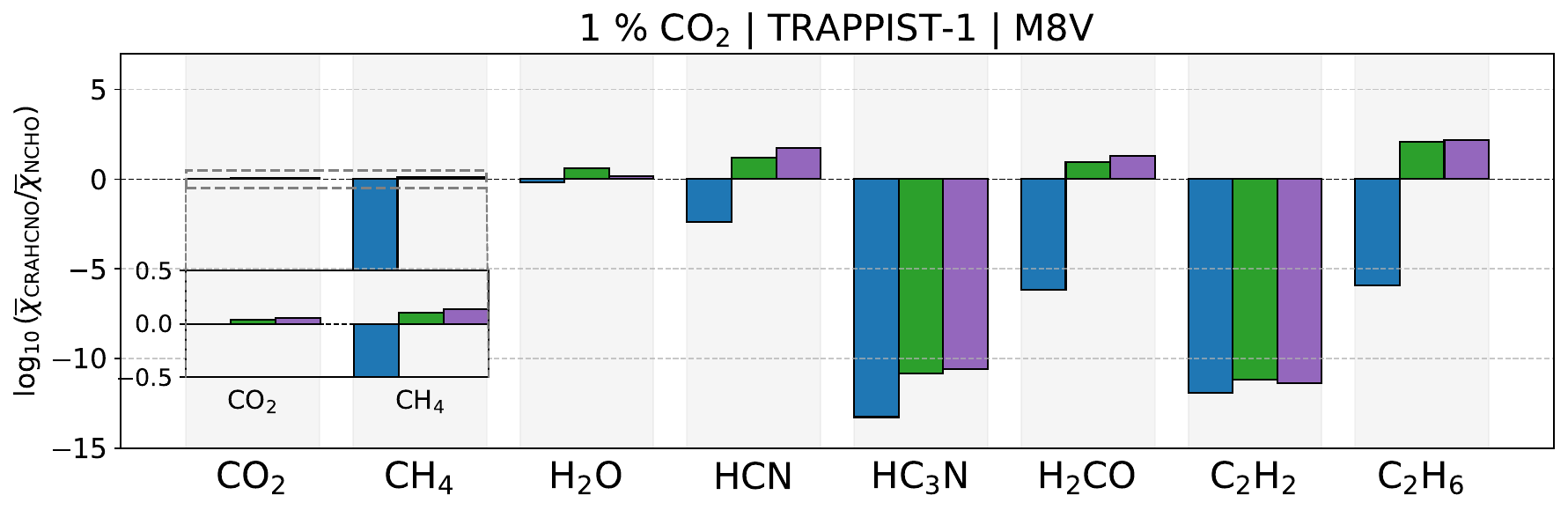}
\caption{Ratios of vertically averaged mixing ratios ($\chi$) from simulations with CRAHCN-O versus N-C-H-O for the 1\% CO$_2$ case, assuming distinct host star spectral energy distributions (see Figure~\ref{methods-fig:pcb_ptkzz_flux}a). We only include C/O ratios of 0.5, 1.0, and 1.5 from Table~\ref{tab:merged_comp} to see if the trends persist for different M stars.}
\label{fig:species_ratios_all_hoststar_dependence}
\end{figure}

Despite distinct spectral energy distributions for other M-type stars (see Figure~\ref{methods-fig:pcb_ptkzz_flux}), the trends with C/O ratio persist, as shown for the 1\% CO$_2$ cases in Figure~\ref{fig:species_ratios_all_hoststar_dependence}. For C/O=0.5, we note the same findings as before that 1) CO$_2$ and H$_2$O are similar, 2) CH$_4$, C$_2$H$_2$, C$_2$H$_6$, and H$_2$CO have large inter-network ratios reflecting their negligible abundances, and 3) HCN consistently forms more efficiently in N-C-H-O. For C/O${>}$0.5, C$_2$H$_6$ starts to accumulate in CRAHCN-O and similarly shields CO$_2$, CH$_4$, HCN, H$_2$CO. Nevertheless, with N-C-H-O, the location of the UV photosphere between 115--145~nm shifts to lower pressures for GJ 676 A and GJ 436 and higher pressures for TRAPPIST-1 (not shown), due to lower (GJ 676 A and GJ 436) and higher (TRAPPIST-1) stellar fluxes in this wavelength range. The photosphere shifts are similar for CRAHCN-O. Lastly, the pathways to HC$_3$N and C$_2$H$_2$ are again slow in CRAHCN-O. 
Importantly, by considering the irradiance from distinct M0V, M2.5V, M5.5V, and M8V stars, we show that the photochemical shielding effects associated with C$_2$H$_6$ accumulation in CRAHCN-O apply to planets orbiting distinct M-type stars.

\subsection{Pathway analyses}\label{sec:pathways}
In this Section, we present the dominant reaction rates (both production and destruction) and the dominant pathways to form feedstock molecules and hydrocarbons from the initial species, to identify the causes of inter-network differences between VULCAN/N-C-H-O and VULCAN/CRAHCN-O. We include the pathways to nitriles (HCN, HC$_3$N), an aldehyde (H$_2$CO), and a subset of hydrocarbons (C$_2$H$_2$, C$_2$H$_6$, C$_3$H$_4$, C$_4$H$_3$, C$_6$H$_5$). By comparing all reactions involving a given species, the dominant reactions are those with the fastest rates. The pathway analysis identifies all the reaction steps and the rate-limiting step in the shortest path between two species, as described in Section~\ref{methods:pathways}. For brevity, we focus on the results for the 1\% CO$_2$ case, but note that the same reaction rates were used in the other simulations. Thus, the results presented here apply more generally, albeit with a different pressure dependence since the UV photosphere moves to higher (400~ppm CO$_2$) or lower (10\% CO$_2$ case) pressures. At the end of the Section, we summarise and compare the rate coefficients for the most important bimolecular and termolecular reactions in Tables~\ref{tab:key_reactions_pathways} and \ref{tab:key_reactions_pathways_termol}, respectively.

\subsubsection{Nitriles}\label{subsec:pathways_nitriles}
\begin{figure}
\centering
\includegraphics[width=0.4\columnwidth]{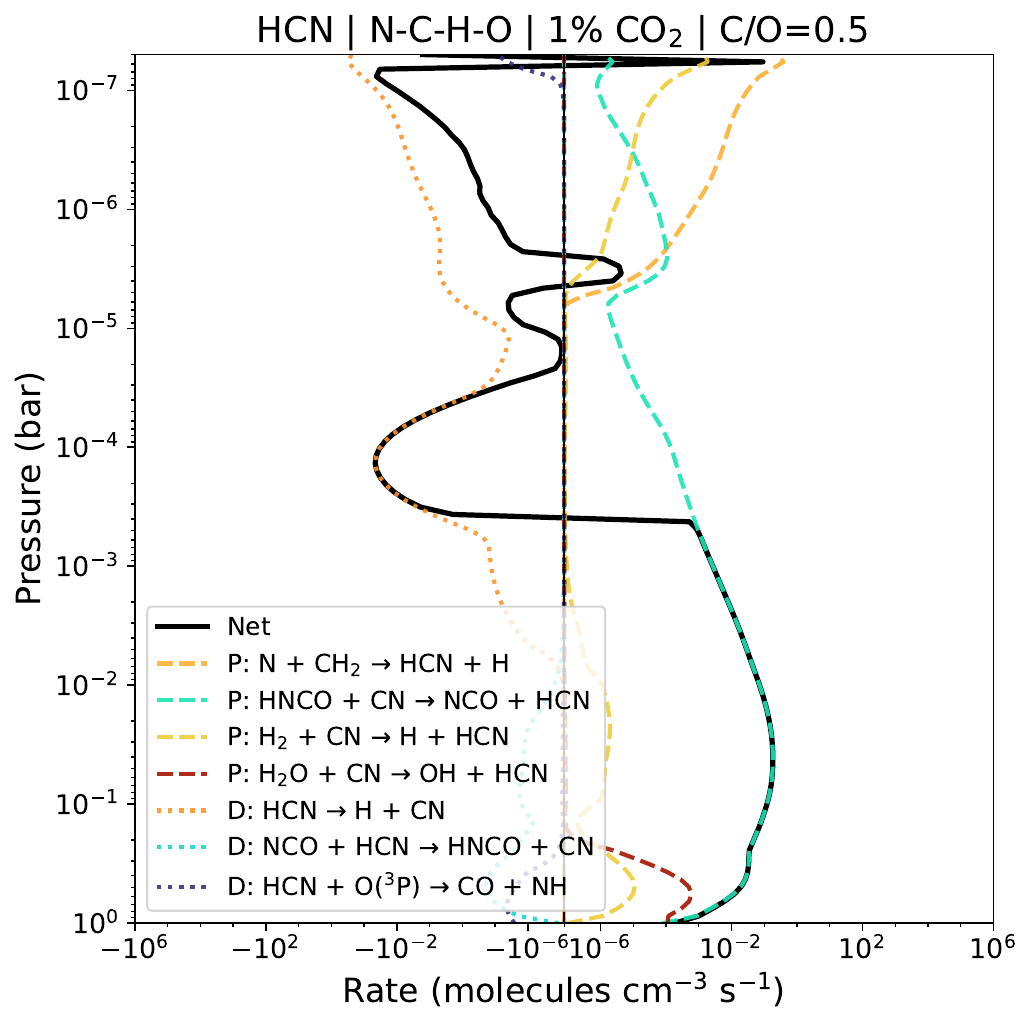}
\includegraphics[width=0.4\columnwidth]{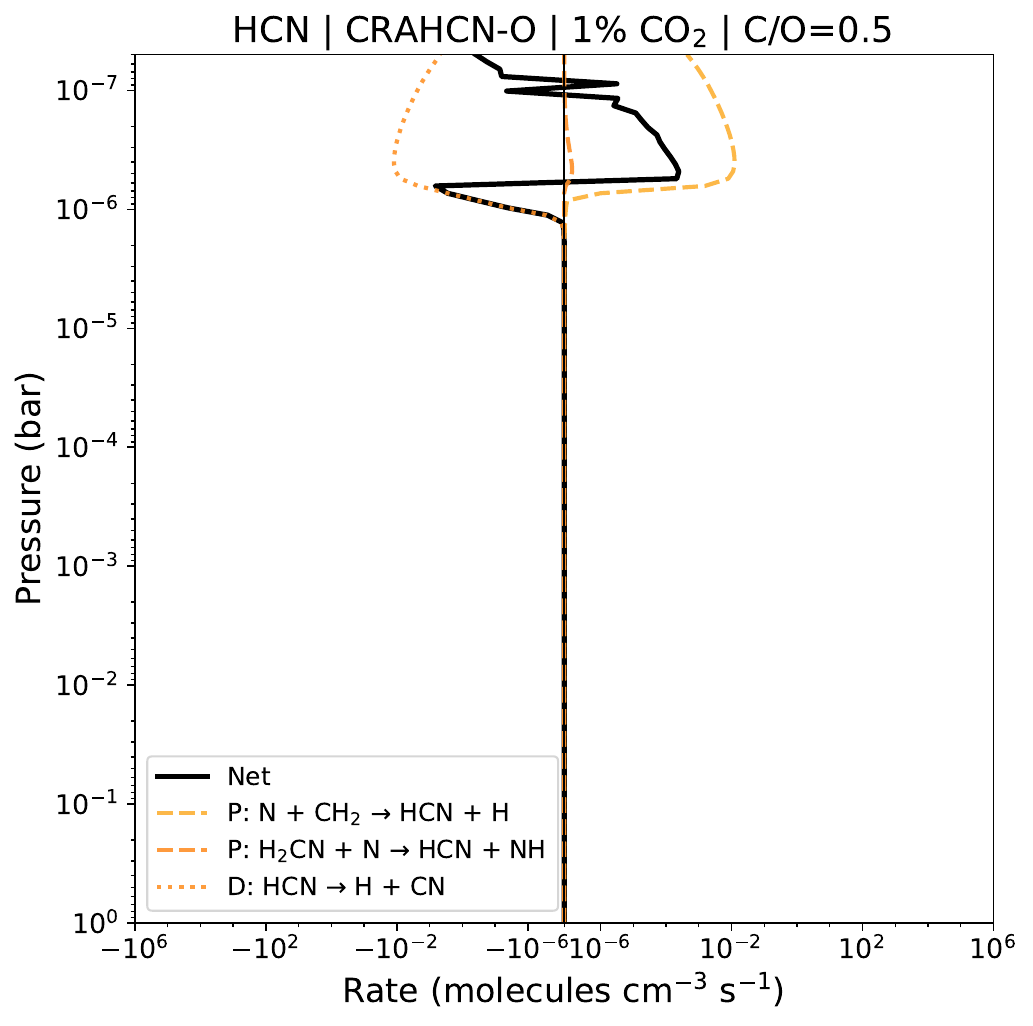}
\caption{Reaction rates for the dominant reactions involving HCN in the 1\% CO$_2$ case, for C/O=0.5 and comparing N-C-H-O (a) and CRAHCN-O (b). The dashed lines represent production reactions, the dotted lines destruction reactions. The net rate shows the difference between all production and all destruction reactions. Reaction colours agree across panels.}
\label{fig:reaction_rates_a05_hcn}
\end{figure}

\begin{figure}
\centering
\includegraphics[width=0.3\columnwidth]{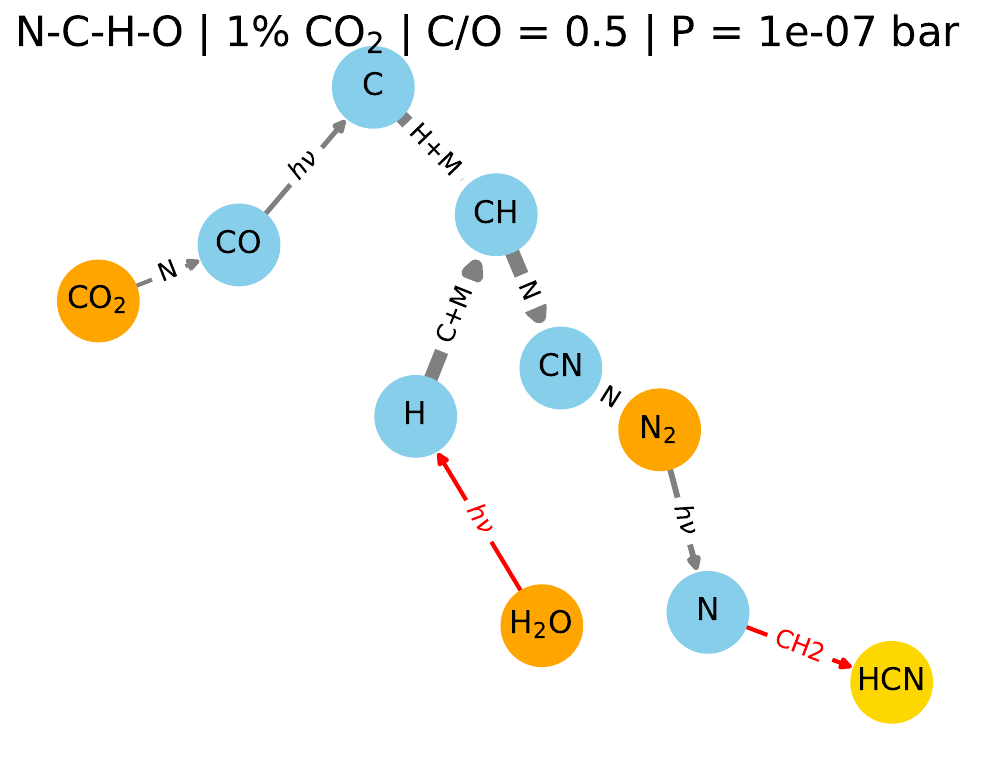}
\includegraphics[width=0.3\columnwidth]{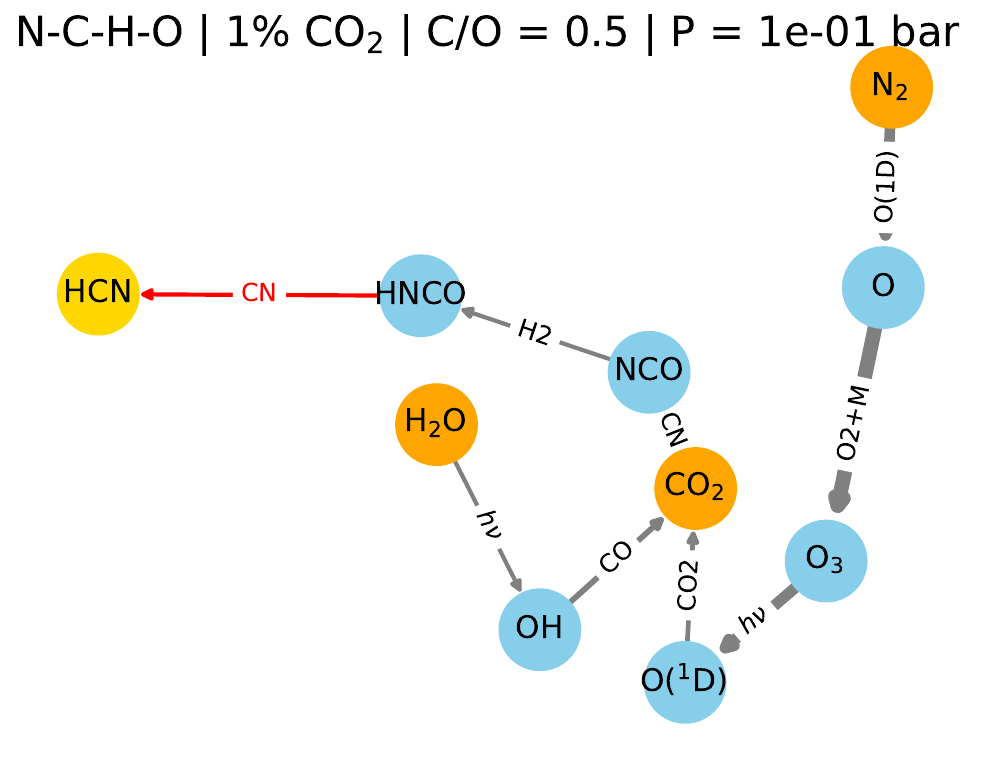}
\includegraphics[width=0.3\columnwidth]{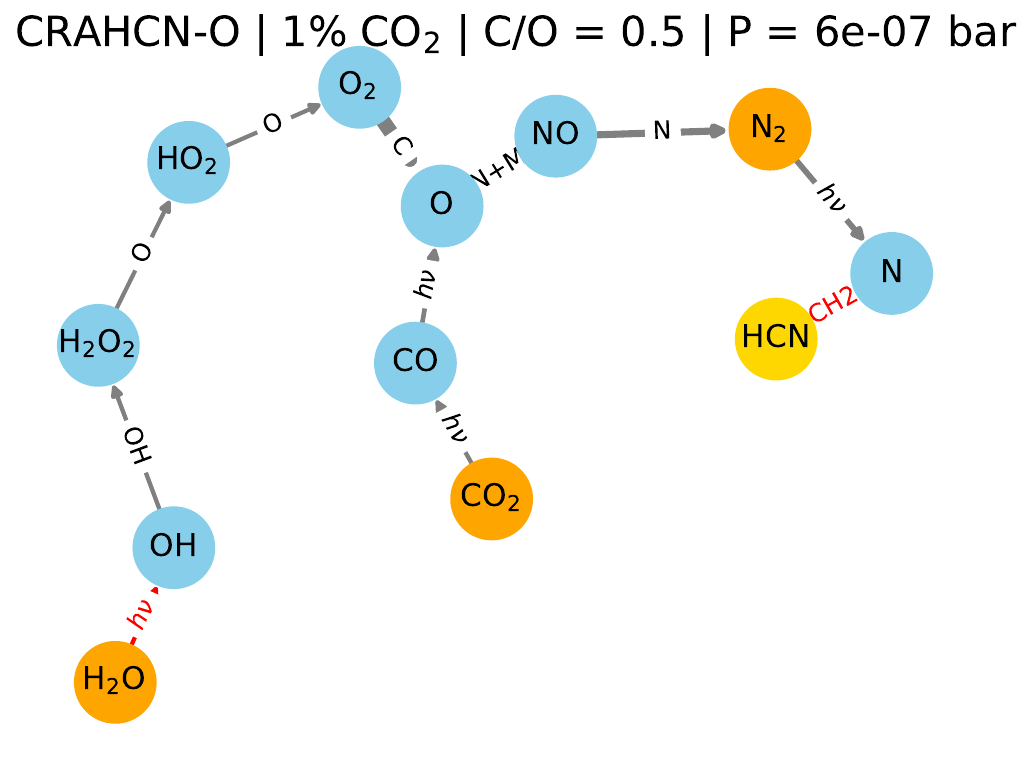}
\caption{Dominant pathways to HCN formation for the N-C-H-O network at 10$^{-7}$ and 10$^{-1}$ bar and CRAHCN-O at $6\times10^{-7}$~bar, as indicated by the titles. Pathways are shown for the 1\% CO$_2$ case at C/O=0.5. Initialisation species are shown in orange, the target species (in this case, HCN) in yellow, and intermediate species in blue. The other reactant species involved in reaching the next node in a pathway are shown as edge labels. Edge widths represent reaction rates, with wider edges indicating faster reactions; the rate-limiting step is shown in red.}
\label{fig:pathways_a05_hcn}
\end{figure}

For C/O=0.5, N-C-H-O has non-zero reaction rates of HCN production at all pressure levels, as we can see from the vertical distribution in Figure~\ref{fig:reaction_rates_a05_hcn}a. Net production regions are seen at low P${<}10^{-6}$~bar and at high P${>}4{\times}10^{-4}$~bar. For CRAHCN-O, we see production of HCN at low pressures (${<}10^{-6}$~bar) but not substantially at high pressures (Figure~\ref{fig:reaction_rates_a05_hcn}b). We investigate the pathways from  N$_2$, CO$_2$, and H$_2$O to HCN, due to the absence of CH$_4$ in the C/O=0.5 setups. In Figure~\ref{fig:pathways_a05_hcn}, we split the pathways for N-C-H-O into a low-pressure ($10^{-7}$~bar) and high-pressure ($10^{-1}$~bar) region. For low pressures, N-C-H-O forms HCN most efficiently via:
\begin{reactions}
    \label{chem:n_hv}
    N2 + h\nu &-> N + N \\
    \label{chem:n_ch2}
    N + CH2 &-> HCN + H 
\end{reactions}
Reaction~R\ref{chem:n_ch2} forms the rate-limiting step. H$_2$O and CO$_2$ photolysis provide CH and thus CH$_2$ for Reaction~R\ref{chem:n_ch2}. For high pressures, there is a pathway starting from CO$_2$:
\begin{reactions}
    \label{chem:cn_co2}
    CN + CO2 &-> CO + NCO  \\
    \label{chem:nco_h2}
    NCO + H2 &-> HNCO + H  \\
    \label{chem:hnco_cn}
    HNCO + CN &-> NCO + HCN 
\end{reactions}
In this pathway, Reaction~R\ref{chem:hnco_cn} is the rate-limiting step. The CN and H$_2$ required in this pathway come from the upper atmosphere following photolysis of N$_2$, CO$_2$, and H$_2$O. In CRAHCN-O, the absence of HNCO means the high-pressure pathway does not occur. The pathway from N$_2$ is the same as in N-C-H-O (Reactions~R\ref{chem:n_hv} and R\ref{chem:n_ch2}), as shown in Figure~\ref{fig:pathways_a05_hcn}. Again, H$_2$O photolysis at higher pressures followed by mixing instead provides H for the production of CH$_2$ in Reaction~R\ref{chem:n_ch2}. The absorption cross sections for reaction R\ref{chem:n_hv} are identical, but the rate coefficients associated with R\ref{chem:n_ch2} are different (see Table~\ref{tab:key_reactions_pathways}). The rate constants at 298~K (k$_{298}$) are similar, but $k$ in CRAHCN-O becomes an order of magnitude smaller with decreasing temperature. Although not shown here for brevity, the pathway differences between both networks are the same for the 400~ppm and 10\% CO$_2$ cases, but with the key pathways shifted to higher and lower pressures, respectively. The pressure shift is due to shielding effects, like in Figure~\ref{fig:all_photosphere}.

\begin{figure}
\centering
\includegraphics[width=0.4\columnwidth]{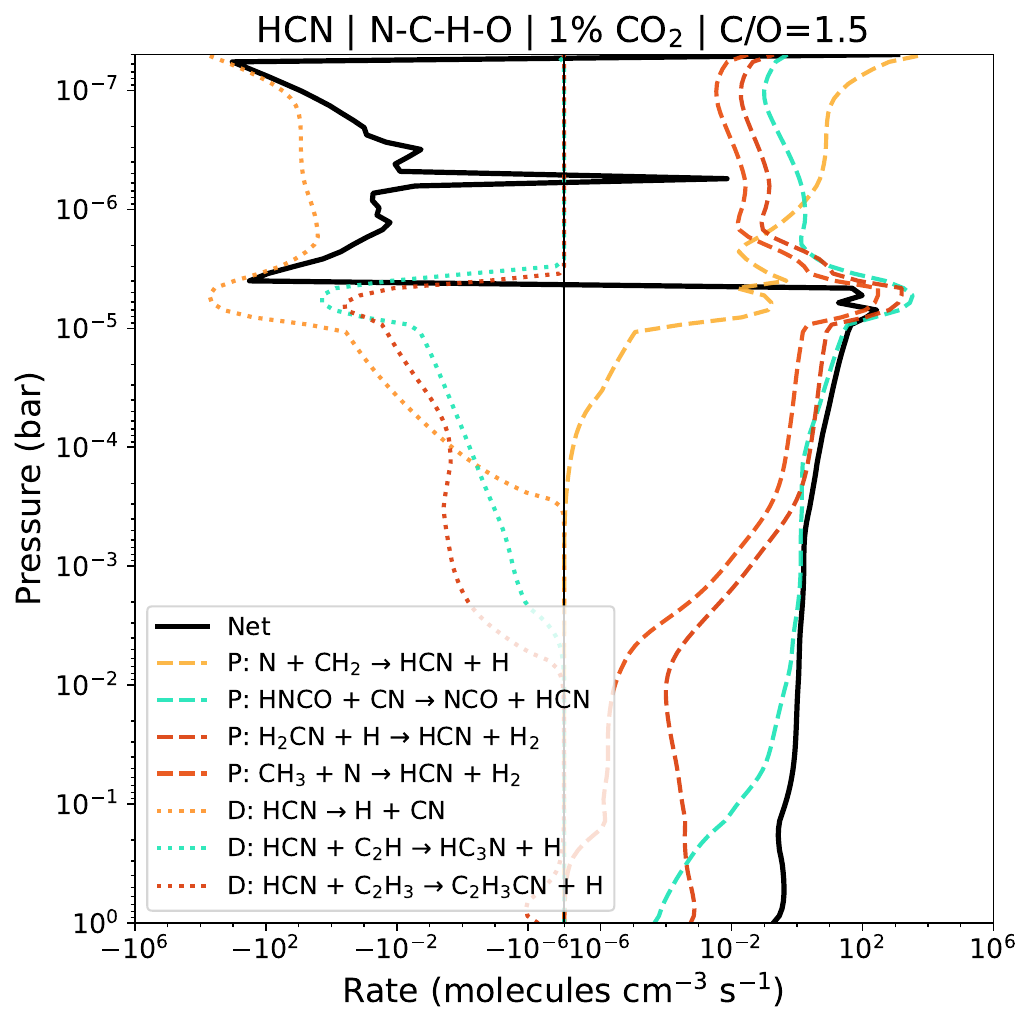}
\includegraphics[width=0.4\columnwidth]{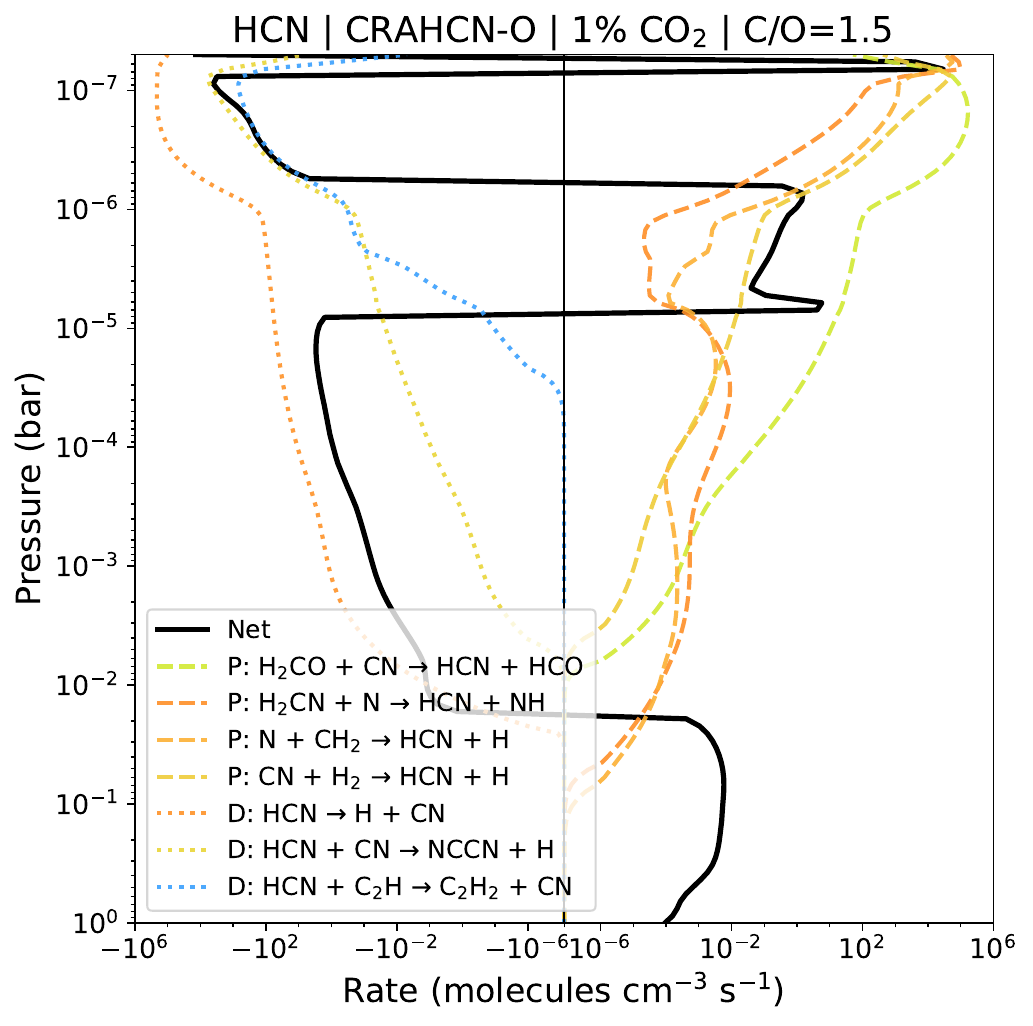}
\caption{Same as Figure~\ref{fig:reaction_rates_a05_hcn}, but for the 1\% CO$_2$ case at C/O=1.5.}
\label{fig:reaction_rates_a15_hcn}
\end{figure}

\begin{figure}
\centering
\includegraphics[width=0.35\columnwidth]{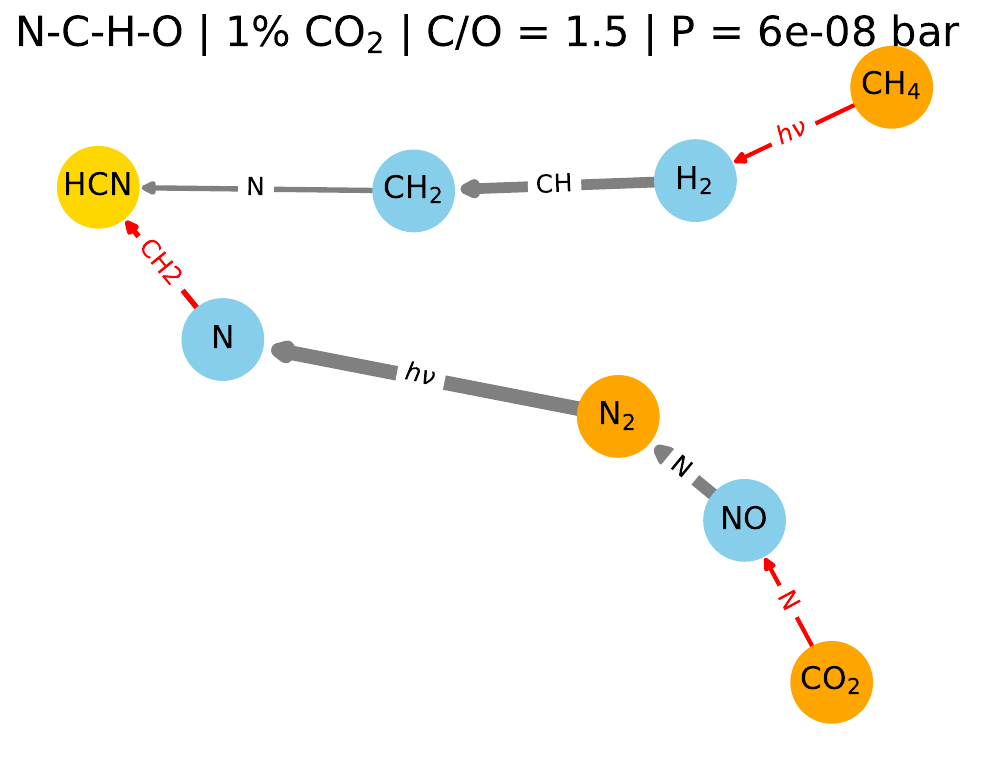}
\includegraphics[width=0.35\columnwidth]{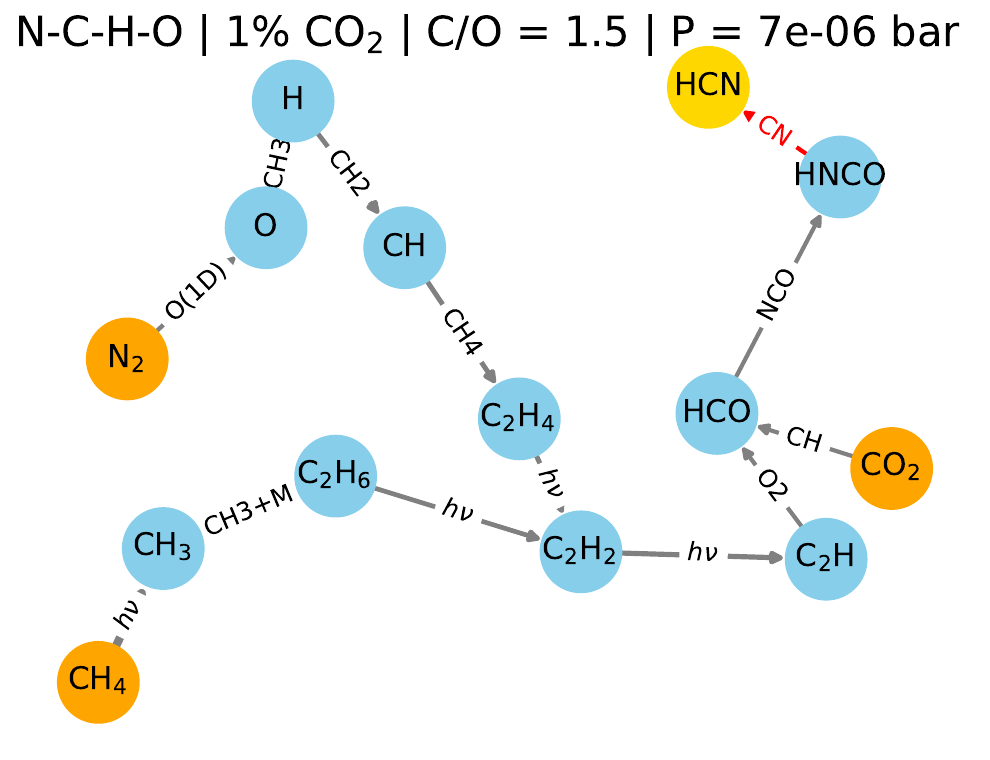}\\
\includegraphics[width=0.35\columnwidth]{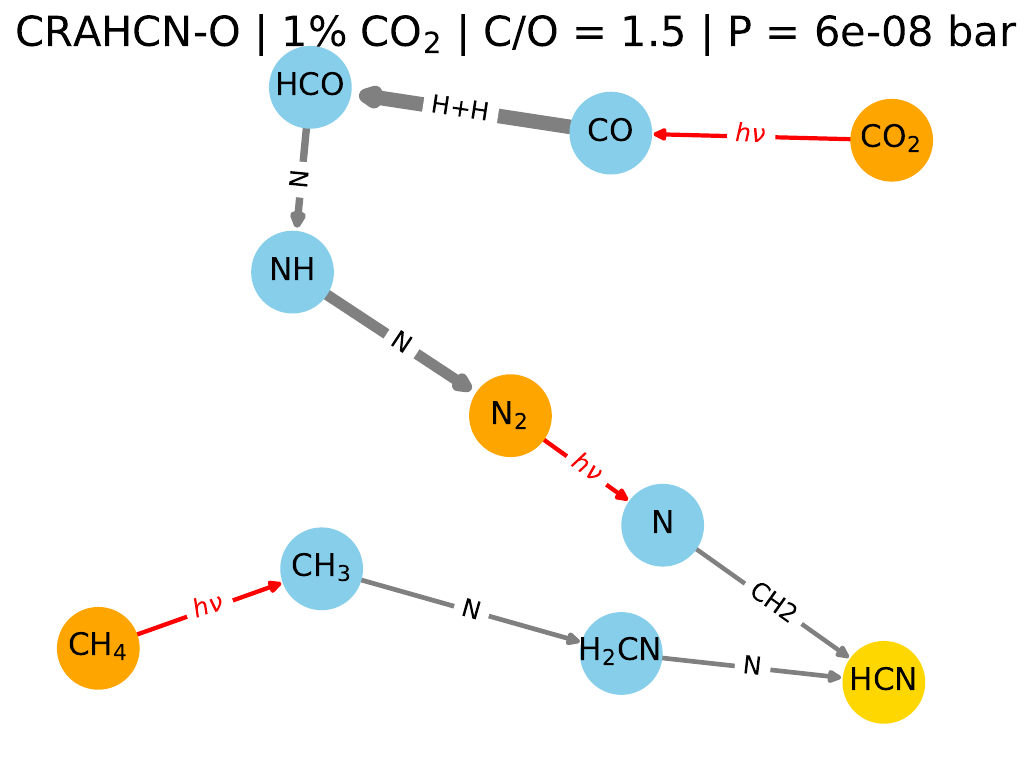}
\includegraphics[width=0.35\columnwidth]{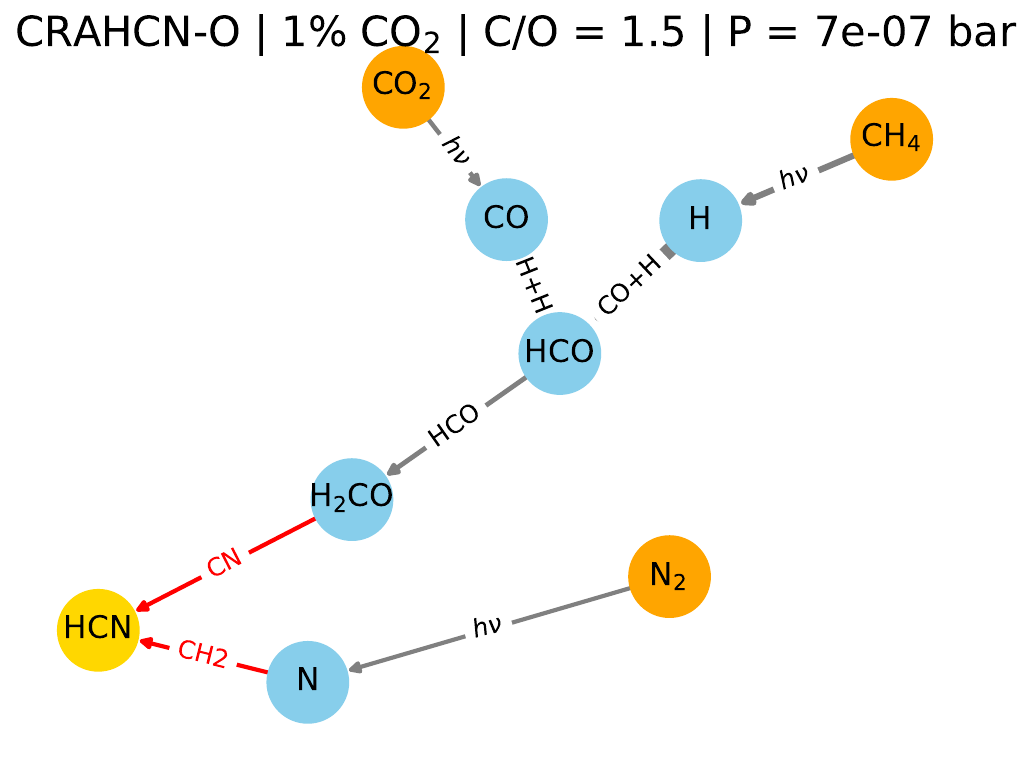}
\caption{Same as Figure~\ref{fig:pathways_a05_hcn}, but for the 1\% CO$_2$ case at C/O=1.5.}
\label{fig:pathways_a15_hcn}
\end{figure}

For C/O${>}$0.5, HCN production through reactions R\ref{chem:n_hv} and R\ref{chem:n_ch2} still occurs close to the top of the atmosphere with N-C-H-O (Figure~\ref{fig:reaction_rates_a15_hcn}a and \ref{fig:pathways_a15_hcn}). Considering the 1\% CO$_2$ case at C/O=1.5, the net HCN production in N-C-H-O peaks around $7{\times}10^{-6}$~bar and remains high for increasing pressures, being dominated by the pathway involving HCNO. The pathway includes Reactions \ref{chem:cn_co2} and \ref{chem:hnco_cn}, but HNCO formation is now dominated by HCO:
\begin{reactions}
    \label{chem:co2_ch}
    CO2 + CH &-> HCO + CO\\
    \label{chem:nco_hco}
    NCO + HCO &-> HNCO + CO 
\end{reactions}
as shown in Figure~\ref{fig:pathways_a15_hcn}. With increasing CO$_2$ and CH$_4$ abundances, the HNCO production regions shift to lower pressures. The lower availability of HNCO then affect the HCN abundances at higher pressures (P${>}10^{-4}$ for the 400~ppm case, ${>}10^{-5}$ for the 1\% case, ${>}10^{-6}$~bar for the 10\% case), explaining why HCN abundances are higher in N-C-H-O than CRAHCN-O for 400~ppm CO$_2$ but not for 1\% and 10\% CO$_2$ (see Figure~\ref{fig:all_vert_distribs}). Although not shown in Figure~\ref{fig:pathways_a15_hcn} for N-C-H-O, pathways involving H$_2$CN then become more dominant. HCN forms at low pressures through:
\begin{reactions}
    \label{chem:ch4_hv_1}
    CH4 + h\nu &-> CH3 + H \\
    \label{chem:ch3_n}
    CH3 + N &-> H2CN + H\\ 
    \label{chem:h2cn_h}
    H2CN + H &-> HCN + H2
\end{reactions}
Reaction~R\ref{chem:ch3_n} is the rate-limiting step. In CRAHCN-O, one of the low-pressure ($6\times10^{-8}$~bar) pathways again consists of Reactions~R\ref{chem:n_hv} and R\ref{chem:n_ch2}, as shown in Figure~\ref{fig:pathways_a15_hcn}. CH$_4$ photolysis initiates another important pathway through Reactions~R\ref{chem:ch4_hv_1} and R\ref{chem:ch3_n}, followed by:
\begin{reactions}
    \label{chem:h2cn_n}
    H2CN + N &-> HCN + NH 
\end{reactions}
Note that CH$_3$ formation also occurs through the oxidation of CH$_4$:
\begin{reactions}
    \label{chem:ch4_oh}
    CH4 + OH &-> H2O + CH3\\
    \label{chem:ch4_o1d}
    CH4 + O(^1D) &-> OH + CH3 
\end{reactions}
For low pressures, we find another pathway to HCN from CH$_4$ (not shown in Figure~\ref{fig:pathways_a15_hcn}):
\begin{reactions}
    \label{chem:ch4_hv_2}
    CH4 + h\nu &-> CH2(^1A_1) + H2 \\
    \label{chem:cn_h2}
    CN + H2 &-> HCN + H 
\end{reactions}
Reaction~R\ref{chem:cn_h2} is the rate-limiting step and CH$_2(^1A_1)$ represents the first excited singlet state of CH$_2$. 

Prominent net production of HCN is also seen around $7\times10^{-7}$~bar, involving H$_2$CO as intermediate species:
\begin{reactions}
    \label{chem:co2_hv_1}
    CO2 + h\nu &-> CO + O(^1D)\\
    \label{chem:co_h_h}
    CO + H + H &-> HCO + H\\ 
    \label{chem:hco_hco}
    HCO + HCO &-> H2CO + CO \\
    \label{chem:h2co_cn}
    H2CO + CN &-> HCN + HCO
\end{reactions}
Reaction~R\ref{chem:h2co_cn} is the rate-limiting step. Note that the fast HCO formation through R\ref{chem:co_h_h} is the important distinction between the networks for this pathway. The high HCO abundances also regenerate CO$_2$ in the upper atmosphere of the 400~ppm case (see Figure~\ref{fig:all_vert_distribs}). For both N-C-H-O and CRAHCN-O, the direct pathway becomes dominant and rate-limiting at high pressures:
\begin{reactions}
    \label{chem:ch4_cn}
    CH4 + CN &-> HCN + CH3 
\end{reactions}
With increasing CH$_4$ abundances, Reaction~R\ref{chem:ch4_cn} starts to dominate a larger vertical extent of the atmosphere (not shown).


\begin{figure}
\centering
\includegraphics[width=0.4\columnwidth]{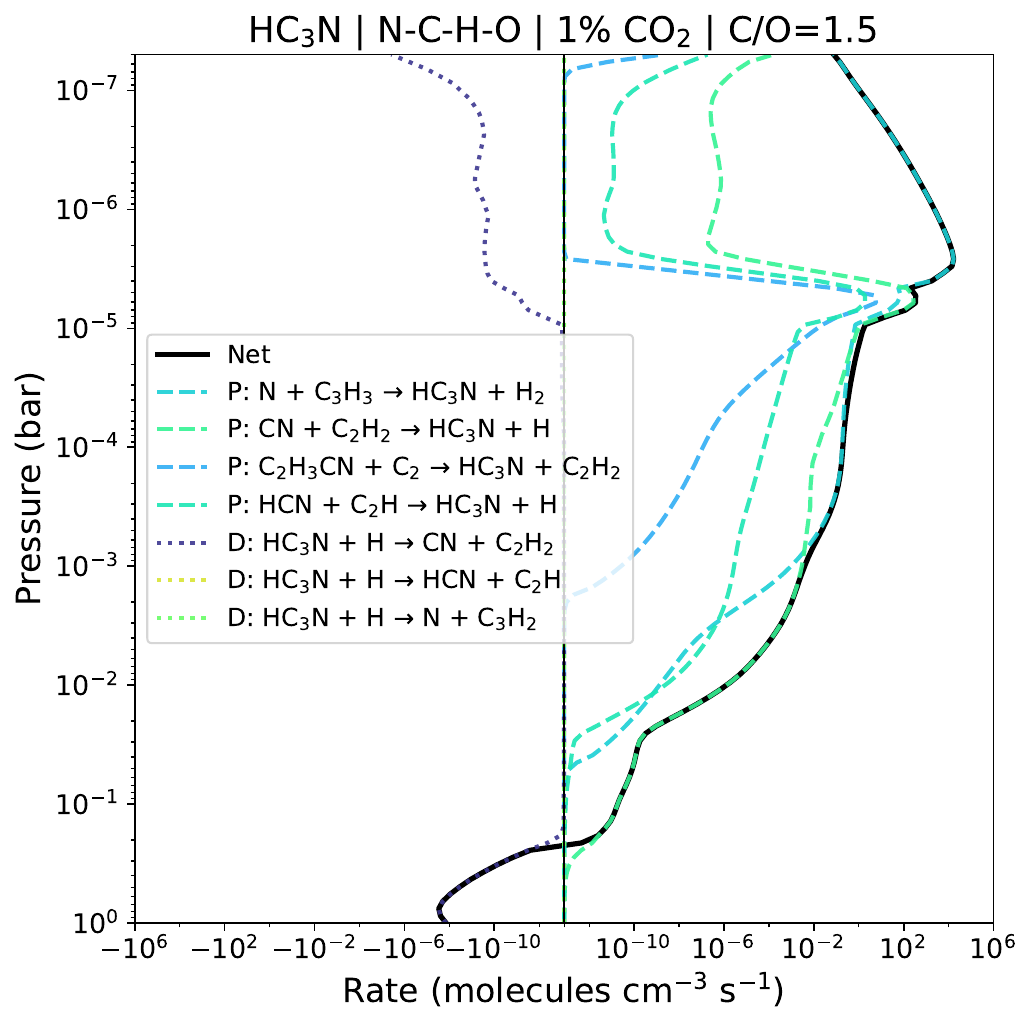}
\includegraphics[width=0.4\columnwidth]{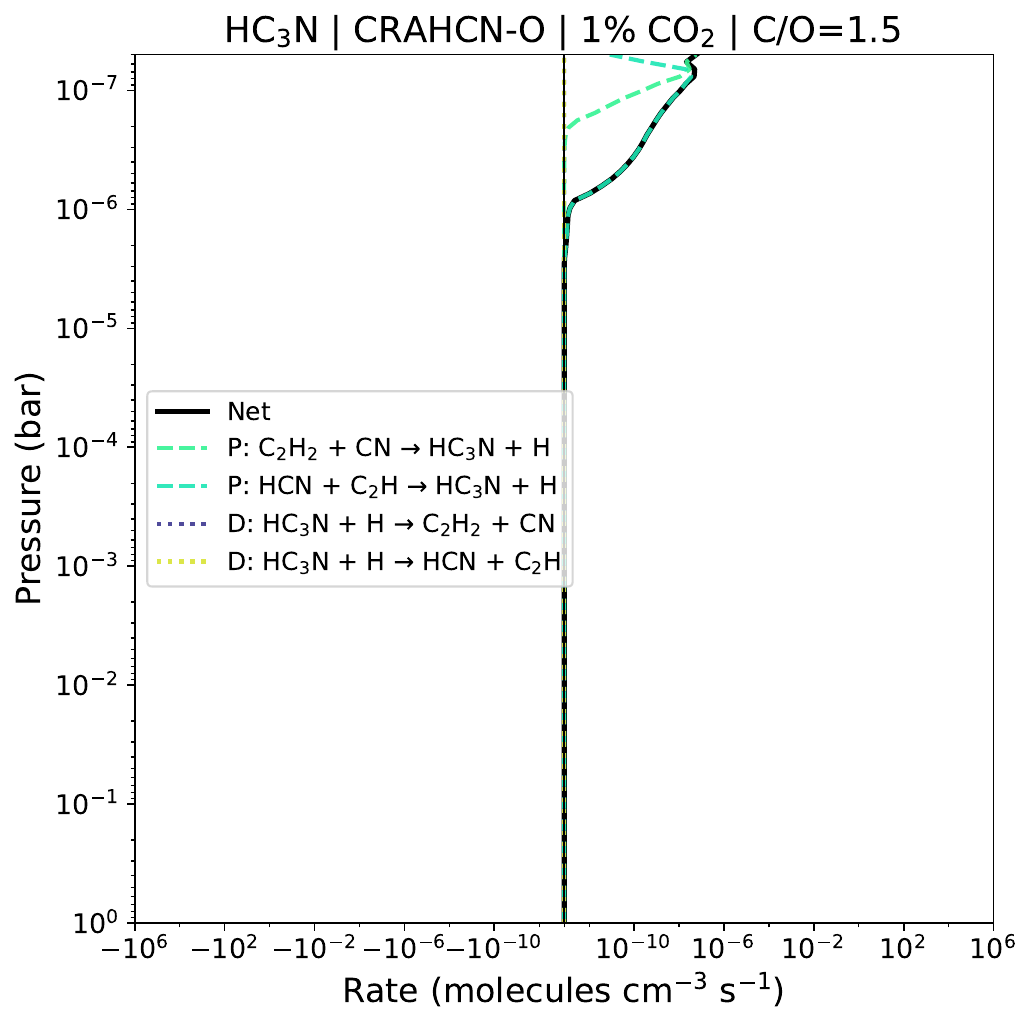}
\caption{Same as Figure~\ref{fig:reaction_rates_a05_hcn}, but showing the dominant rates controlling HC$_3$N in the 1\% CO$_2$ case at C/O=1.5.}
\label{fig:reaction_rates_a15_hc3n}
\end{figure}

\begin{figure}
\centering
\includegraphics[width=0.3\columnwidth]{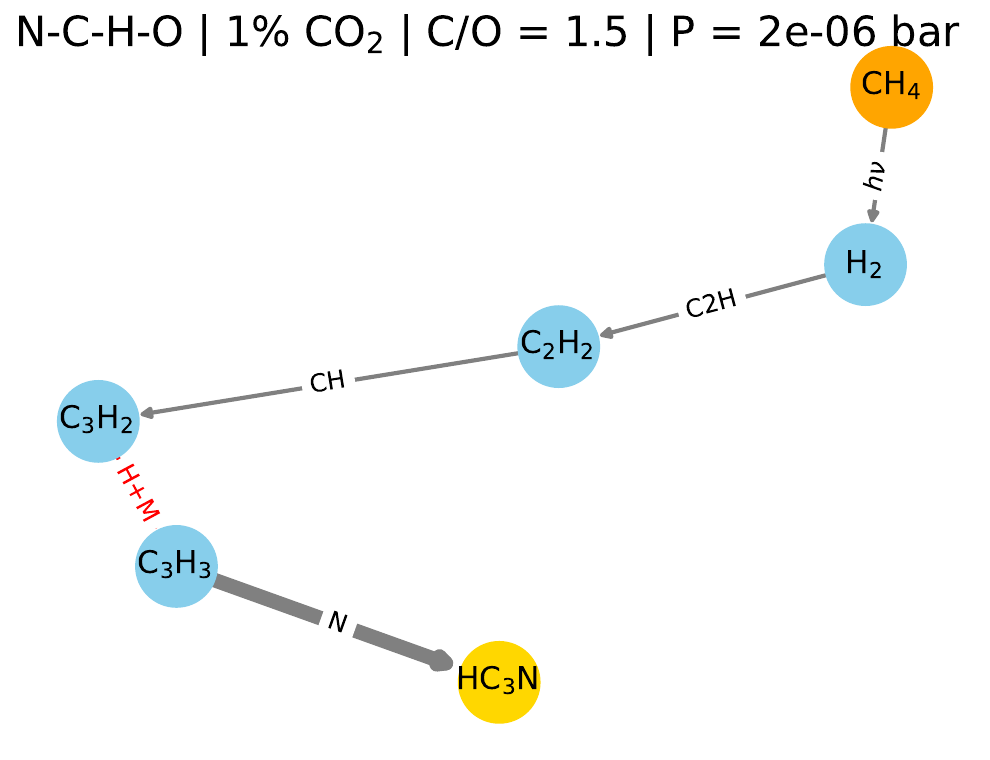}
\includegraphics[width=0.3\columnwidth]{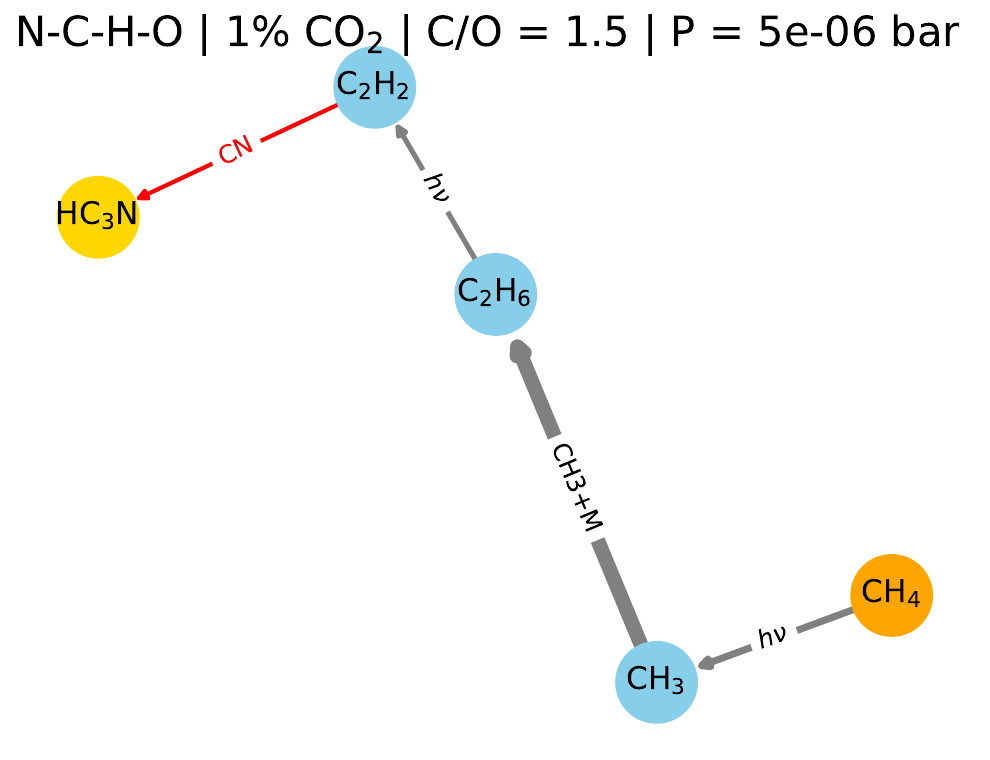}
\includegraphics[width=0.3\columnwidth]{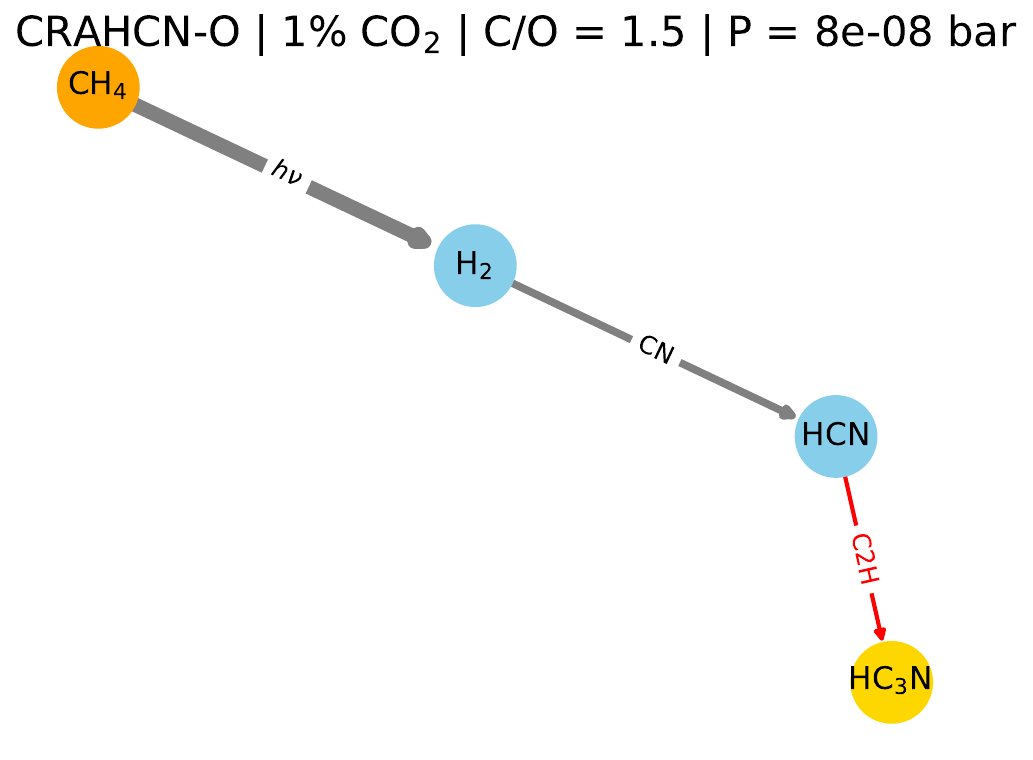}
\caption{Same as Figure~\ref{fig:pathways_a05_hcn}, but showing the dominant pathways to HC$_3$N for the 1\% CO$_2$ case at C/O=1.5.}
\label{fig:pathways_a15_hc3n}
\end{figure}

HC$_3$N formation is much more efficient in the N-C-H-O network. For N-C-H-O, its net production is strongest for low pressures (${<}10^{-4}$~bar, see Figure~\ref{fig:reaction_rates_a15_hc3n}a) and depends on hydrocarbon availability. As shown in Figure~\ref{fig:pathways_a15_hc3n}, the dominant upper atmosphere pathways trace back to CH$_4$ photolysis via Reaction~R\ref{chem:ch4_hv_2} followed by:
\begin{reactions}
    \label{chem:c2h_h2}
    C2H + H2 &-> C2H2 + H \\
    \label{chem:ch_c2h2}
    CH + C2H2 &-> C3H2 + H\\
    \label{chem:c3h2_h_m}
    C3H2 + H + M &-> C3H3 + M\\
    \label{chem:n_c3h3}
    N + C3H3 &-> HC3N + H
\end{reactions}
Around $5\times10^{-6}$~bar, CH$_4$ photolysis via Reaction~R\ref{chem:ch4_hv_1} initiates a second important pathway:
\begin{reactions}
    \label{chem:ch3_ch3_m}
    CH3 + CH3 + M &-> C2H6 + M \\ 
    \label{chem:c2h6_hv}
    C2H6 + h\nu &-> C2H2 + H2 + H2\\
    \label{chem:cn_c2h2}
    CN + C2H2 &-> HC3N + H
\end{reactions}
The rate-limiting steps in these pathways are Reactions~R\ref{chem:c3h2_h_m} and R\ref{chem:cn_c2h2}, respectively.

Figure~\ref{fig:reaction_rates_a15_hc3n}b shows that HC$_3$N forms in CRAHCN-O with Reaction~R\ref{chem:cn_c2h2} as the rate-limiting step, but the process is slower due to low C$_2$H$_2$ abundance, as discussed in Section~\ref{subsec:pathways_hydrocarbons}. The other pathway to HC$_3$N involves HCN formation as described in Section~\ref{subsec:pathways_nitriles} and the rate-limiting step:
\begin{reactions}
    \label{chem:hcn_c2h}
    HCN + C2H &-> HC3N + H
\end{reactions}
CRAHCN-O predicts relatively low C$_2$H abundances, similar to C$_2$H$_2$, which makes this reaction slow. Hence, a clear difference between the networks is caused by the dominant role of hydrocarbons in HC$_3$N formation. Note that CRAHCN-O was not necessarily built to comprehensively simulate HC$_3$N formation, so distinct HC$_3$N profiles between the networks are not unexpected.


\subsubsection{Formaldehyde}\label{subsec:pathways_aldehydes}
\begin{figure}
\centering
\includegraphics[width=0.4\columnwidth]{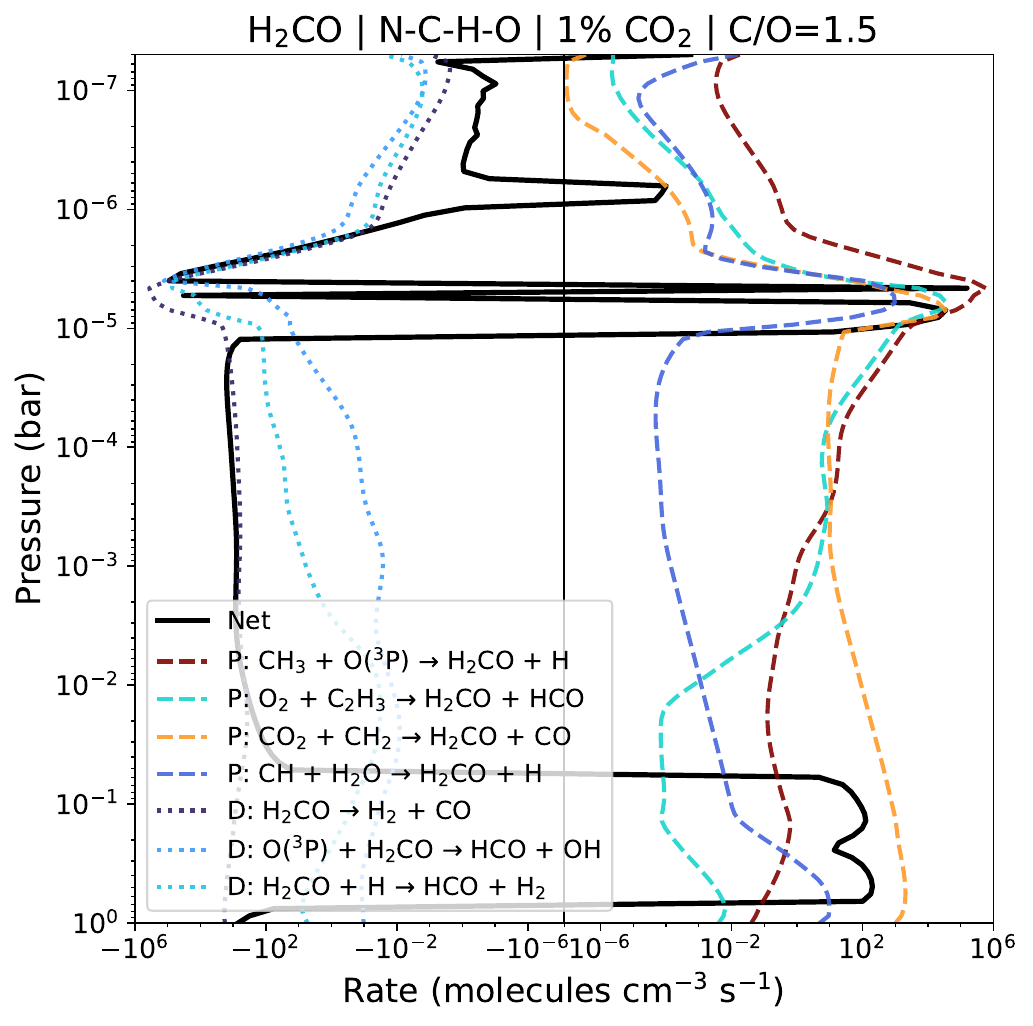}
\includegraphics[width=0.4\columnwidth]{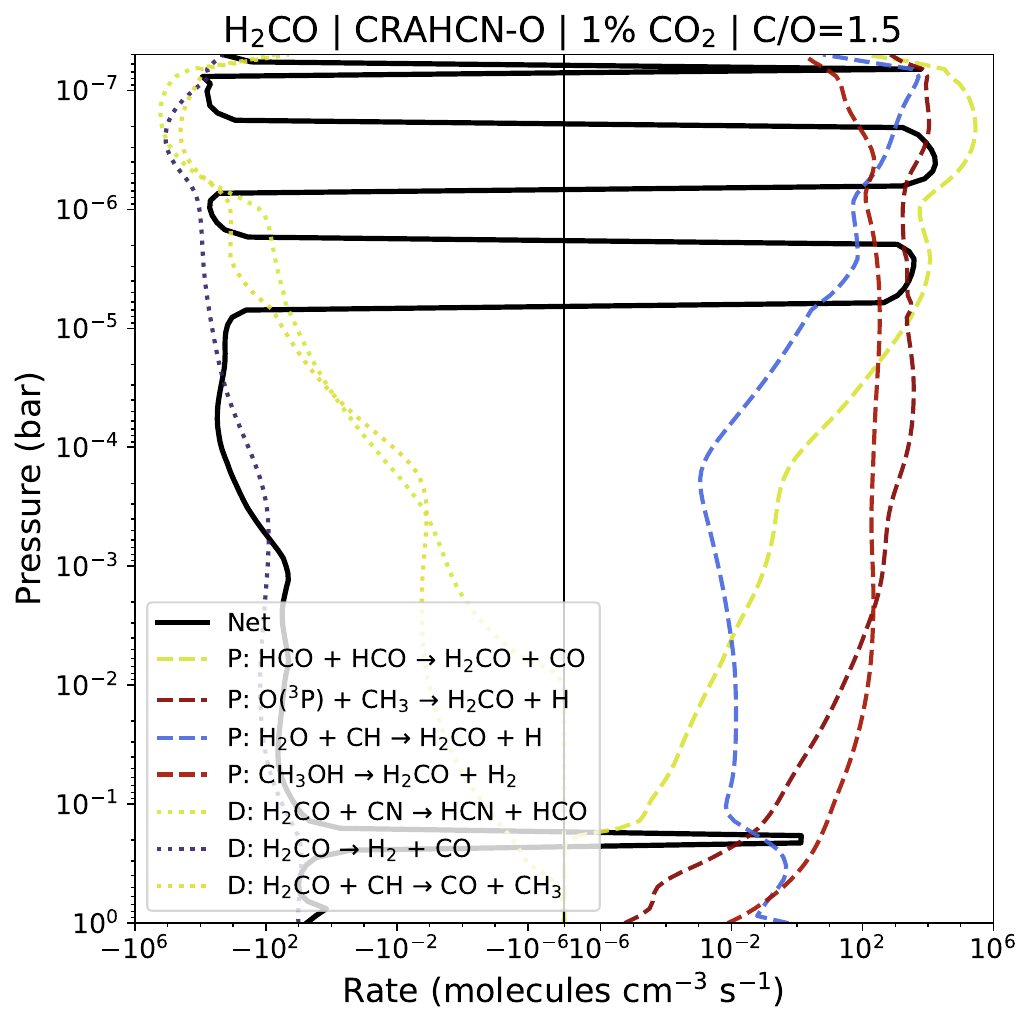}
\caption{Same as Figure~\ref{fig:reaction_rates_a05_hcn}, but showing the dominant rates controlling H$_2$CO in the 1\% CO$_2$ case at C/O=1.5.}
\label{fig:reaction_rates_a15_h2co}
\end{figure}

\begin{figure}
\centering
\includegraphics[width=0.3\columnwidth]{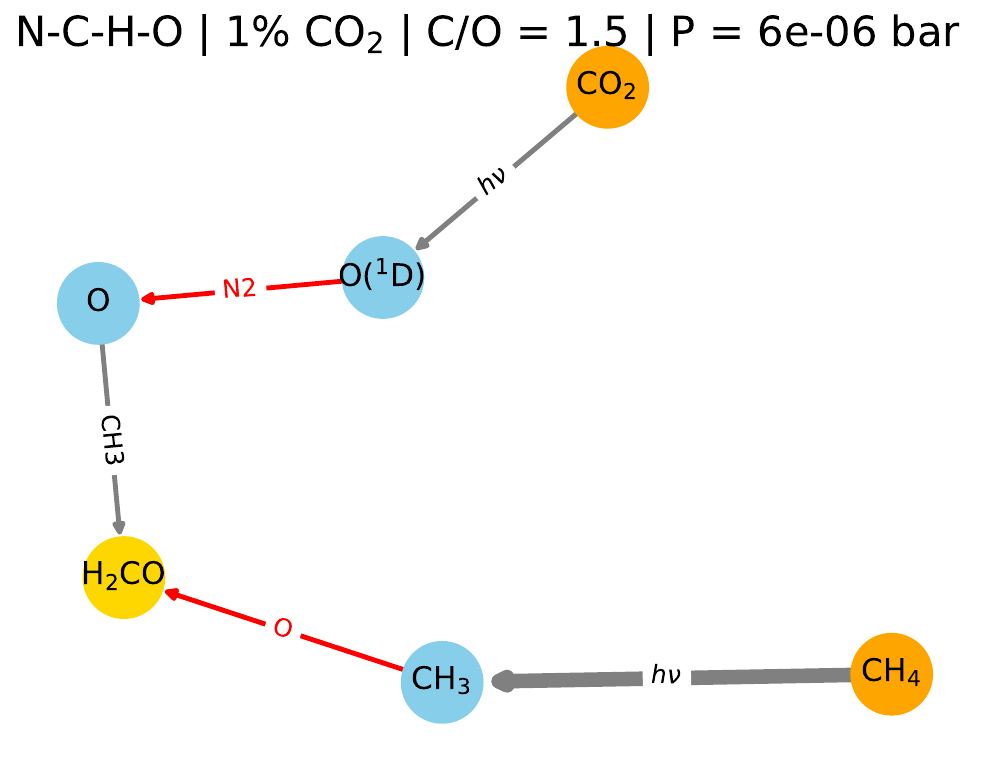}
\includegraphics[width=0.3\columnwidth]{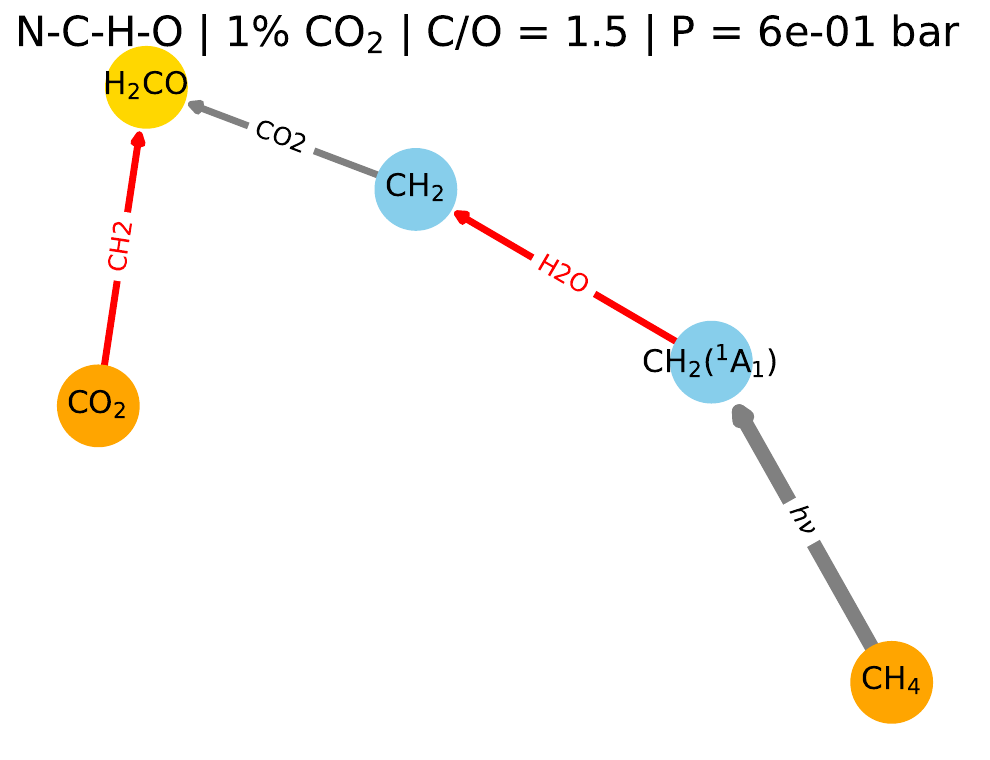}
\includegraphics[width=0.3\columnwidth]{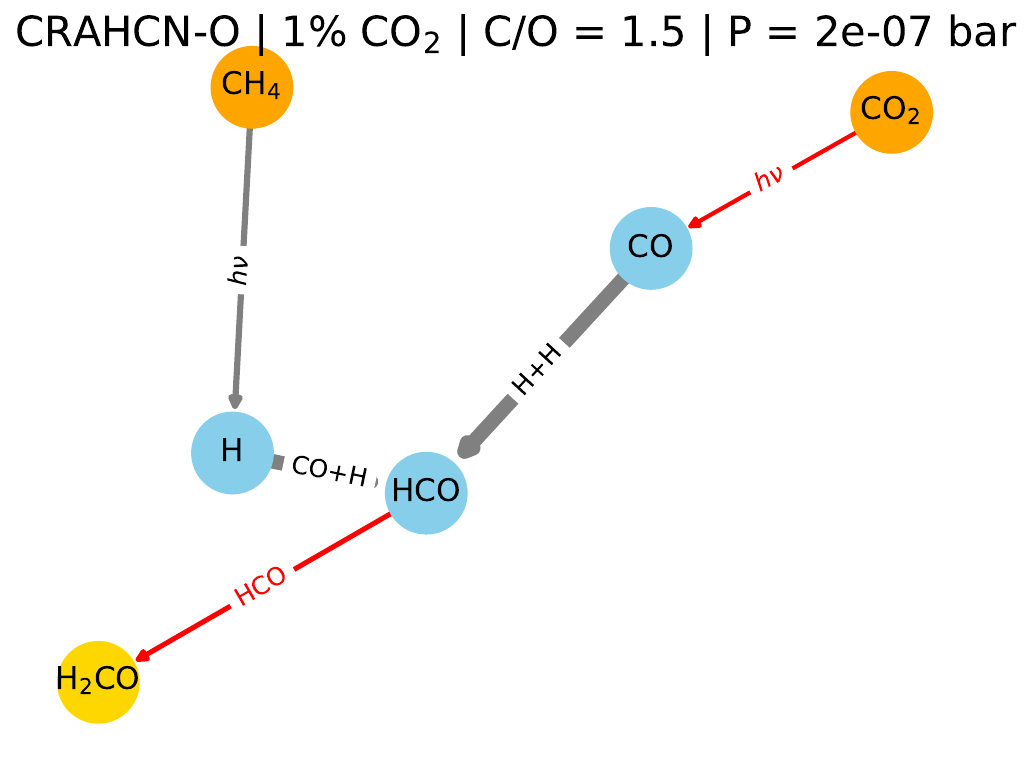}
\caption{Same as Figure~\ref{fig:pathways_a05_hcn}, but showing the dominant pathways to H$_2$CO for the 1\% CO$_2$ case at C/O=1.5.}
\label{fig:pathways_a15_h2co}
\end{figure}

The formation of H$_2$CO traces back to both CH$_4$ and CO$_2$. The net rates in Figure~\ref{fig:reaction_rates_a15_h2co} include strong peaks for both networks, although their vertical distributions differ due to the shielding effects in CRAHCN-O and network differences. The strongest peak in N-C-H-O involves CH$_3$ (${\sim}6\times10^{-6}$~bar), whereas the self-reaction of HCO dominates the strongest peaks in CRAHCN-O (${<}10^{-5}$~bar). For N-C-H-O, the most important pathways to H$_2$CO are represented by four reactions, as shown in Figure~\ref{fig:pathways_a15_h2co}, and start with CH$_4$ (Reaction~R\ref{chem:ch4_hv_1}) and CO$_2$ (Reaction~R\ref{chem:co2_hv_1}) photolysis followed by:
\begin{reactions}
    \label{chem:o1d_n2}
    O(^1D) + N2 &-> O(^3P) + N2 \\
    \label{chem:ch3_o3p}
    CH3 + O(^3P) &-> H2CO + H 
\end{reactions}
Importantly, net production of H$_2$CO is also predicted at high pressures (${>}10^{-1}$~bar) by N-C-H-O (Figure~\ref{fig:reaction_rates_a15_h2co}a), through a pathway that involves CH$_4$ photolysis via Reaction~R\ref{chem:ch4_hv_2} and CO$_2$:
\begin{reactions}
    \label{chem:1ch2_h2o}
    CH2(^1A_1) + H2O &-> CH2 + H2O\\
    \label{chem:ch2_co2}
    CH2 + CO2 &-> H2CO + CO
\end{reactions}
This pathway scales with background CO$_2$ and CH$_4$ levels, as it is less efficient for the 400~ppm CO$_2$ cases. For H$_2$CO formation, Reactions~R\ref{chem:o1d_n2} and R\ref{chem:1ch2_h2o} are the rate-limiting steps. In CRAHCN-O, H$_2$CO formation in the main net production peaks is dominated by the presence of HCO, which is formed through Reactions~R\ref{chem:co2_hv_1}, R\ref{chem:co_h_h} and R\ref{chem:hco_hco} as the rate-limiting step.


\subsubsection{Hydrocarbons}\label{subsec:pathways_hydrocarbons}

\begin{figure}
\centering
\includegraphics[width=0.4\columnwidth]{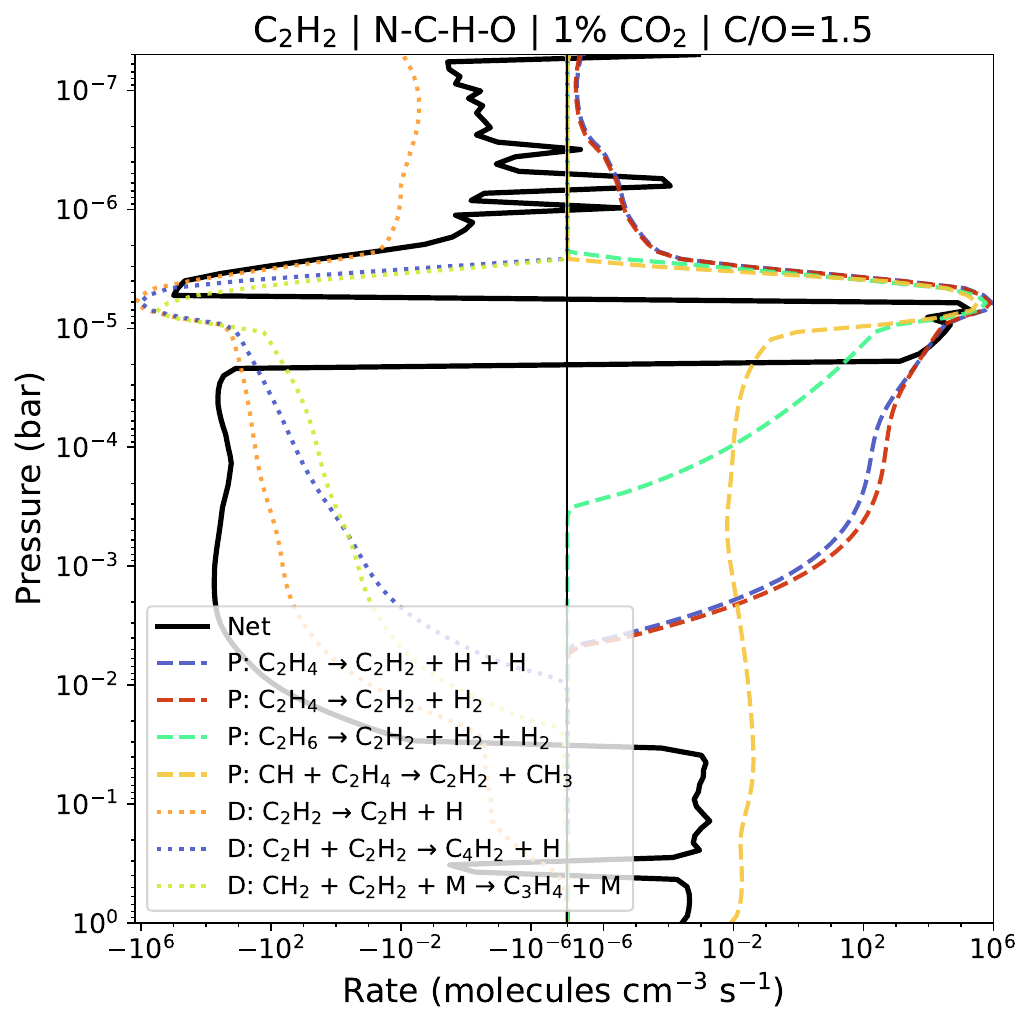}
\includegraphics[width=0.4\columnwidth]{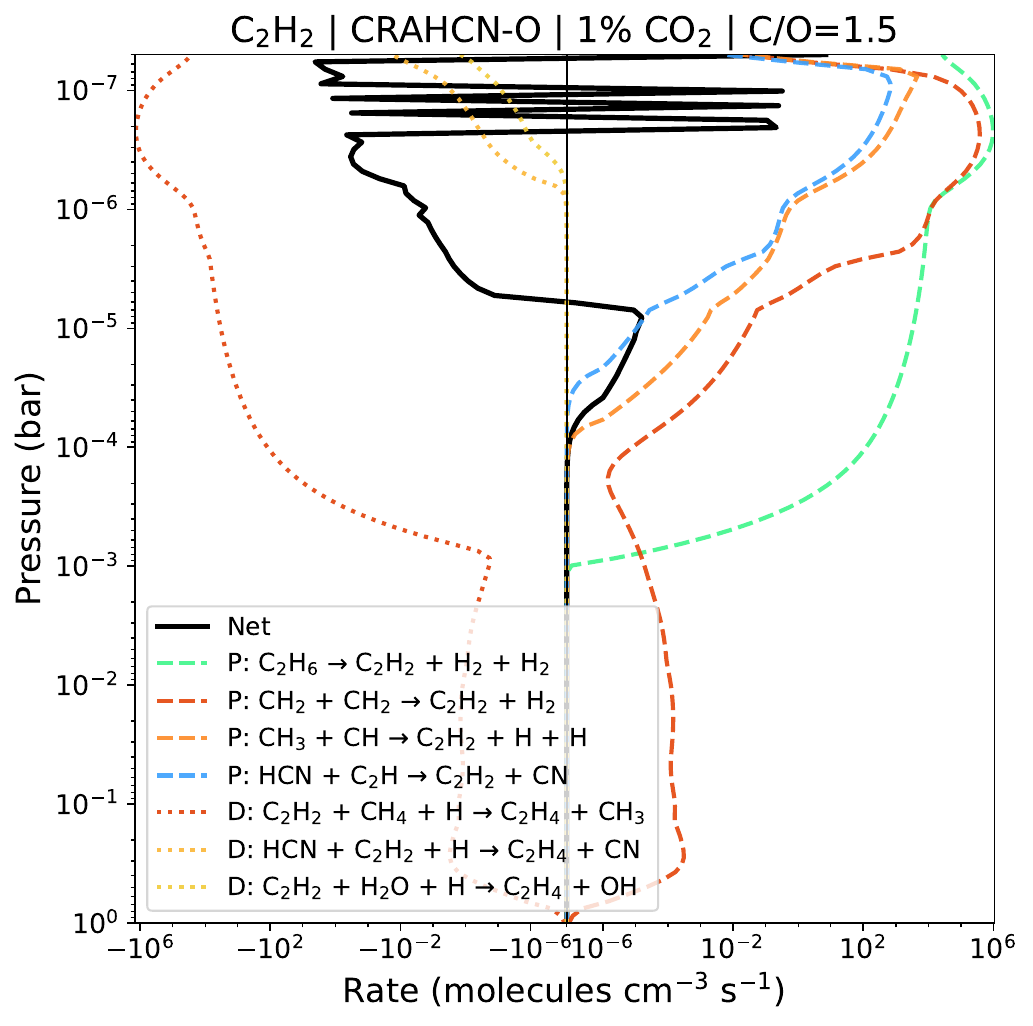}
\includegraphics[width=0.4\columnwidth]{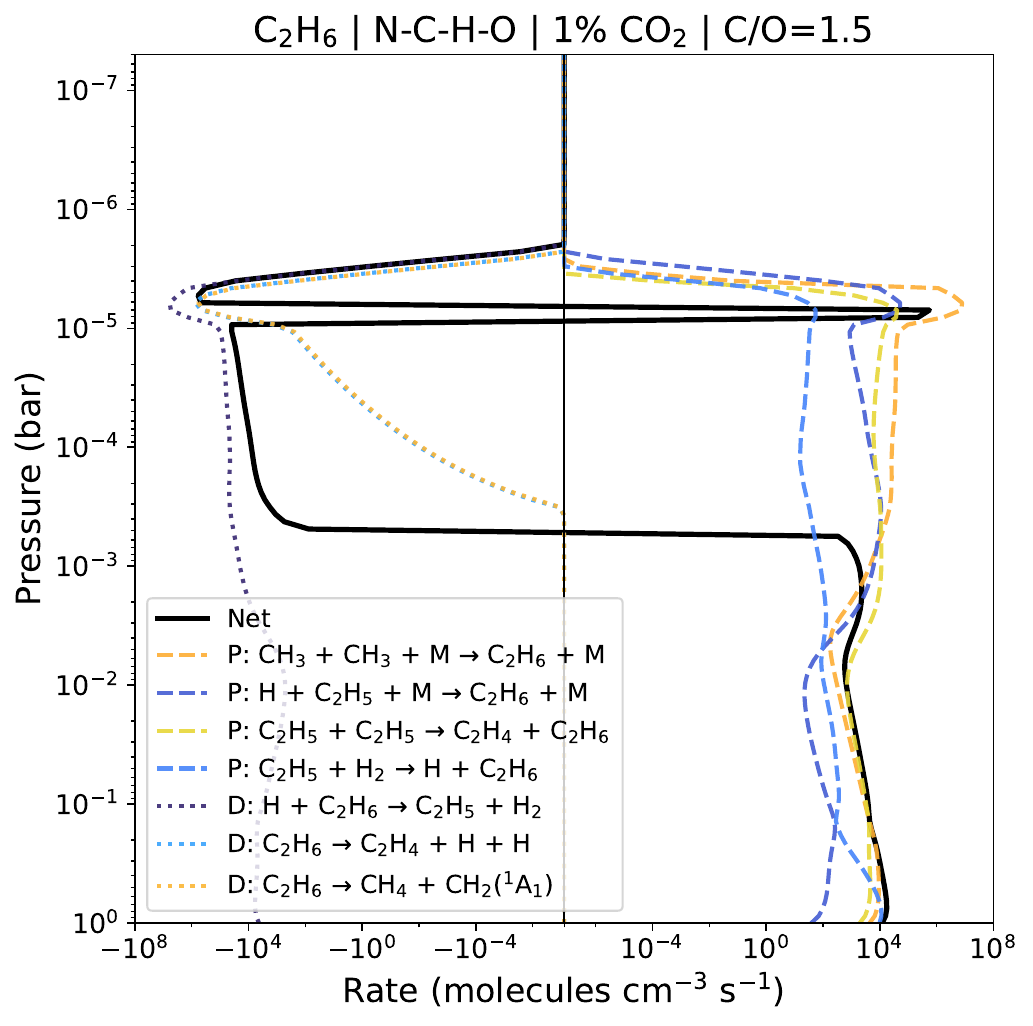}
\includegraphics[width=0.4\columnwidth]{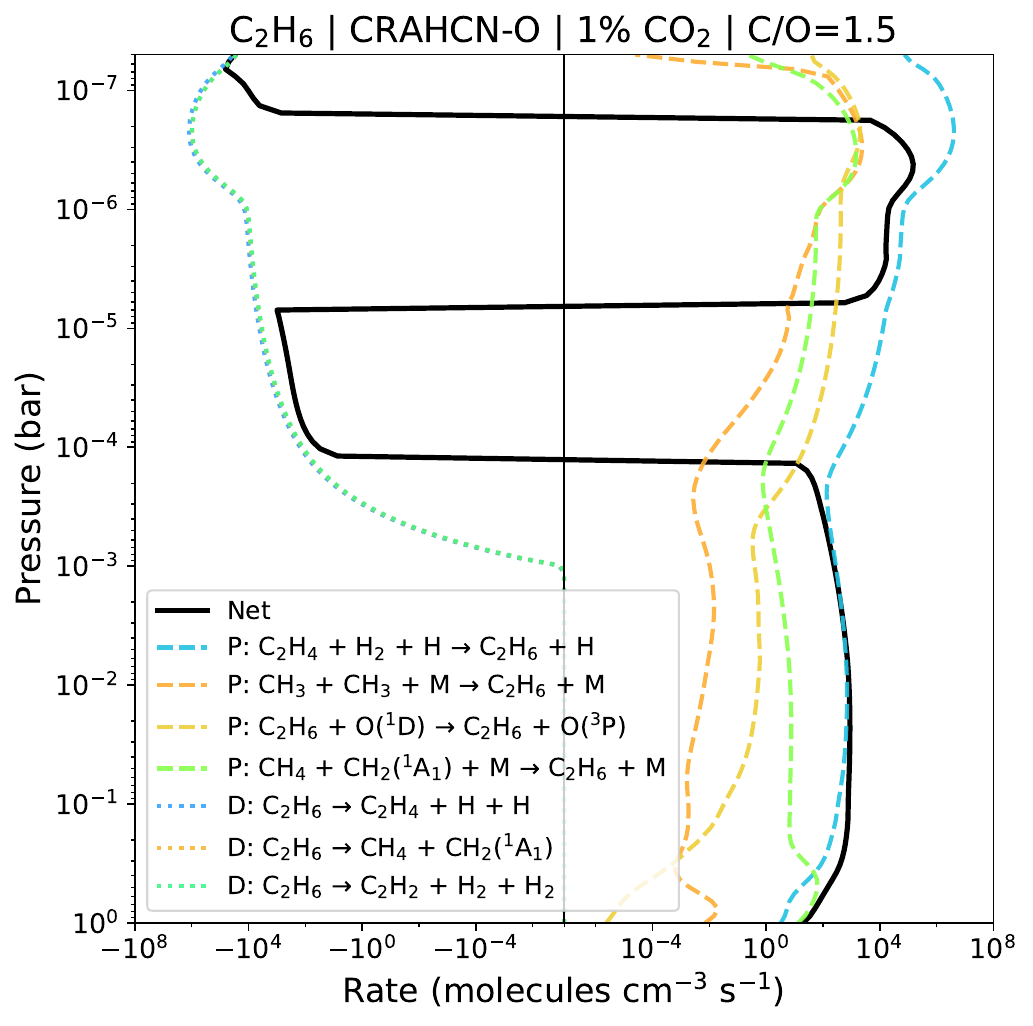}
\caption{Same as Figure~\ref{fig:reaction_rates_a05_hcn}, but showing the dominant rates controlling C$_2$H$_2$ and C$_2$H$_6$ in the 1\% CO$_2$ case at C/O=1.5.}
\label{fig:reaction_rates_a15_hydrocarbons}
\end{figure}

\begin{figure}
\centering
\includegraphics[width=0.35\columnwidth]{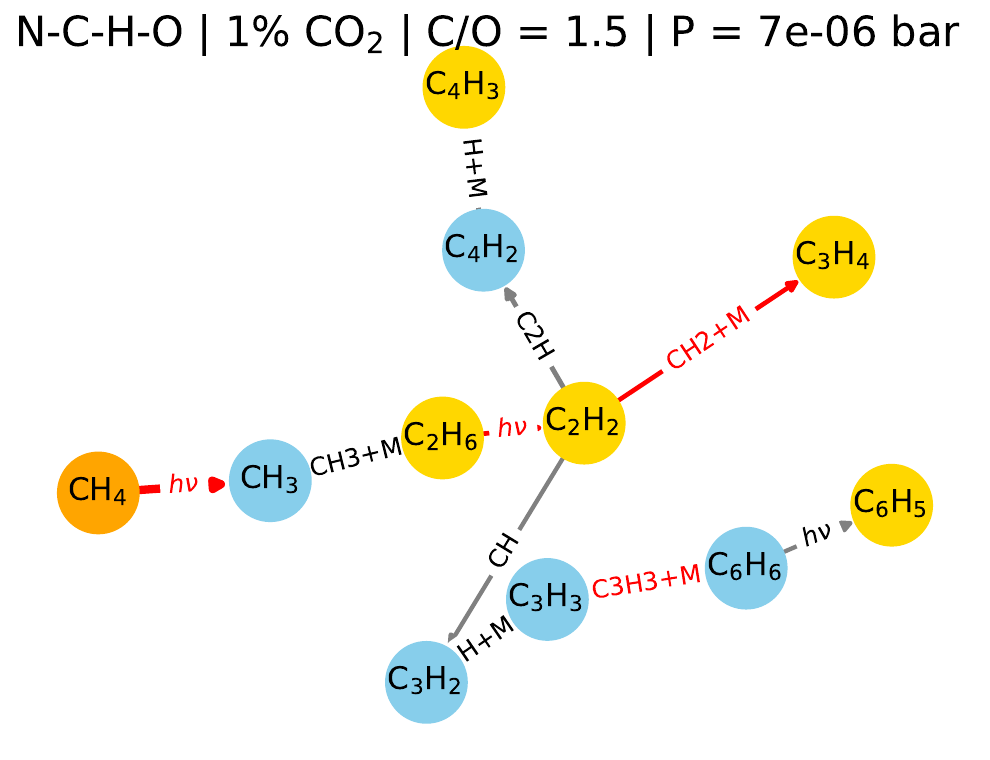}
\includegraphics[width=0.35\columnwidth]{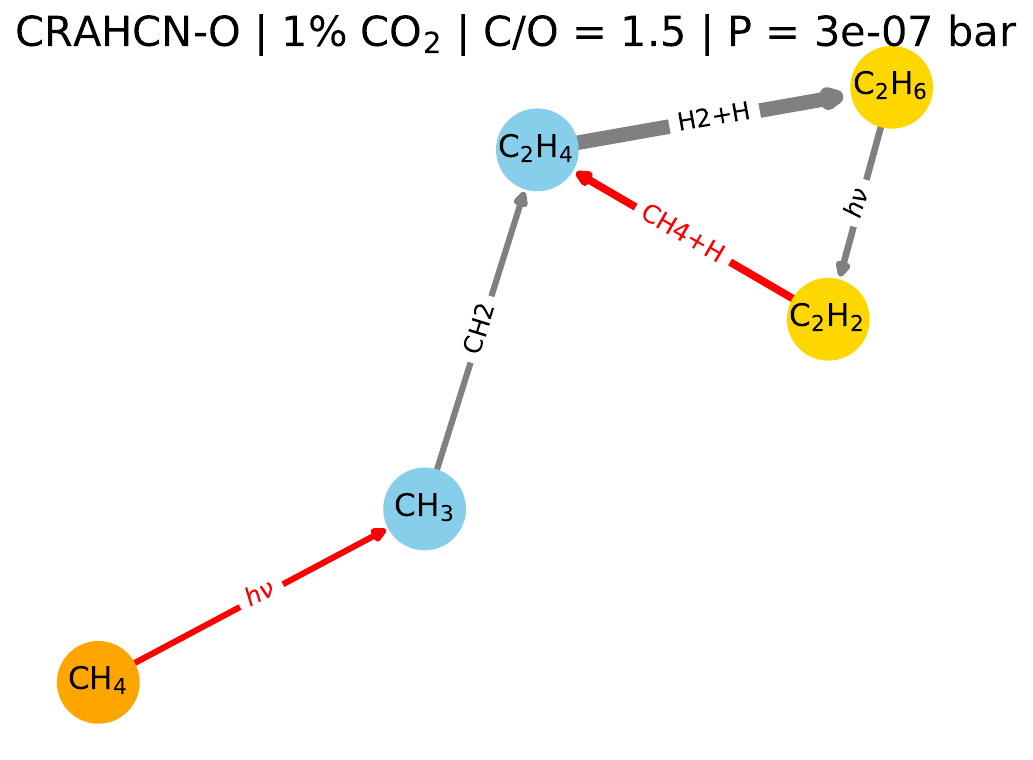}
\caption{Same as Figure~\ref{fig:pathways_a05_hcn}, but showing the dominant pathways to hydrocarbons for the 1\% CO$_2$ case at C/O=1.5.}
\label{fig:pathways_a15_hydrocarbons}
\end{figure}

The dominant production pathways of hydrocarbons like C$_2$H$_2$, C$_2$H$_6$, and C$_4$H$_3$ in the N-C-H-O network are found at the same pressure levels as the peaks in their abundance profiles in Figure~\ref{fig:all_vert_distribs}. For the 1\% CO$_2$ case at C/O=1.5, this corresponds to ${\sim}7\times10^{-6}$~bar, as can also be seen in panels a and c of Figure~\ref{fig:reaction_rates_a15_hydrocarbons}. In N-C-H-O, C$_2$H$_2$ predominantly forms from the photolysis of C$_2$H$_6$, meaning that the dominant production pathways are shared. The other production and destruction reactions show that other hydrocarbons (e.g., C$_2$H$_4$, C$_2$H$_5$, C$_2$H, C$_4$H$_2$, C$_3$H$_4$) are closely related to the abundances of C$_2$H$_2$ and C$_2$H$_6$. Figure~\ref{fig:pathways_a15_hydrocarbons} summarizes the pathways at a pressure of ${\sim}7\times10^{-6}$~bar and how they trace back to CH$_4$. Importantly, N-C-H-O accumulates the final nodes in these pathways, being C$_3$H$_4$, C$_4$H$_3$, and C$_6$H$_5$. Since the production of most hydrocarbons centres around the same pressure region, it is straightforward to construct a complete chemical mechanism from the pathway analysis for a simpler comparison between both networks and derive the net reactions. First, C$_3$H$_4$ traces back to the photolysis of CH$_4$ through two different channels:

\noindent\begin{minipage}{\textwidth}
\begin{reactions*}
   2(CH4 + h\nu &-> CH3 + H)
\tag{\ref{chem:ch4_hv_1}} \\
    CH4 + h\nu &-> CH2(^1A_1) + H2 
\tag{\ref{chem:ch4_hv_2}} \\
    CH2(^1A_1) + H2O &-> CH2 + H2O
\tag{\ref{chem:1ch2_h2o}} \\
    CH3 + CH3 + M &-> C2H6 + M
\tag{\ref{chem:ch3_ch3_m}} \\
    C2H6 + h\nu &-> C2H2 + H2 + H2
\tag{\ref{chem:c2h6_hv}} 
\end{reactions*}
\vspace*{-3em}
\begin{reactions}
    \label{chem:ch2_c2h2_m}
    CH2 + C2H2 + M &-> C3H4 + M
\end{reactions}
\vspace*{-3em}
\begin{reactions*}
\textbf{Net:} 3CH4 + 4h\nu &-> C3H4 + 3H2 + 2H
\end{reactions*}
\end{minipage}

\vspace{\baselineskip}

As shown in Figure~\ref{fig:pathways_a15_hydrocarbons}, the rate-limiting steps include Reactions~R\ref{chem:ch4_hv_1} for C$_2$H$_6$, R\ref{chem:c2h6_hv} for C$_2$H$_2$, and R\ref{chem:ch2_c2h2_m} for C$_3$H$_4$.
C$_4$H$_3$ is formed from photolysis of multiple CH$_4$ molecules in a single channel:

\noindent\begin{minipage}{\textwidth}
\begin{reactions*}
    4(CH4 + h\nu &-> CH3 + H)
\tag{\ref{chem:ch4_hv_1}} \\
    2(CH3 + CH3 + M &-> C2H6 + M)
\tag{\ref{chem:ch3_ch3_m}} \\
   2(C2H6 + h\nu &-> C2H2 + H2 + H2)
\tag{\ref{chem:c2h6_hv}} 
\end{reactions*}
\vspace{-3em}
\begin{reactions}
    \label{chem:c2h2_hv}
    C2H2 + h\nu &-> C2H + H \\
    \label{chem:c2h_c2h2}
    C2H + C2H2 &-> C4H2 + H \\
    \label{chem:c4h2_h_m}
    C4H2 + H + M &-> C4H3 + M
\end{reactions}
\vspace*{-3em}
\begin{reactions*}
    \textbf{Net:} 4CH4 +7h\nu &-> C4H3 + 4H2 + 5H
\end{reactions*}
\end{minipage}

\vspace{\baselineskip}

The rate-limiting step for C$_4$H$_3$ is Reaction~R\ref{chem:c4h2_h_m}. The same photolysis channel of CH$_4$ and a third channel drive the formation of C$_6$H$_5$ and C$_6$H$_6$:

\noindent\begin{minipage}{\textwidth}
\begin{reactions*}
    4(CH4 + h\nu &-> CH3 + H)
\tag{\ref{chem:ch4_hv_1}} \\
    2(CH3 + CH3 + M &-> C2H6 + M)
\tag{\ref{chem:ch3_ch3_m}} \\
   2(C2H6 + h\nu &-> C2H2 + H2 + H2)
\tag{\ref{chem:c2h6_hv}} 
\end{reactions*}
\vspace*{-3em}
\begin{reactions}
    \label{chem:ch4_hv_3}
    2(CH4 + h\nu &-> CH + H2 + H) \\
    \label{chem:c2h2_ch}
    2(C2H2 + CH &-> C3H2 + H) 
\end{reactions}
\vspace*{-3em}
\begin{reactions*}
     2(C3H2 + H + M &-> C3H3 + M)
\tag{\ref{chem:c3h2_h_m}} 
\end{reactions*}
\vspace*{-3em}
\begin{reactions}
    \label{chem:c3h3_c3h3_m}
    C3H3 + C3H3 + M &-> C6H6 + M\\
    \label{chem:c6h6_hv}
    C6H6 + h\nu &-> C6H5 + H
\end{reactions}
\vspace*{-3em}
\begin{reactions*}
    \textbf{Net:} 6CH4 + 9h$\nu$ &-> C6H5 + 6H2 + 7H\\
\end{reactions*}
\end{minipage}

\vspace{\baselineskip}

The rate-limiting steps are Reaction~R\ref{chem:c3h3_c3h3_m} for C$_6$H$_6$ and R\ref{chem:c6h6_hv} for C$_6$H$_5$. The four net reactions above then summarise the photochemical buildup of hydrocarbons in the N-C-H-O network. Another important pathway to C$_2$H$_2$ is initiated by CH$_4$ photolysis to CH (Reaction~R\ref{chem:ch4_hv_3}) and involves C$_2$H$_4$ as an intermediate species:
\begin{reactions}
    \label{chem:ch_ch4}
    CH + CH4 &-> H + C2H4\\
    \label{chem:c2h4_hv}
    C2H4 + h\nu &-> C2H2 + H + H
\end{reactions}
The dominant destruction pathway for C$_3$H$_4$ at all pressure levels and initial conditions forms C$_2$H$_2$ again (not shown):
\begin{reactions}
    \label{chem:c3h4_h}
    C3H4 + H &-> CH3 + C2H2
\end{reactions}
However, formation reactions are persistently faster than destruction ones, thus promoting C$_3$H$_4$ accumulation in N-C-H-O. C$_4$H$_3$ destruction leads to the most complex hydrocarbons in the network (C$_6$H$_5$ and C$_6$H$_6$):
\begin{reactions}
    \label{chem:c2h2_c4h3_m}
    C2H2 + C4H3 + M &-> C6H5 + M\\ 
    \label{chem:c6h5_h_m}
    C6H5 + H + M &-> C6H6 + M
\end{reactions}
Again, the balance between reaction strengths strongly favours C$_4$H$_3$ accumulation. Additionally, C$_3$H$_3$ can directly form C$_3$H$_4$:
\begin{reactions}
    \label{chem:c3h3_h_m}
    C3H3 + H + M &-> C3H4 + M
\end{reactions}
Since the pathways to C$_6$H$_5$ and C$_6$H$_6$ are slower, C$_4$H$_3$ and C$_3$H$_4$ end up as the most abundant hydrocarbon reservoirs in N-C-H-O. As mentioned in Section~\ref{sec:vertdistributions} and shown in Appendix Figure~\ref{fig:ncho_crahcno_argo_prebiotic}, C$_4$H$_3$ abundances are about an order of magnitude higher than C$_3$H$_4$, illustrating the preferential formation of C$_4$H$_3$ due to atomic H (relatively abundant at low pressures) driving its rate-limiting step (Reaction~R\ref{chem:c4h2_h_m}).

For CRAHCN-O, C$_2$H$_2$ and C$_2$H$_6$ production again show a peak as a function of pressure (${\sim}3\times10^{-7}$~bar, see Figure~\ref{fig:reaction_rates_a15_hydrocarbons}b). At this pressure level, we can construct a chemical mechanism for C$_2$H$_6$ formation based on the pathways in Figure~\ref{fig:pathways_a15_hydrocarbons}:

\noindent\begin{minipage}{\columnwidth}
\begin{reactions*}
    CH4 + h\nu &-> CH3 + H
\tag{\ref{chem:ch4_hv_1}} \\
    CH4 + h\nu &-> CH2(^1A_1) + H2 
\tag{\ref{chem:ch4_hv_2}} \\
    CH2(^1A_1) + H2O &-> CH2 + H2O
\tag{\ref{chem:1ch2_h2o}} 
\end{reactions*}
\vspace*{-3em}
\begin{reactions}
  \label{chem:ch3_ch2}
  CH3 + CH2 &-> C2H4 + H \\
  \label{chem:c2h4_h2_h}
  C2H4 + H2 + H &-> C2H6 + H
\end{reactions}
\vspace*{-3em}
\begin{reactions*}
    \textbf{Net:} 2CH4 + 2h$\nu$ &-> C2H6 + 2H
\end{reactions*}
\end{minipage}

\vspace{\baselineskip}

If we keep the five reactions of this mechanism and add the cycle introduced by C$_2$H$_6$ destruction twice, we find that C$_2$H$_6$ is formed along with one of its precursors:

\noindent\begin{minipage}{\textwidth}
\begin{reactions*}
  C2H6 + h\nu &-> C2H2 + H2 + H2 \tag{\ref{chem:c2h6_hv}} 
\end{reactions*}
\vspace*{-3em} \begin{reactions}
  \label{chem:c2h2_ch4_h}
  C2H2 + CH4 + H &-> C2H4 + CH3 
\end{reactions} 
\vspace*{-3em}
\begin{reactions*}
  C2H4 + H2 + H &-> C2H6 + H \tag{\ref{chem:c2h4_h2_h}} \\
  C2H6 + h\nu &-> C2H2 + H2 + H2 \tag{\ref{chem:c2h6_hv}} \\
  C2H2 + CH4 + H &-> C2H4 + CH3 \tag{\ref{chem:c2h2_ch4_h}} \\
  C2H4 + H2 + H &-> C2H6 + H \tag{\ref{chem:c2h4_h2_h}} \\
  \textbf{Net:} 4CH4 + 4h$\nu$ &-> 2C2H6 + CH3 + 2H2 + H
\end{reactions*}
\end{minipage}

\vspace{\baselineskip}

This CH$_3$ can again be used for C$_2$H$_6$ production through Reaction~R\ref{chem:ch3_ch3_m}. However, the tendency to accumulate C$_2$H$_6$ is mainly due to the imbalance in reaction rates in the cycle between C$_2$H$_4$, C$_2$H$_6$, and C$_2$H$_2$ as shown by the width of the edges in Figure~\ref{fig:pathways_a15_hydrocarbons}. The wider edge of Reaction~R\ref{chem:c2h4_h2_h} shows that it is faster than Reactions~R\ref{chem:c2h6_hv} and R\ref{chem:c2h2_ch4_h} and illustrates how the cycle can lead to an accumulation of C$_2$H$_6$. This mechanism subsequently causes the substantial photochemical shielding by C$_2$H$_6$ in CRAHCN-O, further affecting other feedstock molecules. Ultimately, the extent and effects of the C$_2$H$_6$ accumulation in CRAHCN-O are due to the absence of pathways to more complex hydrocarbons like those in N-C-H-O.

For pressures ${>}10^{-4}$~bar, we also see inter-network differences in the net rates in Figure~\ref{fig:reaction_rates_a15_hydrocarbons}, where C$_2$H$_6$ formation in N-C-H-O is dominated by Reaction~R\ref{chem:ch3_ch3_m} and in CRAHCN-O by Reaction~R\ref{chem:c2h4_h2_h} and at the highest pressures by:
\begin{reactions}
    \label{chem:ch4_1ch2_m}
    CH4 + CH2(^1A_1) + M &-> C2H6 + M
\end{reactions}
These partially explain the differences in C$_2$H$_6$ abundances at higher pressures, along with vertical mixing from the peak production regions. Unlike C$_2$H$_6$, net production of C$_2$H$_2$ is negligible at ${>}10^{-4}$~bar in both networks (Figure~\ref{fig:reaction_rates_a15_hydrocarbons}). Therefore, the tropospheric abundances of C$_2$H$_2$ and its derivatives (e.g., C$_3$H$_4$, C$_4$H$_3$, C$_6$H$_5$, C$_6$H$_6$) strongly depend on vertical mixing from the photochemical production regions in the upper atmosphere. This has profound implications for the predicted tropospheric abundances with the N-C-H-O network. With increasing CO$_2$ and CH$_4$ levels, the production regions shift to lower pressures, making it harder for mixing processes to reach the troposphere by crossing the mixing barrier presented by the tropopause (${\sim}10^{-1}$~bar in Figure~\ref{methods-fig:pcb_ptkzz_flux}). Therefore, with increasing CO$_2$ and CH$_4$ levels, the photochemically produced C$_2$H$_2$ and C$_4$H$_3$ start to disappear from the troposphere, as we see for the 10\% CO$_2$ cases in Figure~\ref{fig:all_vert_distribs}f and i. The same reasoning explains the absence of HCN and HC$_3$N in the troposphere for the high CO$_2$ and CH$_4$ scenarios. Since pathways at high pressures still exist for C$_2$H$_6$ (Reaction~R\ref{chem:ch3_ch3_m}) as well as H$_2$CO (Reaction~R\ref{chem:ch2_co2}), their tropospheric abundances are less affected by this inability of vertical mixing to transport the photochemical products down to higher pressures.


\begin{table}
	\centering
	\caption{Key bimolecular reactions in the pathways to feedstock molecules, comparing N-C-H-O and CRAHCN-O. We provide the rate coefficients $A$, $b$, and $E_a$ as well as the rate at 298~K ($k_{298}$). Note that similar rates at 298~K do not guarantee agreement at lower temperatures relevant to the simulated atmosphere (see Figure~\ref{methods-fig:pcb_ptkzz_flux}). In the final column, we recommend using the rate coefficients from CRAHCN-O (C), N-C-H-O (N), or further measurements (M). The three blocks split the reactions controlling nitriles, formaldehyde, and hydrocarbons.}
	\label{tab:key_reactions_pathways}
    \scriptsize
	\begin{tabular}{l|l|l|l|l|l|l} 
		\hline
		R & Reaction & \multicolumn{2}{c|}{[$A$ (cm$^{3}$s$^{-1}$), $b$, $E_a$ (K)]} & \multicolumn{2}{c|}{k$_{298}$ (cm$^{3}$s$^{-1}$)} & Rec. \\
          & & CRAHCN-O & N-C-H-O & CRAHCN-O & N-C-H-O & \\ \hline
        R\ref{chem:n_ch2} & \ce{N + CH2 -> HCN + H} & [1e-10, 0, 0] & [4.3e-10, 0, 420] & 1.0e-10 & 1.05e-10 & M \\
        R\ref{chem:cn_co2} & \ce{CN + CO2 -> CO + NCO} & [N/A] & [6.14e-12, 0, 0] & N/A & 6.14e-12 & M\\    
        R\ref{chem:nco_h2} & \ce{HNCO + H -> NCO + H2} & [N/A] & [2.91e-19, 2.41, 6190] & N/A & 2.54e-22 & N\\    
        R\ref{chem:hnco_cn} & \ce{NCO + HCN -> HNCO + CN} & [N/A] & [2.01e-11, 0, 4460] & N/A & 6.36e-18 & M \\
        R\ref{chem:nco_hco} & \ce{NCO + HCO -> HNCO + CO} & [N/A] & [5.99e-11, 0, 0] & N/A & 5.99e-11 & M \\
        R\ref{chem:ch3_n} & \ce{CH3 + N -> H2CN + H} & [7.7e-11, 0, 0] & [1.01e-9, -0.31, 145] & 7.7e-11 & 1.06e-10 & C \\
        R\ref{chem:h2cn_n} & \ce{H2CN + N -> HCN + NH} & [1e-10, 0, 200] & [N/A] & 5.11e-11 & N/A & C \\
        R\ref{chem:h2co_cn} & \ce{H2CO + CN -> HCN + HCO} & [2.74e-19, 2.72, -718] & [N/A] & 1.64e-11 & N/A & M\\
        R\ref{chem:cn_h2} & \ce{CN + H2 -> HCN + H} & [5.93e-16, 1.55, 1510] & [5.99e-16, 1.55, 1510] & 2.56e-14 & 2.58e-14 & - \\
        R\ref{chem:ch4_cn} & \ce{CH4 + CN -> HCN + CH3} & [1.50e-11, 0, 940.0] & [1.50e-19, 2.64, -150] & 6.40e-13 & 8.45e-13 & C \\
        R\ref{chem:n_c3h3} & \ce{N + C3H3 -> HC3N + H} & [N/A] & [5.00e-11, 0, 0] & N/A & 5.00e-11 & M \\
        R\ref{chem:cn_c2h2} & \ce{CN + C2H2 -> HC3N + H} & [2.27e-10, 0, 0] & [2.27e-10, 0, 0] & 2.27e-10 & 2.27e-10 & - \\
        R\ref{chem:hcn_c2h} & \ce{HCN + C2H -> HC3N + H} & [5.30e-12, 0, 768.4] & [5.30e-12, 0, 769] & 4.02e-13 & 4.01e-13 & M \\ \hline
        R\ref{chem:ch4_oh} & \ce{CH4 + OH -> H2O + CH3} & [2.58e-17, 1.83, 1396] & [1.68e-18, 2.18, 1231] & 8.03e-15 & 6.68e-15 & -- \\ 
        R\ref{chem:ch4_o1d} & \ce{CH4 + O(^1D) -> OH + CH3} & [2.2e-10, 0, 0] & [1.13e-10, 0, 0] & 2.2e-10 & 1.13e-10 & M \\ 
        R\ref{chem:co_h_h} & \ce{HCO + H -> CO + H + H} & [2.4e-11, 0, 0] & [N/A] & 2.4e-11 & N/A & M \\ 
        R\ref{chem:hco_hco} & \ce{HCO + HCO -> H2CO + CO} & [4.20e-11, 0, 0] & [3.01e-11, 0, 0] & 4.2e-11 & 3.01e-11 & -- \\ 
        R\ref{chem:o1d_n2} & \ce{O(^1D) + N2 -> O(^3P) + N2} & [N/A] & [1.79E-11, 0, -107] & N/A & 2.56e-11 & N \\ 
        R\ref{chem:ch3_o3p} & \ce{CH3 + O(^3P) -> H2CO + H} & [1.40e-10, 0, 0] & [1.40e-10, 0, 0] & 1.40e-10 & 1.40e-10 & -- \\ 
        R\ref{chem:1ch2_h2o} & \ce{CH2(^1A_1) + H2O -> CH2 + H2O} & [N/A] & [5.0e-11, 0, 0] & N/A & 5.0e-11 & N \\ 
        R\ref{chem:ch2_co2} & \ce{CH2 + CO2 -> H2CO + CO} & [N/A] & [3.90e-14, 0, 0] & N/A & 3.9e-14 & M \\ \hline
        R\ref{chem:ch_c2h2} & \ce{CH + C2H2 -> C3H2 + H} & [N/A] & [1.5e-10, 0, -252] & N/A & 3.49e-10 & N \\ 
        R\ref{chem:c2h_c2h2} & \ce{C2H + C2H2 -> C4H2 + H} & [N/A] & [3.18e-11, 0.24, -37] & N/A & 1.41e-10 & N \\ 
        R\ref{chem:ch_ch4} & \ce{CH + CH4 -> H + C2H4} & [9.8e-11, 0, 0] & [5e-11, 0, -200] & 9.8e-11 & 9.78e-11 & N \\ 
        R\ref{chem:c3h4_h} & \ce{CH3 + C2H2 -> C3H4 + H} & [N/A] & [8.57e-15, 0.86, 11100] & N/A & 7.66e-29 & M \\
        R\ref{chem:ch3_ch2} & \ce{CH3 + CH2 -> C2H4 + H} & [7e-11, 0, 0] & [7.01e-11, 0, 0] & 7e-11 & 7.01e-11 & --\\ 
        R\ref{chem:c2h4_h2_h} & \ce{C2H6 + H  -> C2H5 + H2} & [2.39e-15, 1.5, 3730] & [9.19E-22, 3.5, 2600] & 6.11e-16 & 6.82e-17 & M \\ 
         & \ce{-> C2H4 + H2 + H} & &  &  &  \\ 
        R\ref{chem:c2h2_ch4_h} & \ce{C2H4 + CH3 -> C2H3 + CH4} & [6.91e-12, 0, 5599] & [N/A] & 4.78e-20 & N/A & M \\ 
        & \ce{-> C2H2 + CH4 + H} & &  &  & \\
        \hline
    \end{tabular}
    
    {References and notes:  
    R\ref{chem:n_ch2}: \citet{pearce_consistent_2019}; estimated by \citet{tsai_comparative_2021},
    R\ref{chem:cn_co2}: \citet{haynes_behavior_1975},
    R\ref{chem:nco_h2}: Inverse; \citet{nguyen_reaction_1996},
    R\ref{chem:hnco_cn}: Inverse; \citet{tsang_chemical_1992},
    R\ref{chem:nco_hco}: \citet{tsang_chemical_1991},
    R\ref{chem:ch3_n}: \citet{pearce_hcn_2020}; \citet{DAVIDSON1991267, dean_combustion_2000},
    R\ref{chem:h2cn_n}: \citet{nesbitt1990kinetic},
    R\ref{chem:h2co_cn}: \citet{yu_kinetics_1993},
    R\ref{chem:cn_h2}: \citet{tsang_chemical_1992},
    R\ref{chem:ch4_cn}: \citet{baulch_evaluated_1992}; \citet{baulch_1994JPCRD..23..847B},
    R\ref{chem:n_c3h3}: \citet{Loison2017},
    R\ref{chem:cn_c2h2}: \citet{gannon_h_2007},
    R\ref{chem:hcn_c2h}: \citet{hoobler_rate_1997},
    R\ref{chem:ch4_oh}: \citet{baulch_evaluated_1992}; \citet{srinivasan_reflected_2005},
    R\ref{chem:ch4_o1d}: \citet{pearce_crahcn-o_2020}; \citet{demore_1994},
    R\ref{chem:co_h_h}: inverse; \citet{pearce_crahcn-o_2020},
    R\ref{chem:hco_hco}: \citet{pearce_crahcn-o_2020}; \citet{tsang1986chemical},
    R\ref{chem:o1d_n2}: \citet{atkinson_evaluated_1997},
    R\ref{chem:ch3_o3p}: \citet{baulch_evaluated_1992},
    R\ref{chem:1ch2_h2o}: \citet{wang_detailed_1997},
    R\ref{chem:ch2_co2}: \citet{tsang1986chemical},
    R\ref{chem:ch_c2h2}: \citet{guadagnini1998ab},
    R\ref{chem:ch_ch4}: \citet{berman_kinetics_1983}; \citet{baulch_evaluated_1992},
    R\ref{chem:c2h_c2h2}: \citet{eiteneer2003experimental},
    R\ref{chem:c3h4_h}: Inverse; \citet{davis1999propyne},
    R\ref{chem:ch3_ch2}: \citet{baulch_evaluated_1992},
    R\ref{chem:c2h4_h2_h}: Inverse, \citet{baulch_evaluated_1992},
    C2H6 + H  $\to$ C2H5 + H2: Inverse; \citet{tsang1986chemical},
    R\ref{chem:c2h2_ch4_h}: Inverse; \citet{baulch_evaluated_1992}.}
\end{table}

\begin{table}
	\centering
	\caption{Key termolecular reactions in the pathways to feedstock molecules (particularly the hydrocarbons), comparing N-C-H-O and CRAHCN-O. We provide the rate coefficients $A_0$, $b_0$, $E_{a,0}$, $A_\infty$, $b_\infty$, and $E_{a,\infty}$ as well as $k_{298}$ at 298~K and [M]=2.5${\times}10^{19}$ molec~cm$^{-3}$. Note that agreement in $k_{298}$ does not necessarily imply agreement at the lower temperatures relevant to much of the simulated atmosphere (see~\ref{methods-fig:pcb_ptkzz_flux}).}
	\label{tab:key_reactions_pathways_termol}
    \scriptsize
	\begin{tabular}{l|l|l|l|l|l} 
		\hline
		R & Reaction (+ M) & \multicolumn{2}{c|}{[$A_0$, $b_0$, $E_{a,0}$]; [$A_\infty$, $b_\infty$, $E_{a,\infty}$]} & \multicolumn{2}{c}{k$_{298}$ (cm$^6$ s$^{-1}$)} \\
          & & CRAHCN-O & N-C-H-O & CRAHCN-O & N-C-H-O \\ \hline
        R\ref{chem:ch3_ch3_m} & \ce{CH3 + CH3 -> C2H6} & [1.7e-26, 0, 0]; [6e-11, 0, 0] & [3.5e-07, -7, 1390]; [1.58e-09, -0.54, 68.6] & 6.00e-11 & 5.78e-11 \\
        R\ref{chem:ch2_c2h2_m} & \ce{CH2 + C2H2 -> C3H4} & [N/A]; [N/A] & [3.09e-15, -3.21, 162]; [1.07e-16, 1.48, -831] & N/A & 7.99e-12 \\
        R\ref{chem:c4h2_h_m} & \ce{C4H2 + H -> C4H3} & [N/A]; [N/A] & [6.16e-03, -8.9, 1261]; [7.08e-14, 1.16, 882] & N/A & 2.72e-12 \\
        R\ref{chem:c2h2_c4h3_m} & \ce{C2H2 + C4H3 -> C6H5} & [N/A]; [N/A] & [3e-23, -1, 0]; [1.16e-09, -0.86, 3210] & N/A & 1.81e-16 \\
        R\ref{chem:c6h5_h_m} & \ce{C6H5 + H -> C6H6} & [N/A]; [N/A] & [1.3e-29, 0, 380]; [6.94e-11, 0.15, 0] & N/A & 5.83e-11 \\
        R\ref{chem:c3h2_h_m} & \ce{C3H2 + H -> C3H3} & [N/A]; [N/A] & [4e-19, -3, 600]; [1.26E-10, 0.102, -43.7] & N/A & 5.06e-10 \\
        R\ref{chem:c3h3_h_m} & \ce{C3H3 + H -> C3H4} & [N/A]; [N/A] & [1.6e-20, -3.3, 0]; [3.33e-11, 0.21, -87.1] & N/A & 1.40e-10 \\
        R\ref{chem:ch4_1ch2_m} & \ce{CH4 + CH2(^1A_1) -> C2H6} & [7.2e-24, 0, 0]; [7.1e-11, 0, 0] & [N/A]; [N/A] & 7.10e-11 & N/A \\
        \hline
    \end{tabular}
    {R\ref{chem:ch3_ch3_m}: \citet{baulch_evaluated_1992}; \citet{1983CPL....94..430M}; \citet{baulch_1994JPCRD..23..847B},  
    R\ref{chem:ch2_c2h2_m}: \citet{laufer1983computations}; \citet{polino2013predictive},  
    R\ref{chem:c4h2_h_m}: \citet{klippenstein2005addition},  
    R\ref{chem:c2h2_c4h3_m}: estimated from \citet{2016ApJ...829...66M}; \citet{westmoreland1989forming},  
    R\ref{chem:c6h5_h_m}: estimated from \citet{harding2005predictive},  
    R\ref{chem:c3h2_h_m}: \citet{harding2007combination}; \citet{2016ApJ...829...66M},  
    R\ref{chem:c3h3_h_m}: \citet{harding2007combination}; \citet{moses_disequilibrium_2011},  
    R\ref{chem:ch4_1ch2_m}: \citet{tsang1986chemical}; \citet{pearce_consistent_2019}.\\}
\end{table}

\section{Discussion and Conclusion}\label{sec:disc_conc}
\subsection{Hydrocarbon complexity}\label{disc_subsec:hydrocarbons}
Our study highlights the strong influence of hydrocarbon complexity on the simulated abundances of feedstock molecules in exoplanet atmospheres. The photochemical environment depends critically on which hydrocarbons accumulate and how effectively they provide photochemical shielding. Current photochemical models are restricted by the availability and accuracy of kinetic data for thermochemical reactions and absorption cross-section data for photodissociation reactions. Developing a framework to evaluate their coupled impacts on further chemical processes is therefore essential.

\begin{figure}
\centering
\includegraphics[width=0.32\columnwidth]{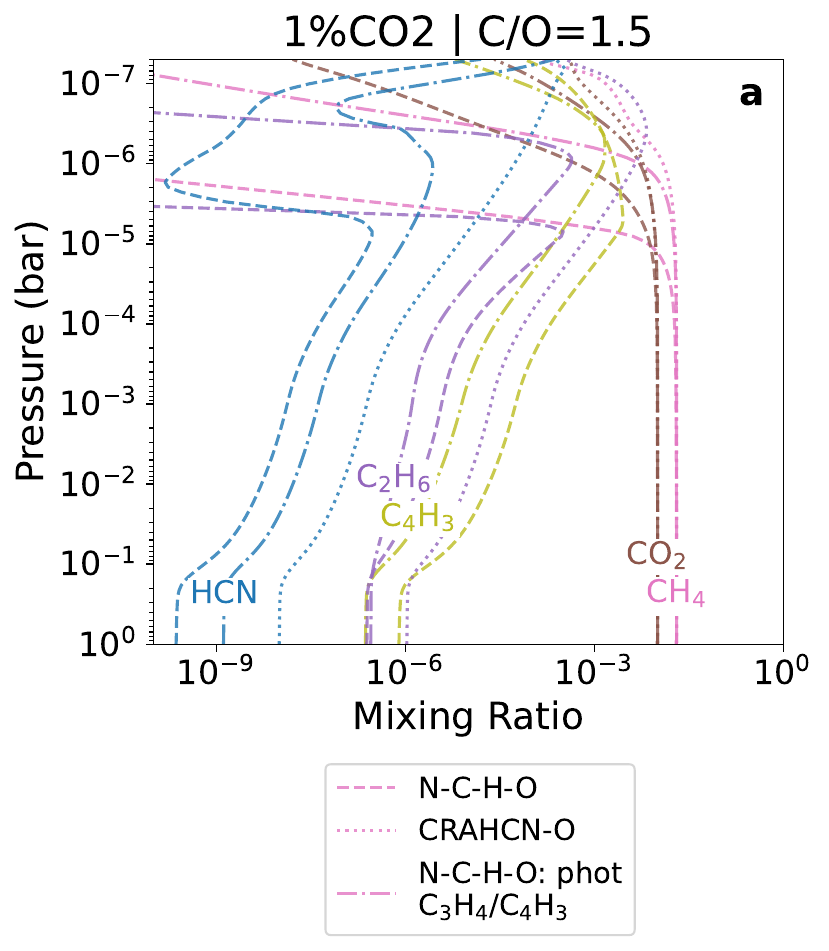}
\includegraphics[width=0.32\columnwidth]{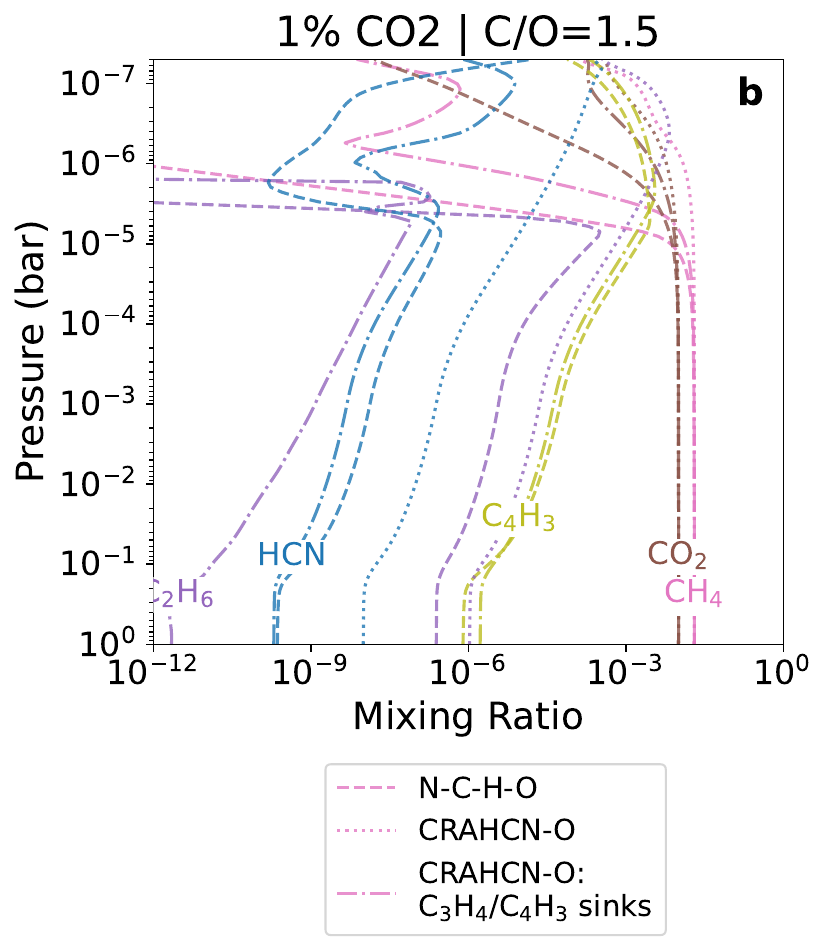}
\includegraphics[width=0.32\columnwidth]{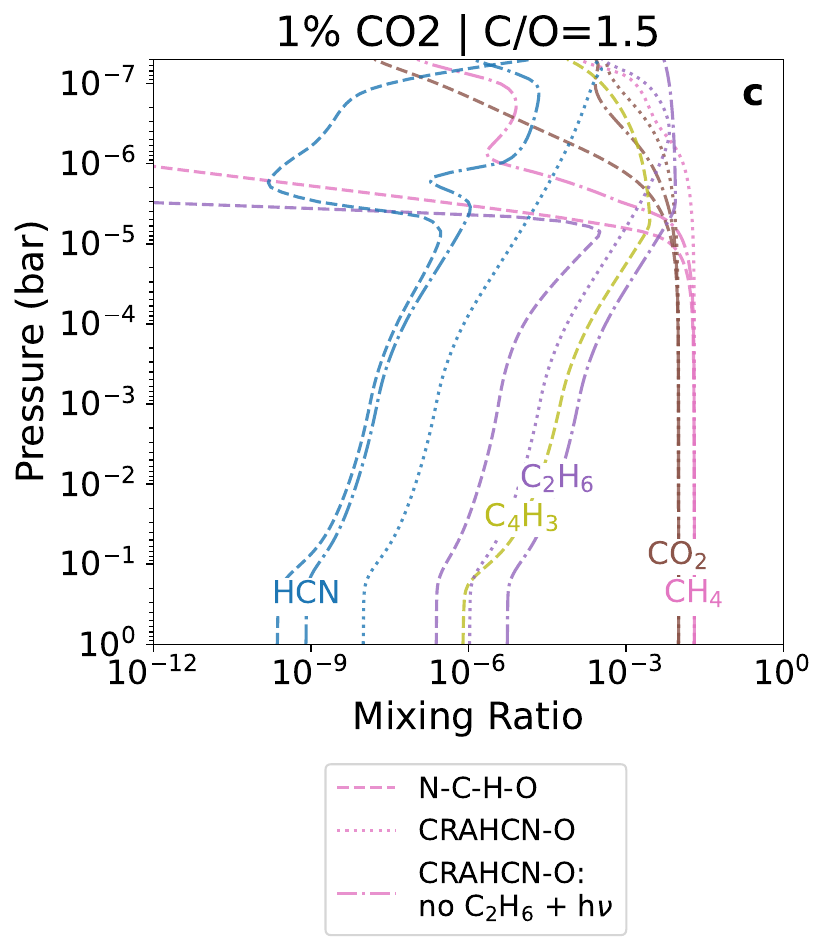}
\caption{Three avenues for reconciliation: (a) photochemically active C$_3$H$_4$ and C$_4$H$_3$ in the N-C-H-O network provide shielding effects, whereas (b) artificial destruction of C$_2$H$_6$ into not photochemically active C$_3$H$_4$ and C$_4$H$_3$ in CRAHCN-O and (c) no photodissociation of C$_2$H$_6$ in CRAHCN-O weaken the shielding effects.}
\label{fig:network_reconciliation}
\end{figure}

We perform three additional tests focused on the shielding effects of hydrocarbons, shown in Figure~\ref{fig:network_reconciliation}. First, we follow \citet{lavvas_coupling_2008} and estimate the photolysis of C$_3$H$_4$ into C$_3$H$_3$ and C$_3$H$_2$, with branching ratios of 0.64 and 0.36, respectively, and cross-section data from \citet{palmer_electronic_1999}. Unable to find cross-section data for C$_4$H$_3$, we used the same cross sections and consider only the destruction into C$_3$H$_3$. The inclusion of C$_3$H$_4$ and C$_4$H$_3$ photolysis adds photochemical shielding to N-C-H-O, protecting CH$_4$ and CO$_2$ abundances to lower pressures and bringing them closer to CRAHCN-O (Figure~\ref{fig:network_reconciliation}a). The HCN profile also moves closer to that in CRAHCN-O. As a second test, we added artificial destruction reactions from C$_2$H$_6$ to C$_3$H$_4$ and C$_4$H$_3$ into CRAHCN-O, without treating the latter two as photochemically active. This decreases the C$_2$H$_6$ shielding effects, bringing photochemical destruction of CH$_4$ and CO$_2$ to higher pressures (Figure~\ref{fig:network_reconciliation}b). Already, this simple artificial C$_2$H$_6$ destruction brings HCN and C$_4$H$_3$ remarkably close to the profiles predicted by N-C-H-O. Lastly, we switch off all photodissociation channels of C$_2$H$_6$ in CRAHCN-O. Again, the HCN profile moves towards that of N-C-H-O and shielding of CH$_4$ and CO$_2$ also visibly weakens (Figure~\ref{fig:network_reconciliation}c). Since C$_4$H$_3$ is a radical species, its accumulation is not expected chemically, highlighting the need to expand on the hydrocarbon chemistry. A comprehensive reconciliation is beyond the scope of this paper, but will be attempted as part of future work that builds upon the key pathways presented here. As the lack of C$_4$H$_3$ cross-section data demonstrates, for many of the higher-order hydrocarbon reactions, we need more measurements of rate coefficients, thermodynamic data (for reversing reactions), and cross sections.

Previous studies have investigated the atmospheres of Titan-like exoplanets (N$_2$-dominated with ${\sim}1.4$\% CH$_4$), including hydrocarbons. \citet{lora_atmospheric_2018} explored both atmospheric circulation and photochemistry, whilst incorporating hydrocarbons up to C$_6$ species. Their predictions of peak C$_2$H$_6$ abundances at $10^{-7}$--$10^{-8}$~bar and increasing HCN abundances \citep[see Figure 6 in][]{lora_atmospheric_2018} agree with CRAHCN-O profiles the 1\% and 10\% CO$_2$ cases. C$_2$H$_6$ abundances from \citet{lora_atmospheric_2018} are considerably lower, but compensated by C$_2$H$_2$ and C$_2$H$_4$ at mixing ratios of $10^{-3}$. We note that our atmospheres are much warmer than the Titan-like exoplanets in \citet{lora_atmospheric_2018, adams_hydrocarbon_2022} and are on the order of early Earth and similar exoplanets \citep[e.g.,][]{arney_pale_2016, arney_pale_2017}. Likely due to the warmer temperatures, reaction rates in our work are 1-2 orders of magnitude faster than in \citet{lora_atmospheric_2018}. We also note that our simulations contain substantial CO$_2$, unlike those of \citet{lora_atmospheric_2018}. Their upper atmospheric boundary is at lower pressures ($10^{-11}$~bar), warranting detailed studies of the effects of upper boundary choices. The pathways to C$_2$H$_6$ are similarly dominated by Reaction~R\ref{chem:ch3_ch3_m}, with Reaction~R\ref{chem:ch3_n} competing to form nitriles instead. \citet{adams_hydrocarbon_2022} expand the study of Titan-like atmospheres around different stars by considering close-in orbits and H$_2$-dominated atmospheres. Similarly, nitriles and hydrocarbons exhibit high mixing ratios in the upper atmosphere for Titan-like conditions, agreeing with our results using CRAHCN-O. For warmer atmospheres and for H$_2$-dominated conditions, CH$_3$ recycles back into CH$_4$ due to reactions with H in the upper atmosphere. Instead, CH$_2$ and CH are the main CH$_4$ sinks, and hydrocarbons do not accumulate at low pressures. At higher pressures, CH$_3$ forms to promote the formation of hydrocarbons and nitriles \citep[][]{adams_hydrocarbon_2022}. This is in better agreement with our N-C-H-O results. 

A hydrocarbon shielding mechanism for Early Earth was discovered by \citet{yoshida_self-shielding_2024}, showing how C$_2$H$_2$ and C$_3$H$_4$ isomers (CH$_2$CCH$_2$, allene and CH$_3$C$_2$H, propyne) suppress H$_2$O photolysis, producing fewer oxidants and thus enhancing the production of feedstock molecules. Their chemical network includes hydrocarbons up to C3 and predicts substantial C$_2$H$_6$ abundances. However, the cross-sections of C$_2$H$_6$ fall below those of CH$_2$CCH$_2$ and CH$_3$C$_2$H \citep[][]{yoshida_self-shielding_2024}, emphasising the potential shielding role of C$_3$H$_4$ isomers. In N-C-H-O, C$_3$H$_4$ represents CH$_2$CCH$_2$, and we also demonstrate its shielding effect on predicted N-C-H-O abundances in Figure~\ref{fig:network_reconciliation}. The added presence of CH$_3$C$_2$H as a photochemically active species in the N-C-H-O network will likely strengthen the photochemical shielding mechanism, making this an important future addition. Moreover, on Titan, CH$_3$C$_2$H has been found at ten times higher abundances than CH$_2$CCH$_2$ \citep{lombardo_detection_2019}, motivating the joint inclusion of both C$_3$H$_4$ isomers. Evidently, any upper limit on hydrocarbon complexity should include a quantification of the potential photochemical impact in the context of reducing atmospheres on rocky exoplanets.

Once reasonably complex hydrocarbons form, we know from Titan that an inevitable consequence is the formation of hydrocarbon hazes in such a reducing environment \citep[e.g.,][]{khare_optical_1984, trainer_organic_2006, tomasko_model_2008, lavvas_coupling_2008, horst_titans_2017, nixon_composition_2024}. Absorption of UV photons in Titan's atmosphere is dominated by CH$_4$ at wavelengths ${<}145$~nm, by C$_2$H$_2$ and C$_2$H$_4$ until $180$~nm, and by hazes beyond 180~nm \citep[][]{vuitton_simulating_2019}, although the expected size, composition, and optical properties of hazes can vary significantly \citep[e.g.,][]{he_photochemical_2018}. We similarly expect potential shielding by hazes to affect the climate and chemistry of the reducing environment of Early Earth \citep[e.g.,][]{clarke_chemical_1997, trainer_organic_2006, domagal-goldman_organic_2008}. Examples include enhanced NH$_3$ abundances due to hydrocarbon haze shielding that potentially explain the Faint Young Sun paradox \citep[][]{sagan_early_1997, pavlov_uv_2001, wolf_fractal_2010}. \citet{arney_pale_2016} use a coupled 1D climate-chemistry model to explore the haze formation process and its climate impacts, showing that hazes can cool the surface of the Early Earth and provide UV shielding effects. However, due to self-shielding by hazes, catastrophic cooling of the planet is not predicted. For M-stars, the cooling effects are reduced, but a significant haze layer forms in the upper atmosphere with UV shielding effects \citep[][]{arney_pale_2017}. As both previous studies predict haze formation for atmospheric temperatures similar to ours \citep{arney_pale_2016, arney_pale_2017}, we also expect haze production. However, haze formation here is parameterised based on gaseous hydrocarbon abundances, following approaches used for Titan \citep{lavvas_coupling_2008} and Titan-like exoplanets \citep{lora_atmospheric_2018, adams_hydrocarbon_2022}. 3D climate simulations using a prescribed haze layer confirm the UV shielding effects as well as the strong decrease of the surface temperature for Early Earth and a tidally locked exoplanet \citep[][]{mak_3d_2023, mak_3d_2024}, along with potential circulation regime changes for the latter. 

\citet{lora_atmospheric_2018} also parameterise haze production rates as combined rates of the reactions forming hydrocarbons with 10 or more C atoms, following \citet{lavvas_coupling_2008} for Titan. For Titan-like exoplanets around M-dwarfs, significant haze formation rates are predicted between $10^{-3}$ and $10^{-10}$~bar. For these pressures, hazes might actually provide shielding. However, further work should explore haze formation, their persistence (compared to advection, settling, rainout), and their potential shielding effects in (rocky) exoplanet settings. The recent detections of CH$_4$ mixing ratios at percent levels in sub-Neptune atmospheres \citep[e.g.,][]{madhusudhan_carbon-bearing_2023, schmidt_comprehensive_2025}, along with potential hints of hydrocarbons such as C$_3$H$_4$ \citep[][]{welbanks_challenges_2025, pica-ciamarra_systematic_2025}, highlights the importance of understanding hydrocarbon chemistry in the appropriate environmental settings and warrants further studies in warmer and hydrogen-rich atmospheric settings \citep[e.g.,][]{adams_hydrocarbon_2022, cooke_considerations_2024}.

\subsection{Considerations for prebiotic chemical processes}
In the context of rocky planets, \citet{rimmer_hydrogen_2019} use the ARGO/STAND2020 photochemical kinetics model to investigate how HCN formation depends on C/O ratio and the particular atmospheric composition, reporting HCN mixing ratios varying between 1--1000 ppm for 0.5${<}$C/O${\leq}$1.5. Our distribution of mixing ratios for C/O${>}0.5$ in Figure~\ref{fig:species_avg_mixrat_all} with CRAHCN-O similarly ranges between 0.1--1000~ppm, depending on both the CO$_2$ background case and the C/O ratio. N-C-H-O predictions remain at the lower end of this range (0.05--3~ppm) and, for a given C/O, are relatively insensitive to the availability of CH$_4$ at the order of hundreds of ppm, per cent, or tens of per cent. A comparison with ARGO/STAND2020 in Appendix~\ref{app:ARGO_comp} confirms that HCN abundances in ARGO are similar to those with CRAHCN-O and sometimes even exceed them. The photochemical shielding effects are less pronounced, although this might also be due to ion-neutral chemistry in the upper atmosphere with ARGO \citep[][]{rimmer_chemical_2016}. The impact of stellar radiation on HCN photochemistry is also summarised by \citet{rimmer_hydrogen_2019}: 50--100~nm and 100--200~nm photons drive HCN production (N$_2$ photolysis) and destruction (H$_2$O photolysis), respectively. In our study, N$_2$ photolysis is relatively unaffected by C$_2$H$_6$ shielding in CRAHCN-O, but H$_2$O and CO$_2$ shielding limit the oxidising capacity of the upper atmosphere and thus allow HCN to form more abundantly. In a separate study, we further explore how the physical parameter space of rocky exoplanets affects atmospheric synthesis of HCN and subsequent rainout \citep[][]{friss_unique_2025}. Using the N-C-H-O network, we confirm trends with C/O and show how semi-major axes and host star type affect prebiotic synthesis. Once HCN accumulates in a shallow pond at the surface, a model of aqueous chemistry can simulate the formation of more complex prebiotic molecules \citep[e.g.,][]{pearce_towards_2022}. Since the prebiotic complexity achieved is dictated by UV irradiation \citep[][]{ranjan_influence_2016}, photochemical shielding is an important consideration. \citet{ranjan_influence_2016} demonstrated photochemical shielding of feedstock molecules like CH$_4$ and HCN by atmospheric CO$_2$. Evidently, various hydrocarbons can potentially provide similar shielding mechanisms \citep[][and this work]{yoshida_self-shielding_2024}. Understanding such shielding mechanisms in the coupled physical and (photo)chemical environments is crucial to predicting the potential for prebiotic synthesis.


Nevertheless, planetary atmospheres are inherently 3D, with several physical and chemical implications for the distribution of feedstock molecules that are not captured in 1D. This study and the related work presented in \citet{friss_unique_2025} pave the way for an implementation of prebiotic chemistry in a 3D climate-chemistry model, complex models that are currently limited by the number of reactions they can feasibly simulate. By identifying the key pathways in the specific environmental setting, as in Section~\ref{sec:pathways}, we demonstrate which reactions are crucial for a given chemical compound. Furthermore, the approximation of these pathways by net reactions such as shown for hydrocarbons in Section~\ref{subsec:pathways_hydrocarbons}, allows for a considerable reduction of the complexity of a chemical network \citep[][]{tsai_mini-chemical_2022, lee_mini-chemical_2023}. The gas-phase chemistry of these hydrocarbons can be used to construct a chemically consistent haze formation mechanism \citep[e.g.,][]{arney_pale_2016} to further explore the 3D effects of radiatively active hazes \citep[][]{mak_3d_2023, mak_3d_2024}. 3D models accurately capture the effects of spatial and temporal variations in stellar radiation and climate from orbital configurations \citep[][]{linsenmeier_climate_2015, turbet2016habitability, boutle_exploring_2017, del_genio_habitable_2019}, associated chemical and wet-dry cycles \citep{braam_earth-like_2025}, and the effects of energetic events such as stellar flares \citep[][]{chen_persistence_2021, ridgway_3d_2023} and lightning \citep[][]{braam_lightning-induced_2022, sergeev_lightning_2025}, all impacting the potential production of feedstock molecules. We note that results on feedstock molecules as well as hydrocarbon shielding might change when accounting for stellar evolution \citep[e.g.,][]{claire_evolution_2012}. Future work will be directed to simulations of prebiotic chemistry that account for the evolution of stellar UV emission from M stars \citep[e.g.,][]{shkolnik_hazmat_2014, peacock_hazmat_2020}.

\subsection{Chemical network recommendations}
In Section~\ref{sec:pathways}, we present the key pathways to important feedstock molecules (HCN, HC$_3$N, H$_2$CO, and hydrocarbons), comparing two independently constructed chemical networks. Key bimolecular reactions are listed and compared in Table~\ref{tab:key_reactions_pathways}, along with a recommendation on using the rate coefficient from CRAHCN-O (C), N-C-H-O (N), or further rate coefficient determination (M). The recommendations are based on a comparison to the reported rate coefficient measurements. For the key termolecular reactions in Table~\ref{tab:key_reactions_pathways_termol}, Reaction~R\ref{chem:ch3_ch3_m} is well-constrained by multiple independent measurements, whereas the other hydrocarbon reactions in Table~\ref{tab:key_reactions_pathways_termol} are typically based on a single experimental measurement, thus justifying further rate coefficient determinations. As noted in Section~\ref{disc_subsec:hydrocarbons}, the pathways and key reactions can be used to reconcile chemical network differences. In what follows, we discuss recommended network improvements, focusing on feedstocks and hydrocarbons, as informed by the key pathways, shielding effects, and rate coefficient comparisons of this work.

A fundamental difference in HCN formation is its dependence on HNCO through Reactions~R\ref{chem:cn_co2}--R\ref{chem:nco_hco}, present in the N-C-H-O but not in the CRAHCN-O network. The role of HNCO in hot Jupiter atmospheres is also discussed by \citet{moses_disequilibrium_2011} and \citet{tsai_comparative_2021}. The HNCO rate coefficients in Tables~\ref{tab:key_reactions_pathways} are valid for temperatures exceeding the atmospheric conditions in this work: 1830--2400~K for R\ref{chem:cn_co2}, 500--2500~K for R\ref{chem:hnco_cn} and R\ref{chem:nco_hco}, and 300--3300~K for R\ref{chem:nco_h2}. This extrapolation to low temperatures also applies to many of the rate coefficients normally used for Titan \citep[e.g.,][]{lavvas_coupling_2008}. Whilst the ranges for R\ref{chem:nco_hco}, R\ref{chem:hnco_cn}, and R\ref{chem:nco_h2} may be acceptable, our simulations are far from the valid temperature range of R\ref{chem:cn_co2}, which also lacks temperature dependence in its rate coefficient in N-C-H-O (as A=0). Therefore, an inclusion of HCN production through the HNCO reactions requires a low-temperature constraint of the rate coefficients for Reactions~R\ref{chem:cn_co2}--R\ref{chem:nco_hco}, in particular R\ref{chem:cn_co2}. The coefficients for the other important HCN pathway, Reaction~R\ref{chem:n_ch2}, agree at 298~K (Table~\ref{tab:key_reactions_pathways}), but do decrease for lower temperatures in N-C-H-O. However, as the N-C-H-O rate coefficient is an estimate based on the \ce{N + CH3 -> HCN + H2} coefficients from \citet{marston_temperature_1989}, we recommend additional measurements. CRAHCN-O has a fast pathway to HCN with H$_2$CO as an intermediate species via Reaction~R\ref{chem:h2co_cn}. Since the rate coefficient of Reaction~R\ref{chem:h2co_cn} in CRAHCN-O originates from a single experimental determination, we recommend additional measurements. Since HCN and HNC are one of the two known isomer pairs in Titan's atmosphere \citep{moreno_first_2011}, HNC chemistry should be added to the N-C-H-O network. 

HC$_3$N formation is relatively fast in N-C-H-O, due to its dependence on hydrocarbons such as C$_2$H (R\ref{chem:hcn_c2h}), C$_2$H$_2$ (R\ref{chem:cn_c2h2}), and C$_3$H$_3$ (R\ref{chem:n_c3h3}). However, the rate coefficients for Reactions~R\ref{chem:hcn_c2h} and R\ref{chem:n_c3h3} need further validation. In CRAHCN-O, HC$_3$N formation is much slower due to low hydrocarbon abundances and the missing Reaction~R\ref{chem:n_c3h3}. Since HC$_3$N plays an important role in several steps of prebiotic synthesis \citep{powner2007prebiotic, powner_synthesis_2009, patel_common_2015}, we recommend the inclusion of additional HC$_3$N pathways in CRAHCN-O, thus also motivating a more comprehensive treatment of C$_2$H$_2$ and C$_3$H$_3$. The measured abundances of HC$_3$N on Titan can be used to validate the pathways \citep{thelen_abundance_2019}. 

The higher H$_2$CO abundances in CRAHCN-O follow from fast HCO production via Reaction~R\ref{chem:co_h_h}, which quickly self-reacts (R\ref{chem:hco_hco}) to form H$_2$CO in CRAHCN-O. We note that for Reaction~R\ref{chem:co_h_h}, \citet{pearce_crahcn-o_2020} present the rate coefficients for the forward reaction \ce{HCO + H -> CO + H + H}, which is reversed in VULCAN using the equilibrium constant \citep{tsai_vulcan_2017}. As such, additional constraints on the rate coefficients of the reverse reaction are valuable. At high pressures for N-C-H-O, H$_2$CO production through Reaction~\ref{chem:ch2_co2} follows rate coefficients in the database presented by \citet{tsang1986chemical}, resulting in enhanced H$_2$CO abundances compared to CRAHCN-O. However, the single experimental determination for Reaction~\ref{chem:ch2_co2} motivates additional measurements.

The treatment of hydrocarbons in N-C-H-O generally focuses on C2 hydrocarbons. Higher-order hydrocarbons were included in a simplified mechanism to form benzene as a haze precursor \citep{tsai_comparative_2021}. Purposefully, this includes only the main formation pathways to benzene to limit the network size, although this is also due to limitations of the available kinetics data \citep{venot_new_2015}. Among the key bimolecular reactions in the hydrocarbon pathways, Reactions~R\ref{chem:ch_c2h2}, R\ref{chem:c2h_c2h2}, R\ref{chem:ch_ch4}, and R\ref{chem:ch3_ch2} are relatively well-constrained by multiple measurements and/or reviews across the relevant temperature range. The destruction of C$_3$H$_4$ via Reaction~R\ref{chem:c3h4_h} has several measured sets of rate coefficients, but no dedicated review, and is generally reported for temperatures exceeding 600~K. The termolecular reactions in Table~\ref{tab:key_reactions_pathways_termol} are usually based on a single measurement (Reactions~R\ref{chem:ch2_c2h2_m} and R\ref{chem:c4h2_h_m}) or estimated from other reactions (Reaction~R\ref{chem:c2h2_c4h3_m}, R\ref{chem:c6h5_h_m}, R\ref{chem:c3h2_h_m}, R\ref{chem:c3h3_h_m}, R\ref{chem:ch4_1ch2_m}), motivating additional rate coefficient determination. Due to the dominance of C$_2$H$_6$ formation in CRAHCN-O by Reactions~R\ref{chem:c2h4_h2_h} and R\ref{chem:c2h2_ch4_h}, we suggest further determination of the reverse reactions. Since C3 and C4 molecules are likely also photochemically active, we recommend including their cross sections in N-C-H-O (like for C$_3$H$_4$ in Section~\ref{disc_subsec:hydrocarbons}). Since allene and propyne are the other known isomer pair in Titan's atmosphere \citep{lombardo_detection_2019}, the networks have to be expanded with an isomeric treatment \citep{hebrard_photochemistry_2013, vuitton_simulating_2019}. Ultimately, the hydrocarbon partitioning should be determined by a more complete expansion of hydrocarbon interactions with other species and, eventually, by parameterising haze formation \citep[e.g.][]{lavvas_coupling_2008, lora_atmospheric_2018, adams_hydrocarbon_2022}. Based on this, any simplified hydrocarbon scheme should include an assessment of the photochemical effects.

An alternative approach to quantify network sensitivity and efficient network reduction is to apply a genetic algorithm such as DARWEN \citep{lira-barria_darwen_2024}. In our VULCAN configurations, the chemistry is mainly driven by the photodissociation reactions. As we know from Titan, energetic particles and cosmic rays also drive a plethora of photochemical processes and the comparison to ARGO hints at ion-neutral effects (see Section~\ref{app:ARGO_comp}). Future work can build upon the wealth of knowledge in models of Titan photochemistry \citep{loison_neutral_2015, dobrijevic_1d-coupled_2016, vuitton_simulating_2019} and consider the effect of the upper boundary, considering photodissociation, electron impacts, and ion-neutral interactions. Nevertheless, the model predictions for any chemical networks will have to be tested against lab experiments, quantum chemical calculations, and observations of planetary environments. For the networks considered here, N-C-H-O was benchmarked against Earth, Jupiter, and previous models of hot Jupiters \citep[][]{tsai_comparative_2021}. For Earth, the network reproduces the main photochemical cycles and also validates VULCAN’s inclusion of boundary conditions and H$_2$O condensation. Jupiter observations are used to validate the hydrocarbon predictions, providing good agreement for CH$_4$ and C2 hydrocarbons. However, \citet{tsai_comparative_2021} note uncertainties in the rate constants for hydrocarbon reactions at temperatures below 200~K and in the higher-order hydrocarbons, notably C$_4$H$_2$ and C$_6$H$_6$. The hot Jupiter results are compared to earlier model results by \citet{moses_disequilibrium_2011} and \citet{venot_chemical_2012}, validating the effects of photochemistry and vertical mixing. CRAHCN-O was benchmarked against HCN on Titan \citep[][]{pearce_crahcn-o_2020}, but does not reproduce other small prebiotic feedstocks or hydrocarbons like the Titan photochemical models commonly do \citep{loison_neutral_2015, dobrijevic_1d-coupled_2016, willacy_new_2016, vuitton_simulating_2019}. \citet{pearce_hcn_2020} show how the CRAHCN-O predictions for the lower atmosphere of Titan fall in the middle of the Cassini measurements and also match the in-situ upper atmosphere Cassini INMS data point. However, like most photochemical models, it fails to reproduce the observed variability in upper-atmospheric HCN profiles from Cassini and ground-based measurements \citep[see also][]{lellouch_intense_2019, thelen_variability_2022}. Accurately modelling this variability may require higher-dimensional photochemical models to account for seasonal and dynamical effects \citep{de_la_haye_coupled_2008, dobrijevic_one_2021}. The simulated network differences illustrate the environmental dependence: benchmarking in one scenario does not inherently imply agreement in another. Moreover, beyond differences in chemical networks, the environmental conditions ultimately determine the chemical pathways. Future plans include a Titan configuration in VULCAN, likely based on the CRAHCN-O chemical network. In turn, this should also advance further modelling of reducing atmospheric conditions on exoplanets.

Investigating prebiotic chemistry in exoplanet atmospheres ultimately feeds into finding the likelihood of certain planetary conditions leading to a specific set of feedstock molecules \citep[][]{sutherland_opinion_2017, benner_prebiotic_2019, rimmer_origins_2023, white_-nothing_2025}. The dependence on boundary conditions and the likelihood of specific reactions occurring in sequence necessitate the use of photochemical kinetics models with associated chemical networks. However, a chemical network is more an expression of our knowledge of the system than a complete representation of reality \citep[][]{white_-nothing_2025}, emphasising the importance of network comparisons for distinct environments. Associated pathway analyses, such as in Section~\ref{sec:pathways}, can quantify the likelihood of specific reactions occurring in sequence and help determine whether a network is comprehensive enough to represent the prebiotic scenario of interest in the environmental conditions. Understanding net reactions and chemical mechanisms allows for network comparison and guides explorations of the chemical processes in more complex atmospheric models \citep[][]{tsai_mini-chemical_2022, lee_mini-chemical_2023}. By determining the most important reactions per network, we identify those that may require further experimental rate constant determination and the conditions under which these measurements would be most useful. For example, our results emphasise the crucial role of hydrocarbon reactions in predicting the production of feedstock molecules for M-star radiation. Both the net and important reactions provide a base for connecting global environmental properties to a given prebiotic scenario and, specifically for exoplanets, quantify the so-called Abiogenesis Zone based on known exoplanet properties such as stellar irradiation and orbital distance \citep[][]{rimmer_origin_2018, rimmer_origins_2023}. Ultimately, this knowledge of chemical kinetics will help us understand whether the emergence of life is more or less likely in a given planetary environment. 

\subsection{Conclusions}
In summary, we present the implementation of the CRAHCN-O chemical network into the VULCAN 1D photochemical kinetics code \citep[][]{tsai_comparative_2021}. CRAHCN-O was specifically constructed to accurately simulate HCN and H$_2$CO, simple hydrocarbons, NH$_3$, and other small organic species in various atmospheric compositions at temperatures of 50--400~K \citep[][]{pearce_crahcn-o_2020}. We test the production of prebiotic feedstock molecules driven by irradiation from M-star hosts and compare the results to predictions by the N-C-H-O network in VULCAN. Various initial atmospheric compositions for the simulations cover C/O ratios of 0.5--1.5 and C/N ratios between $2{\times}10^{-4}$--$2{\times}10^{-1}$. We find agreement between the vertical distributions of mixing ratios in the absence of CH$_4$ for C/O=0.5, but diverging distributions when CH$_4$ is introduced for C/O${>}$0.5. Inter-network differences in hydrocarbon complexity and subsequent photochemical shielding drive the diverging mixing ratio profiles. The most complex hydrocarbon in CRAHCN-O is C$_2$H$_6$, which accumulates in the upper atmosphere. Importantly, C$_2$H$_6$ has known photochemical shielding effects, and its high abundance shields CH$_4$ and CO$_2$ from photodissociation, allowing their existence at lower pressures and reducing the production of reactive oxygen species that destroy HCN and H$_2$CO. The N-C-H-O network simulates more complex hydrocarbons, including C$_4$H$_3$, C$_3$H$_4$, and C$_6$H$_6$, that instead accumulate in the upper atmosphere. However, they are not photochemically active and are thus unable to provide the same photochemical shielding. Other feedstock molecules, such as HC$_3$N and C$_2$H$_2$, form much more efficiently in N-C-H-O. The shielding mechanism and weak formation of HC$_3$N and C$_2$H$_2$ persist across M-star types. We determine the key pathways to feedstock molecules in both networks and derive formation mechanisms to explain the distinct hydrocarbon profiles and shielding effects. The results demonstrate the crucial role of chemical kinetics in understanding prebiotic chemistry in exoplanet atmospheres, including important considerations for constructing chemical networks. In a related paper, we explore how the atmospheric feedstock molecules are transferred to the planetary surface \citep[][]{friss_unique_2025}, where aqueous chemistry can drive further complexity. We identify several paths forward: to expand the hydrocarbon complexity and their photochemical properties, to test the effect of hazes in the shielding scenario, and to use the identified pathways and net reactions to explore prebiotic chemistry in 3D climate-chemistry models. Lastly, we discuss recommended network improvements, based on the network comparison and known mechanisms of importance for Titan's atmospheric chemistry.

\section*{Acknowledgements}
MB thanks Arturo Lira-Barria, Jeremy Harvey, and Susanne Wampfler for helpful discussions about rate coefficient uncertainties, chemical mechanisms, and previous astrochemical network analyses. We thank Ben K. D. Pearce and the anonymous reviewer whose comments helped to significantly improve the manuscript. MB appreciates support from a CSH Fellowship. Part of this work has been carried out within the framework of the NCCR PlanetS supported by the Swiss National Science Foundation under grants 51NF40\_182901 and 51NF40\_205606. EG acknowledges a Summer Internship supported by the Centre for Exoplanet Science, University of Edinburgh. GF acknowledges his studentship by the Science and Technology Facilities Council (project reference: 2902875). PIP acknowledges funding from the STFC consolidator grant ST/V000594/1. MB and PIP were part of the CHAMELEON MC ITN EJD, which received funding from the European Union's Horizon 2020 research and innovation program under the Marie Sklodowska-Curie grant agreement No. 860470. 

\appendix

\section{ARGO/STAND2020 Comparison}\label{app:ARGO_comp}
To further expand our comparison of chemical networks in relatively unexplored environments, we conduct an additional set of six photochemical simulations using ARGO -- a Lagrangian code that solves the photochemistry-transport equation -- and the extensive STAND2020 chemical network that includes over 6000 reactions for 480 species \citep[][]{rimmer_chemical_2016, 2021PSJ.....2..133R}. Our simulation sets include the 1\% and 10\% CO$_2$ scenarios for C/O ratios of 0.5, 1.0, and 1.5. Except for the photochemical kinetics solver and the chemical network, the initialisation (Table~\ref{tab:merged_comp}) and boundary conditions (Figure~\ref{methods-fig:pcb_ptkzz_flux}) remain the same. For the ARGO simulations in this comparison, the surface mixing ratio of H$_2$O is not fixed at 0.00894 but instead initialised as 1~ppm on all vertical levels.

\begin{figure}
        \centering
        \includegraphics[width=0.6\textwidth]{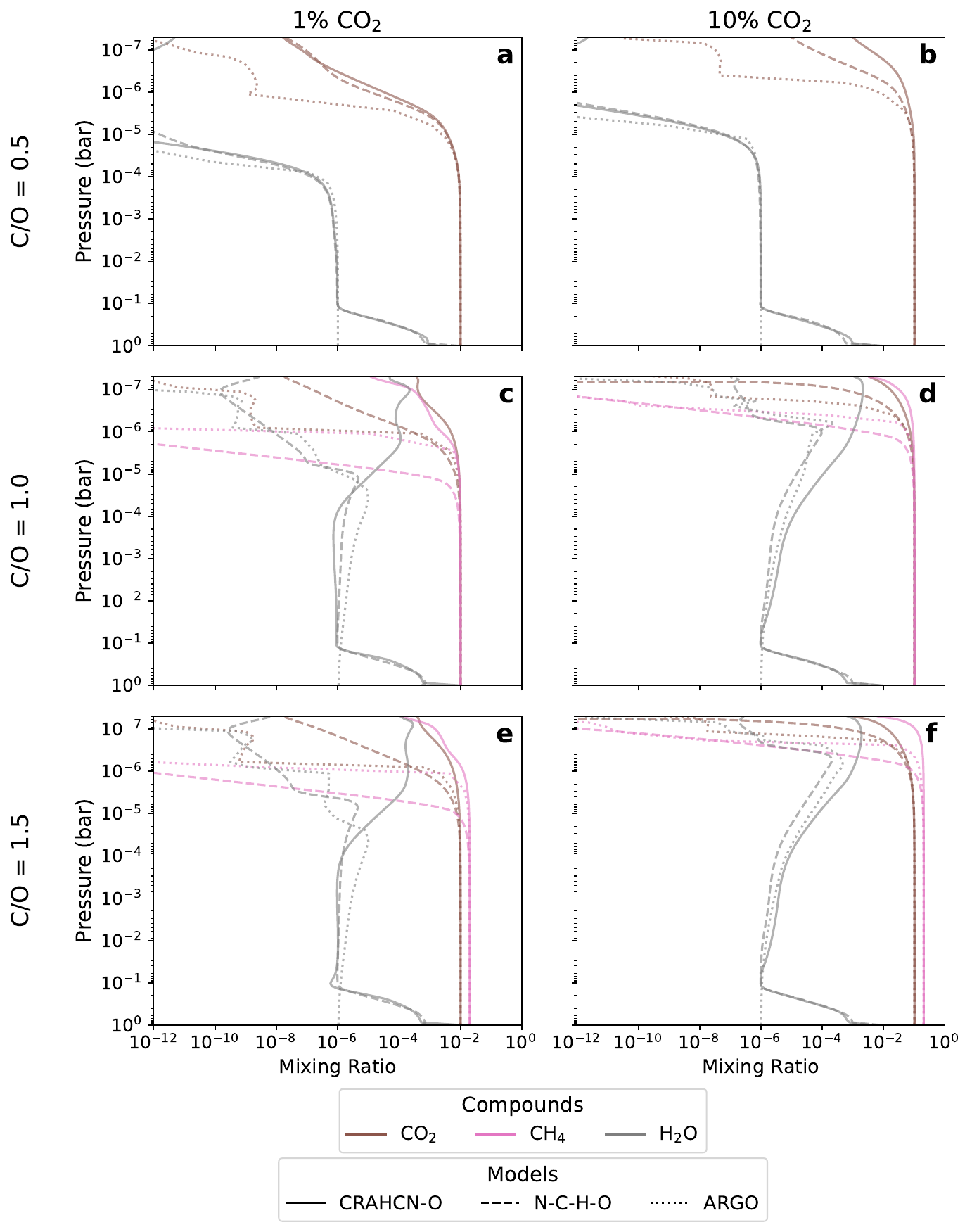}
    \caption{Grid comparing the chemical abundance profiles of initialisation species versus atmospheric pressure for the 1~\% (left), and 10\% (right) CO$_2$ cases to predictions made by ARGO/STAND2021 code. The horizontal rows correspond to C/O=0.5, 1.0, and 1.5, respectively, whereas solid lines correspond to the VULCAN/CRAHCN-O, dashed lines to the VULCAN/N-C-H-O, and dotted lines to ARGO/STAND2021 simulations. }
    \label{fig:ncho_crahcno_argo_init}
\end{figure}

Figure~\ref{fig:ncho_crahcno_argo_init} shows the distribution of initialisation species CH$_4$, CO$_2$, and H$_2$O after each of the three models (VULCAN/CRAHCN-O, VULCAN/N-C-H-O, ARGO/STAND2021) has reached steady state. For C/O=0.5, we again see good agreement between models for CO$_2$ and H$_2$O. Destruction of CO$_2$ in ARGO is stronger for P${<}10^{-6}$~bar, most likely as a consequence of the inclusion of ionisation and resulting ion-neutral reactions in the network. Ionisation and ion-neutral reactions normally drive chemical processes at much lower pressures (to as low as ${\sim}10^{-12}$~bar). Due to our upper boundary at $5{\times}10^{-8}$~bar, the ionisation processes are instead simulated at higher pressures, affecting the predicted vertical profiles in the photochemical region as seen for the CO$_2$ destruction. Moving to higher C/O ratios, we see again that CO$_2$ destruction is more efficient than in both networks in VULCAN. H$_2$O is in good agreement with N-C-H-O. Notably, Figures~\ref{fig:ncho_crahcno_argo_init}c--f, we see that CH$_4$ is stable to lower pressures using ARGO compared to N-C-H-O, potentially pointing to differences in hydrocarbons and photochemical shielding. In no simulation is CH$_4$ shielding as efficient as the predictions by CRAHCN-O.

\begin{figure}
        \centering
        \includegraphics[width=0.9\textwidth]{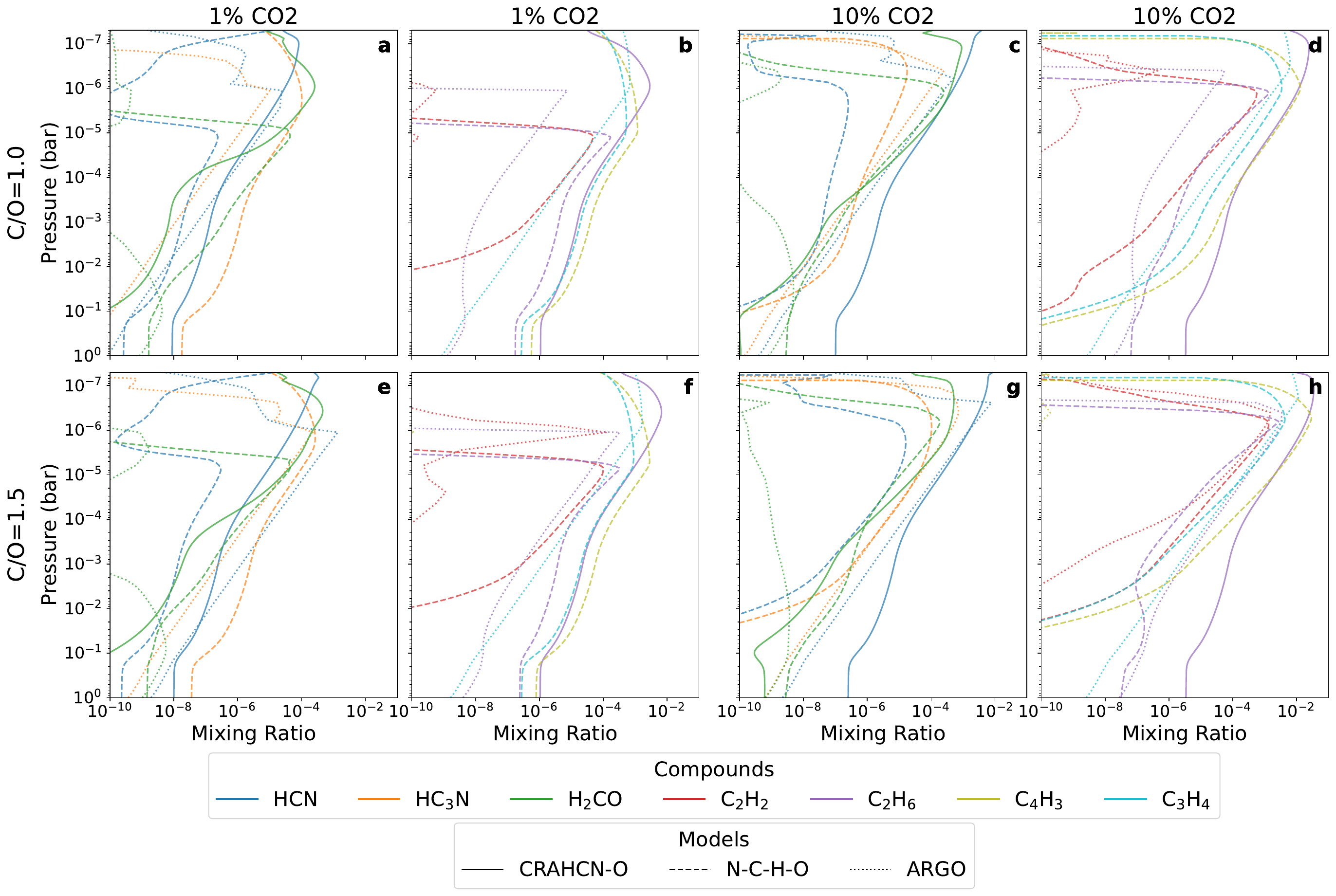}
    \caption{Grid comparing the vertical profiles of feedstock molecules for the 1~\% (panels a, b, e, f), and 10\% (panels c, d, g, h) CO$_2$ cases to predictions made by ARGO/STAND2021 photochemical kinetics code. For visual clarity, we separate HCN, HC$_3$N, and H$_2$CO in the left column for each scenario and C$_2$H$_2$, C$_2$H$_6$, C$_3$H$_4$, and C$_4$H$_3$ in the right column. The horizontal rows correspond to C/O=1.0 and 1.5, respectively, as C/O=0.5 barely produces these feedstock molecules (see Figure~\ref{fig:all_vert_distribs}). Solid lines correspond to the VULCAN/CRAHCN-O, dashed lines to the VULCAN/N-C-H-O, and dotted lines to ARGO/STAND2021 simulations.}
    \label{fig:ncho_crahcno_argo_prebiotic}
\end{figure}

We separately plot the feedstock molecules and a selection of hydrocarbons (C$_2$H$_2$, C$_2$H$_6$, C$_3$H$_4$, C$_4$H$_3$) for C/O=1.0 and 1.5 in Figure~\ref{fig:ncho_crahcno_argo_prebiotic}. Panels a, b, e, and f correspond to the 1\%~CO$_2$ and panels c, d, g, and h to the 10\%~CO$_2$ cases, respectively. We omit the results for C/O=0.5 here because of the negligible feedstock and hydrocarbon abundances (like in Figure~\ref{fig:all_vert_distribs}a and b). Especially for the feedstock molecules (Figures~\ref{fig:ncho_crahcno_argo_prebiotic}a,e for 1\%~CO$_2$ and Figures~\ref{fig:ncho_crahcno_argo_prebiotic}c,g for 10\%~CO$_2$), the ARGO comparison adds further ambiguity. Whilst HCN in ARGO peaks at abundances similar to and occasionally higher than CRAHCN-O (Figure~\ref{fig:ncho_crahcno_argo_prebiotic}e), the HC$_3$N dotted agrees more with the N-C-H-O predictions. H$_2$CO abundances are considerably lower than either of the two networks in VULCAN, hinting at slower production or additional destruction pathways in ARGO. 

In terms of hydrocarbons, ARGO predicts lower abundances for C$_2$H$_2$ than N-C-H-O in a spiky profile (Figures~\ref{fig:ncho_crahcno_argo_prebiotic}b, d, f) but reasonably similar abundances for the C/O=1.5 case at 10\% CO$_2$ (Figure~\ref{fig:ncho_crahcno_argo_prebiotic}h). C$_2$H$_6$ abundances with ARGO are also two to three orders of magnitude lower than N-C-H-O at C/O=1.0  (Figure~\ref{fig:ncho_crahcno_argo_prebiotic}b, d). The C$_2$H$_6$ profile in ARGO also goes down to lower pressures, and its shielding effects lift HCN profiles (Figure~\ref{fig:ncho_crahcno_argo_prebiotic}a, b, c, g) and the regions subject to CH$_4$ photochemistry (Figure~\ref{fig:ncho_crahcno_argo_init}) to lower pressures. In Figure~\ref{fig:ncho_crahcno_argo_prebiotic}h, we see that N-C-H-O and ARGO also predict comparable C$_2$H$_6$ profiles. C$_3$H$_4$ and C$_4$H$_3$ are included in this comparison as sinks for lower-order hydrocarbons that are notably not photochemically active in VULCAN. Peak abundances of C$_3$H$_4$ in ARGO are similar to N-C-H-O, but these abundance peaks go up to the top boundary of the atmosphere. Also, like N-C-H-O, C$_3$H$_4$ is not treated as photochemically active in ARGO. C$_4$H$_3$ is predicted at low (${<}10^{-10}$) mixing ratios in ARGO. Therefore, in ARGO, we also suspect hydrocarbon accumulation in C$_3$H$_4$ without providing a photochemical shielding mechanism.

The comparison highlights some important discussion points. First, networks that have been benchmarked for one set of environmental conditions can predict differences for another set of environmental conditions. Second, the effect of neutral-ion chemistry in ARGO illustrates the importance of the correct upper boundary given the processes considered, e.g. to as low as ${\sim}10^{-12}$~bar when neutral-ion chemistry is included. Third, hydrocarbon complexity in a chemical network and its potential shielding effects are key to predicting the production of feedstock molecules. Finally, as different panels in Figures~\ref{fig:all_vert_distribs}, \ref{fig:ncho_crahcno_argo_init}, and \ref{fig:ncho_crahcno_argo_prebiotic} demonstrate, exact abundance differences due to chemical networks are also sensitive to the chosen boundary and initial conditions.

\printcredits


\bibliography{prebiotic.bib}




\end{document}